\documentclass[a4paper,11pt]{article}
\usepackage{jheppub} 
 \usepackage{amssymb,amsthm}
 \usepackage{amsmath,amssymb,amsfonts,bm,amscd}
 \usepackage{graphicx}
 
 \usepackage{caption, subcaption}
 
 \usepackage{tikz}
 
 \usepackage{float}
 \restylefloat{table}

\def\CB{{\cal B}}

\def\CI{{\cal I}}

\def\CO{{\cal O}}

\def\tr{\mathop{\rm tr}}

\def\beq#1\eeq{\begin{align}#1\end{align}}

\usepackage{enumitem}

\def\fg{\mathfrak{g}}

\def\fsl{\mathfrak{sl}}
\def\fsu{\mathfrak{su}}

\def\half{\frac{1}{2}}

\def\ch{\mathrm{ch}}

\def\bbC{\mathbb{C}}

\newcommand{\pe}[1]{\mathrm{PE}\left[#1\right]}

\title{4d $\mathcal{N}=2$ SCFTs and  lisse W-algebras}

\author[a,b]{Dan Xie}
\author[a]{ Wenbin Yan}

\affiliation[a]{Yau Mathematics Science center, Tsinghua University, Beijing, 10084, China}
\affiliation[b]{Department of Mathematics, Tsinghua University, Beijing, 10084, China}

\abstract{We continue our studies of the correspondence between  4d $\mathcal{N}=2$ SCFTs and 2d W-algebras. The purpose of this paper is to study the relationship between 2d \textbf{lisse} W-algebras and their 4d SCFT partners. 
The lisse  W-algebra is the  W-algebra  whose associated Zhu's $C_2$ algebra is finite dimensional. As the associated variety of Zhu's $C_2$ algebra is identified with the 
Higgs branch in the 4d/2d correspondence, the lisse condition is equivalent to the absence of the Higgs branch on the 4d side. We classify 4d $\mathcal{N}=2$ SCFTs which 
do not admit Higgs branch, then these theories would give lisse W-algebras through the 4d/2d correspondence. In particular, we predict the existence of a large class of new 
non-admissible lisse W-algebras, which have not been studied before. The 4d theories corresponding to lisse W-algebra can appear in the  Higgs branches of  generic 4d 
$\mathcal{N}=2$ SCFTs,  therefore they are crucial to understand the Higgs branches of  $\mathcal{N}=2$ SCFTs. }

\begin{document} 
\maketitle
\flushbottom
\section{Introduction}
The correspondence between the Schur sector of 4d $\mathcal{N}=2$ superconformal field theories (SCFTs) and 2d vertex operator algebras (VOAs) is very useful 
in understanding both 4d and 2d theories \cite{Beem:2013sza}.  It is now clear that results from one side of the correspondence can provide invaluable insights to 
the other side.  In this paper, we continue our studies between   4d $\mathcal{N}=2$  Argyres-Douglas (AD) theories and 2d W-algebras (See \cite{Xie:2016evu, Song:2017oew, Xie:2019yds, Xie:2019zlb} for our previous studies), and provide a systematic description of the so-called lisse W-algebras and 4d $\mathcal{N}=2$ SCFTs.

Rational VOAs have attracted lots of attentions in mathematics and physics literature \cite{moore1989classical,kac2008rationality}. Rational VOAs arise from the study of 2d rational conformal field theories \cite{moore1989classical}, and 
 studies of such VOAs turn out to be quite useful in understanding the general structure of VOAs and conformal field theories. Rational VOAs also have interesting applications in other branches of mathematics 
 and physics \cite{witten1989quantum}, i.e. three dimensional topological field theory, condensed matter physics, and etc.  Given a VOA, it is in general difficult to determine whether it is rational or not.  Zhu constructed a non-commutative algebra called Zhu's algebra \cite{zhu1996modular} and reduced the study of 
rationality of VOAs to representation properties of Zhu's algebra, which is simpler.  Zhu also defined a commutative algebra called Zhu's $C_2$ algebra, and a VOA is called $C_2$ co-finite if the corresponding Zhu's $C_2$ algebra is finite dimensional. $C_2$ co-finiteness is closed related to rationality \footnote{It is an open question to prove that $C_2$ co-finiteness implies rationality.} and is much easier to check in practice. A $C_2$ co-finite VOA is also called lisse \cite{arakawa2012remark}.

Now in the context of  the 4d/2d correspondence \cite{Beem:2013sza}, Zhu's $C_2$ algebra also  plays a crucial role: the associated variety of Zhu's $C_2$ algebra is identified with the Higgs branch of 4d $\mathcal{N}=2$ theories \cite{Song:2017oew, Beem:2017ooy,Arakawa:2017fdq}. Since the associated variety of the lisse VOA is trivial,
 the \textbf{lisse} condition 
of a 2d VOA is just equivalent to the \textbf{absence} of the Higgs branch of the corresponding 4d $\mathcal{N}=2$ theory. While it is in general quite complicated to compute Zhu's $C_2$ algebra in the VOA context, it is a lot easier
to check whether a 4d $\mathcal{N}=2$ theory has a Higgs branch by using 4d methods, i.e. the geometry of Coulomb branch. 

We perform a systematic search for 4d theories which do not admit a Higgs branch.
The two major classes of 4d $\mathcal{N}=2$ SCFTs are class-${\cal S}$ theories \cite{Gaiotto:2009we,Xie:2012hs} constructed using 6d $(2,0)$ SCFTs and theories constructed from three dimensional canonical singularities \cite{Xie:2015rpa}. 
Class-${\cal S}$ theories  always have a Higgs branch if they are constructed using regular singularities only \cite{Gaiotto:2009we}, so we are forced to consider AD theories 
which are constructed by using an  irregular singularity and a possible regular singularity \cite{Xie:2012hs}.  In the paper we first classify irregular singularities which do not contribute to the Higgs branch using the identification 
between the irregular singularity and the 3-fold singularity \cite{Wang:2015mra,Wang:2018gvb}, then we study theories engineered using an irregular singularity and a regular singularity, and it is much harder to classify the ones
without Higgs branches. To accomplish this task, we develop new methods to compute the associated variety of non-admissible affine Kac-Moody (AKM) algebra,  then complete 
the classification of lisse W-algebras arising from 4d theories constructed using 6d $(2,0)$ theory. 

For theories constructed using three fold singularities, the Higgs branch is identified as the crepant resolution of the three dimensional canonical singularity \cite{Chen:2017wkw}. 
So the classification of theory without Higgs branch is reduced to finding all 3-fold singularity with a $\bbC^*$ action, and without the crepant resolution. Such singularity is also called a Q-factorial terminal singularity. While 
it might not be easy to give a classification of such singularities using algebraic geometrical method, it is much easier to check using the input of 4d physics. For example, a Gorenstein terminal singularity is 
Q-factorial if the corresponding 4d theory has no mass deformation, and this can be easily verified using the Coulomb branch solution.

The main purpose of this paper is to classify 4d $\mathcal{N}=2$ SCFTs which do not have a Higgs branch. According to the identification between Zhu's $C_2$ algebra and Higgs branch, each such 4d SCFT would give a 2d lisse VOA. 
The list of  principal lisse W-algebras \footnote{The principal lisse W-algebra means that we use principal nilpotent element to define our W-algebra.} are listed in table \ref{table:admissibleprinciple}, \ref{table:lisseprinciple1}, and \ref{table:lisseexceptional}.  Some more examples are found by adding one regular singularity to the irregular singularity, see table \ref{table:admissibledata}, \ref{table:admissibledata2}, \ref{table:charactersClassical}, \ref{table:associated}, \ref{table:e6e7lisse} and \ref{table:e8lisse}.  We would like to highlight  
two interesting aspects of our results:
\begin{itemize}
\item Previous studies in VOA literature mainly focuses on studies of so-called admissible lisse W-algebras \cite{MR3456698}. Our results show that there are actually a large class of non-admissible lisse 
W-algebras. It is definitely interesting to study the mathematical and physical consequences of these new lisse W-algebras.
\item The 4d theory corresponding to the lisse W-algebra has no Higgs branch, so they can be the IR theory in the Higgs branch deformation of a $\mathcal{N}=2$ theory. Therefore, the classification of these 4d theories are crucial to understand the Higgs branch behavior of generic 4d 
$\mathcal{N}=2$ SCFTs. More physical consequences of these theories on space of 4d $\mathcal{N}=2$ SCFT would be discussed somewhere else. 
\end{itemize}

This paper is organized as follows: section \ref{sec:2dVOAand4dSCFTs} reviews basic facts of lisse VOAs  and their relations to the absence of Higgs branches of 4d $\mathcal{N}=2$ SCFTs. Section \ref{sec:ADandWk} reviews the 
space of 4d $\mathcal{N}=2$ SCFTs and how their Higgs branch can be computed. Section \ref{sec:nomass} classifies 4d theories which do not have a Higgs branch. Section \ref{sec:admissble} studies 
4d theories whose 2d VOAs are admissible lisse W-algebra. Section \ref{sec:nonadm:classical} studies non-admissible lisse W-algebra with classical Lie algebra.  Section \ref{sec:nonadm:exceptional} studies non-admissible lisse W-algebra with 
exceptional Lie algebra. Finally a conclusion is given in section \ref{sec:conclusion}.

\section{Review of VOAs and 4d $\mathcal{N}=2$ SCFTs}
\label{sec:2dVOAand4dSCFTs}

\subsection{Lisse VOAs}
A VOA arises as the chiral part of a 2d conformal field theory (CFT). 
In math literature, a VOA is a vector space $V$ with following properties ($V$ can be thought as the vacuum module of the chiral part of a 2d CFT) \cite{kac1998vertex}:
\begin{itemize}
	\item A vacuum vector $|0\rangle$. 
	\item A linear map
	\begin{equation}
	Y:V\rightarrow {\cal F}(V),~~a\rightarrow Y(a,z)=\sum_n a_{n} z^{-n-1}=a(z),
	\end{equation}
	where $a_{n} \in \mathrm{End}(V)$. This is just the state-operator correspondence \footnote{In physics literature, the mode expansion of a field takes the form $\sum a_n z^{-n-h}$ with $h$ being the scaling dimension. In VOA literature, however, the above convention of mode expansion without further shift by the scaling dimension $h$ is used so that
 VOAs without the definition of scaling dimension can be considered. }. Given an operator $a(z)$, one can recover the corresponding state $|a(z)\rangle=\lim_{z\rightarrow 0} a(z)|0\rangle$.
\end{itemize}
For our purpose, we need to consider the VOA with a conformal vector $\omega$, which is nothing but the chiral part $T(z)$ of the stress tensor. Modes in the expansion of 
$T(z)=\sum L_n z^{-n-2}$ satisfy the Virasoro algebra (by using the standard contour integral and OPE of $T(z)$)
\begin{equation}
[L_n,L_m]=(n-m)L_{n+m}+{c(n^3-n)\over 12}\delta_{n+m,0}.
\end{equation}
Here $c$ is the central charge, an important quantity associated with a VOA.
The normal order product of two fields $a(z)$ and $b(z)$ is denoted as $:ab:(z)$, and its modes are
\begin{equation}
(:ab:(z))_m=\sum_{n\leq -h_a} a_n b_{m-n}+\sum_{n>-h_a}b_{m-n}a_n
\end{equation}
In current convention we set $h_a=1$. Other properties of VOAs can be found in \cite{kac1998vertex}. 

A VOA is called \textbf{rational} if 
\begin{itemize}
\item V has finite number of irreducible representations $M_j$.
\item The character $\ch_j=\tr_{M_j}(e^{2\pi i \tau (L_0-{c\over 24})})$ converges to a holomorphic function on upper half plane $\mathbb{C}^+$ \footnote{We use $\ch$ to denote the character of a VOA and $\chi$ to denote the character of a finite Lie algebra.}.
\item Functions $\ch_j$ span an $\fsl_2(\mathbb{Z})$ invariant space.
\end{itemize}
A VOA is called \textbf{finitely strongly generated} if there is finite number of elements $a_i \in V,~~i=1,\ldots,s$ such that the whole VOA is spanned by  normal order products of the following form
\begin{equation}
:\partial^{k_1} a_1 \ldots \partial^{k_s} a_s:.
\end{equation}
Notice that the choice of generators may not  be unique and in general there are relations among the above basis. It is an interesting problem to find a minimal generating set given a finitely strongly generated VOA.

For a VOA $V$, there exists a Li's filtration \cite{Li2005} which is a decreasing filtration
\begin{equation}
F^0\supset F^1\supset F^2\supset \ldots,
\end{equation}
in which each $F^p$ is spanned by  following states
\begin{equation}
F^p(V)=\{a^{i_1}_{-n_1-1}a^{i_2}_{-n_2-1}\ldots |0\rangle,~~~\sum n_i\geq p \},
\end{equation}
then one can define a graded sum of the VOA
\begin{equation}
Gr(V)=\oplus_p  {F^p \over F^{p+1}}.
\end{equation}
It is obvious that $F^0=V$, and $F^{1}$ is generated by $\{ a_{-2} b | a\in V,~b\in V\}$.  
The Zhu's $C_2$ algebra is defined as 
\begin{equation}
R_V={F^0(V)\over F^1(V)}.
\end{equation} 
$R_V$ is a Poisson algebra and is finitely generated if and only if $V$ is strongly finitely generated. 
Moreover the image of generators of $V$ in $R_V$ generates $R_V$ as well. Notice that $R_V$ is in general 
not reduced, namely the ideal defining it would contain a nilpotent element \footnote{A nilpotent element $x$ of an ideal is an element not in $I$ but $x^n\in I$ for some $n$.}.
Once we determine $R_V$, 
we have now an associated scheme and an associated variety $X_V$ defined from the Zhu's $C_2$ algebra
\begin{equation}
\tilde{X}_V=\mathrm{spec}(R_V),~~~~~~X_V=\mathrm{spec}((R_V)_{red}).	
\end{equation}	
A \textbf{lisse} VOA is defined as the VOA such that $\dim(X_V)=0$, which also implies that Zhu's $C_2$ algebra is finite dimensional. 
A rational VOA has to be lisse, but it is an open problem that whether a lisse VOA has to be rational. 
A quasi-lisse VOA is defined as the VOA whose associated 
variety $X_V$ has finite number of symplectic leaves. 
Quasi-lisse VOAs have many interesting properties \cite{Arakawa:2016hkg, Beem:2017ooy}: 
\begin{itemize}
	\item The VOA is strongly finitely generated.
	\item The Virasoro vector $\omega_V$ is nilpotent in $R_V$.
	\item There are finite number of ordinary modules, and they transform nicely under modular transformations. A weak $V$-module $(M, Y_M)$ is called \emph{ordinary} if $L_0$ acts semi-simply on $M$, which means that any $L_0$-eigenspace $M_\Delta$ of $M$ of eigenvalue $\Delta\in\mathbb{C}$ is finite-dimensional, and for any $\Delta\in\mathbb{C}$, $M_{\Delta-n}=0$ for all sufficiently large $n\in \mathbb{Z}$.
	\item Characters are solutions of a modular differential equation.  
\end{itemize}

\subsubsection{Lisse W-algebras}
A class of interesting VOAs are constructed from quantum Drinfeld-Sokolov (qDS) reduction of affine Kac-Moody (AKM) algebras. Namely, we start with an AKM algebra $V^k(\mathfrak{g})$, where $k$ is the level, and $\mathfrak{g}$ is 
a simple Lie algebra, and then choose a nilpotent element $f$ of $\fg$ to perform the qDS reduction. The W-algebra obtained in this way is labeled as 
\begin{equation}
W^{k}(\mathfrak{g},f).
\end{equation}

Now an AKM algebra is called principal admissible  if the level $k$ takes the following value 
\begin{equation}
k=-h^\vee+{p\over q},~~~p\in\mathbb{Z},~~q\in\mathbb{Z}_{>0},~~p\geq h^\vee,~~(p,q)=1,~(p,r^\vee)=1,
\end{equation}
where $r^{\vee}$ is the lacety for the Lie algebra $\mathfrak{g}$ \footnote{$r^\vee$ =1 for ADE Lie algebra,  $r^\vee=2$ for $BCF$ Lie algebra, and  $r^\vee=3$ for $G_2$ algebra.}, and $h^\vee$ is the dual Coxeter number. If we start with an admissible AKM algebra, the
resulting W-algebra after qDS reduction  is called admissible W-algebra.  If $p=h^\vee$, the AKM and the corresponding W-algebra is called \textbf{boundary admissible} \cite{kac2017remark}. 
Other cases are called non-admissible. 

A very interesting question is to classify all possible lisse W-algebras. The admissible case has been studied in \cite{kac2008rationality, 2019arXiv190511473A}.

\subsection{4d/2d correspondence}
It was proposed in \cite{Beem:2013sza} that 
one can get a 2d VOA from the Schur sector of a 4d $\mathcal{N}=2$ SCFT (see also in  \cite{Beem:2014rza, Lemos:2014lua, Buican:2015hsa, Cordova:2015nma, Buican:2015tda,  Song:2015wta, Lemos:2015awa, Cecotti:2015lab, Lemos:2015orc, Nishinaka:2016hbw,
Buican:2016arp, Xie:2016evu, Cordova:2016uwk, Arakawa:2016hkg, 
Lemos:2016xke, Bonetti:2016nma, Song:2016yfd, Creutzig:2017qyf, Fredrickson:2017yka, Cordova:2017ohl, Cordova:2017mhb, Dedushenko:2017tdw, Buican:2017uka, Song:2017oew, Buican:2017fiq, Beem:2017ooy, Pan:2017zie, Fluder:2017oxm, Buican:2017rya, Choi:2017nur, Arakawa:2017fdq, Kozcaz:2018usv, Costello:2018fnz, Wang:2018gvb, Feigin:2018bkf, Niarchos:2018mvl, Creutzig:2018lbc,  Dedushenko:2018bpp, Bonetti:2018fqz, Arakawa:2018egx, Costello:2018swh, Nishinaka:2018zwq,  Agarwal:2018zqi, Beem:2018duj,  Kiyoshige:2018wol, Xie:2019yds, Buican:2019huq, Pan:2019bor, Beem:2019tfp, Oh:2019bgz,Dedushenko:2019yiw, Xie:2019zlb} for recent developments), and the basic 4d/2d dictionary used in current paper is \cite{Beem:2013sza}:
\begin{itemize}
	\item There is an AKM sub-algebra ($V^{k_{2d}}(\fg)$) in 2d VOA, where $\fg$ is the Lie algebra of 4d flavor symmetry $G_F$.
	\item The 2d central charge $c_{2d}$ and the level of AKM algebra $k_{2d}$ are related to the 4d central charge $c_{4d}$ and the flavor central charge $k_F$ as
	\begin{equation}
	c_{2d}=-12 c_{4d},~~k_{2d}=-k_F \footnote{Our normalization of $k_F$ is half of that of \cite{Beem:2013sza,Beem:2014rza}.}.
	\label{eq:centralchargerelation}
	\end{equation}
	\item The (normalized) vacuum character of 2d VOA is the 4d Schur index $\CI(q)$.
	\item The associated variety is the Higgs branch of the 4d $\mathcal{N}=2$ SCFT \cite{Song:2017oew,Beem:2017ooy, arakawa2018chiral}. 
\end{itemize}

\subsection{Lisse VOAs and the absence of Higgs branches}

Given a 2d VOA corresponding to a 4d $\mathcal{N}=2$ SCFT, it is conjectured that the associated variety is identified with the Higgs branch of the 4d theory \cite{Song:2017oew,Beem:2017ooy, arakawa2018chiral}. We are particularly interested 
in the lisse VOA whose Zhu's $C_2$ algebra is finite dimensional, and would like to explore what the exact meaning of the lisse property of 2d VOAs for 4d $\mathcal{N}=2$ SCFTs.

A 4d $\mathcal{N}=2$ SCFT has a bosonic symmetry group $SO(2,4)\times SU(2)_R\times U(1)_r \times G_F$, where $SO(2,4)$ is the 4d conformal group, $SU(2)_R\times U(1)_r$ is the R symmetry group which exists for every $\mathcal{N}=2$ SCFT, and $G_F$ is the flavor symmetry group which might be absent for some theories. The representation theory of 4d $\mathcal{N}=2$ superconformal algebra is studied in \cite{Dolan:2002zh}, in which short representations  (where the primary states in the supermultiplet are annihilated by a fraction of supercharges) were classified. Important half-BPS operators include primary operators of multiplets ${\cal E}_{r}$ and $\hat{{\cal B}}_R$.

The moduli space of vacuum of a 4d $\mathcal{N}=2$ SCFT is extremely rich. It consists of various branches. The Coulomb
 branch is parameterized by expectation values of  the primaries of ${\cal E}_{r}$ multiplets, and the low energy effective theory  is described by a Seiberg-Witten (SW) geometry \cite{Seiberg:1994rs,Seiberg:1994aj}.
  The set of rational numbers $[r_1,\ldots, r_s]$ of $U(1)_r$ charges of ${\cal E}_{r}$ (unitarity implies that $r_i>1$) is an important set associated to a 4d $\mathcal{N}=2$ SCFT. The $U(1)_r$ symmetry acts non-trivially on the Coulomb branch while the $SU(2)_R\times G_F$ symmetry acts trivially.  The set of Coulomb branch operators ${\cal E}_{r}$  can be found from 
 the SW geometry.
   ${\cal E}_{r}$ operators form a ring which might be called the Coulomb branch chiral ring, and it is freely generated for almost all  theories we know.

The Higgs branch is parameterized by expectation values of the primaries of half-BPS operators $\hat{{\cal B}}_R$. These operators also form 
a ring which might be called the Higgs branch chiral ring. It is believed that the Higgs branch chiral ring is finitely generated, and there are in general complicated relations among generators.  It is in general difficult  to 
determine the full set of $\hat{{\cal B}}_R$ operators, and relations among them. The $\hat{{\cal B}}_1$  type operators, on the other hand, are easily determined by the SW geometry. 
The $\hat{{\cal B}}_1$ multiplet consists of 
conserved currents of the flavor symmetry, and one can use them to deform the 4d theory as mass deformations
\begin{equation}
\delta S=m^2\int d^2\theta \hat{{\cal B}}_1+c.c,
\end{equation}
which can be detected from the SW geometry of the Coulomb branch: they are dimension one parameters of the SW curve.  Therefore if 
we find a mass deformation in the SW geometry, we know for sure the existence of at least one $\hat{{\cal B}}_1$ type operator, therefore the theory has a Higgs branch. 
However, if there is no mass deformation in the SW geometry, one can not claim that the theory has no Higgs branch!  As it is still quite possible that other $\hat{{\cal B}}_R$ type operators with $R>1$ exist. 
There are many such examples in class-${\cal S}$ theories as we will review in later sections. 

More generally, we could  have a branch parameterized by expectation values of both ${\cal E}_{r}$ and $\hat{{\cal B}}_R$ type operators. In general, this branch is a direct product of 
a component parameterized by  ${\cal E}_{r}$ type operators and a component parameterized by $\hat{{\cal B}}_R$ type operators.  This type of branch is called the mixed branch.

 Now we  discuss the low energy behavior in various branches. The Coulomb branch is parameterized by expectation values of ${\cal E}_r$ type operators, and the low energy effective theory at generic point  consists of 
 free vector multiplets, free hypermultiplets, and interacting theories whose Coulomb branch is trivial (see \cite{Chen:2017wkw} for examples of this type). In the literature, one mainly consider theories whose Coulomb branch consists of only free vectormultiplets. More general cases whose Coulomb branch consist of free hypermultiplets and free vector multiplets are considered in \cite{Argyres:2015ffa}. 
 
 The Higgs branch is parameterized by  
 expectation values of $\hat{{\cal B}}_R$ type operators, and the low energy effective theory at generic point would also involve free hypermultiplets, free vector multiplets, and an interacting theory which {\bf does not} have a Higgs branch.
 In literature, people mainly consider theories whose Higgs branch only consists of free hypermultiplets, and our studies in this paper shows that it is quite common that the Higgs branch of a general 4d $\mathcal{N}=2$ SCFT
 consists of an interacting theory and/or free vector multiplets.

Let us  further clarify some definitions used in the literature: a branch of $\mathcal{N}=2$ moduli space is an irreducible component of the variety defined by full chiral ring of an $\mathcal{N}=2$ theory.  A Coulomb branch 
 is called \textbf{pure} Coulomb branch if the low energy effective theory on the generic point consists of only abelian gauge theory. A  Higgs branch is called \textbf{pure} Higgs branch if the low energy effective theory on 
 the generic point  consists of only free hypermultiplets.

We now state the exact meaning of the lisse property for 4d $\mathcal{N}=2$ SCFTs. The corresponding 2d VOA counts Schur operators and in particular count $\hat{\mathcal{B}}_R$ operators, and the reduced Zhu's $C_2$ algebra is conjectured to be the chiral ring for $\hat{\CB}_R$ operators. Assuming that the Higgs branch chiral ring is reduced, we have the following conjecture:
\begin{itemize}
\item A 2d VOA corresponding to a 4d $\mathcal{N}=2$ SCFT is \textbf{lisse} if the moduli space of the 4d theory has no Higgs branch, or equivalently the 4d theory has no $\hat{\CB}_R$ type operators.
\end{itemize}

\section{Space of $\mathcal{N}=2$ SCFTs and their Higgs branch}
\label{sec:ADandWk}

\subsection{A Review on various constructions}
\subsubsection{4d $\mathcal{N}=2$ SCFT from 6d $(2,0)$ theory}
One can engineer a large class of 4d $\mathcal{N}=2$ SCFTs by starting with a 6d $(2,0)$ theory of type $\mathfrak{j}=ADE$ on a sphere with an irregular singularity and at most one regular singularity \cite{Gaiotto:2009we,Gaiotto:2009hg,Xie:2012hs,Wang:2015mra,Wang:2018gvb} \footnote{See the appendix A of \cite{Xie:2019yds} for relations between this construction and other constructions.}. 
Its Coulomb branch is captured by a Hitchin system with singular boundary conditions near singularities. The Higgs field $\Phi$ of the Hitchin system near the irregular singularity takes the following form
\begin{equation}
\Phi={T\over z^{2+{k\over b}}}+\ldots.
\end{equation}
Here $T$ is determined by a positive principal grading of Lie algebra $\mathfrak{j}$ \cite{reeder2012gradings}, and is a regular semi-simple element of $\mathfrak{j}$. $k$  is an integer greater than $-b$. Subsequent terms are chosen 
such that they are compatible with the leading order term (essentially the grading determines the choice of these terms). We call them $J^{(b)}[k]$ type irregular punctures, and summarize allowed values of $b$ in table \ref{table:sing:b} ($k$ and $b$ are {\bf not} required to be coprime). Theories constructed using the above irregular singularity alone can also be engineered by using a three (complex) dimensional singularity in type IIB string theory which are also summarized in table \ref{table:sing:b} \cite{Xie:2015rpa}. 
One can add another regular singularity which is labeled by a nilpotent orbit $f$ of $\mathfrak{j}$ (We use Nahm labels such that the trivial orbit corresponding to the regular puncture with maximal flavor symmetry). Detailed discussion about these defects can be found in  \cite{Chacaltana:2012zy}. All in all, the theory we consider could be labeled by a pair $(J^{(b)}[k],f)$.

\begin{table}[!htb]
\begin{center}
	\begin{tabular}{ |c|c|c|c| }
		\hline
		$ \mathfrak{j}$& $b$  & Singularity   \\ \hline
		$A_{N-1}$&$N$ &$x_1^2+x_2^2+x_3^N+z^k=0$\\ \hline
		$~$  & $N-1$ & $x_1^2+x_2^2+x_3^N+x_3 z^k=0$\\ \hline
		
		$D_N$  &$2N-2$ & $x_1^2+x_2^{N-1}+x_2x_3^2+z^k=0$ \\     \hline
		$~$  & $N$ &$x_1^2+x_2^{N-1}+x_2x_3^2+z^k x_3=0$ \\     \hline
		
		$E_6$&12  & $x_1^2+x_2^3+x_3^4+z^k=0$   \\     \hline
		$~$ &9 & $x_1^2+x_2^3+x_3^4+z^k x_3=0$   \\     \hline
		$~$  &8 & $x_1^2+x_2^3+x_3^4+z^k x_2=0$    \\     \hline
		
		$E_7$& 18  & $x_1^2+x_2^3+x_2x_3^3+z^k=0$   \\     \hline
		$~$&14   & $x_1^2+x_2^3+x_2x_3^3+z^kx_3=0$    \\     \hline

		$E_8$ &30   & $x_1^2+x_2^3+x_3^5+z^k=0$  \\     \hline
		$~$  &24  & $x_1^2+x_2^3+x_3^5+z^k x_3=0$  \\     \hline
		$~$  & 20  & $x_1^2+x_2^3+x_3^5+z^k x_2=0$  \\     \hline
	\end{tabular}
\end{center}
\caption{Three-fold isolated quasi-homogenous singularities of cDV type corresponding to the $J^{(b)}[k]$  irregular punctures of the regular-semisimple type in \cite{Wang:2015mra}. These 3d singularity is very useful in extracting the Coulomb branch spectrum, see \cite{Xie:2015rpa}. }
\label{table:sing:b}
\end{table}

\begin{table}[htb]
	\begin{center}
		\begin{tabular}{ |c|c| c|c|c|c| }
			\hline
			$  j $ ~&$A_{2N}$ &$A_{2N-1}$ & $D_{N+1}$  &$E_6$&$D_4$ \\ \hline
			Outer-automorphism $o$  &$Z_2$ &$Z_2$& $Z_2$  & $Z_2$&$Z_3$\\     \hline
			Invariant subalgebra  $\fg^\vee$ &$B_N$&$C_N$& $B_{N}$  & $F_4$&$G_2$\\     \hline
			Flavor symmetry $\fg$ &$C_N^{(1)}$&$B_N$& $C_{N}^{(2)}$  & $F_4$&$G_2$\\     \hline
		\end{tabular}
	\end{center}
	\caption{Outer-automorphisms of simple Lie algebras $ j$, its invariant sub-algebra $ g^\vee$ and flavor symmetry $ g$ from the Langlands dual of $g^\vee$.}
	\label{table:outm}
\end{table}

\begin{table}[!htb]
	\begin{center}
		\begin{tabular}{|c|c|c|c|}
			\hline
			$j$ with twist & $b_t$ & SW geometry at SCFT point & $\Delta[z]$ \\ \hline
			$A_{2N}/Z_2$ & $4N+2$ &$x_1^2+x_2^2+x^{2N+1}+z^{k+{1\over2}}=0$ & ${4N+2\over 4N+2k+3}$  \\ \hline
			~& $2N$ & $x_1^2+x_2^2+x^{2N+1}+xz^k=0$ & ${2N\over k+2N}$ \\ \hline
			$A_{2N-1}/Z_2$& $4N-2$ & $x_1^2+x_2^2+x^{2N}+xz^{k+{1\over2}}=0$ & ${4N-2\over 4N+2k-1}$ \\ \hline
			~ & $2N$ &$x_1^2+x_2^2+x^{2N}+z^{k}=0$ & ${2N\over 2N+k}$  \\ \hline
			$D_{N+1}/Z_2$& $2N+2$ & $x_1^2+x_2^{N}+x_2x_3^2+x_3z^{k+{1\over2}}=0$ & ${2N+2\over 2k +2N+3}$ \\ \hline
			~ & $2N$ &$x_1^2+x_2^{N}+x_2x_3^2+z^{k}=0$ & ${2N\over k+2N}$  \\ \hline
			$D_4/Z_3$ & $12$ &$x_1^2+x_2^{3}+x_2x_3^2+x_3z^{k\pm {1\over3}}=0$ & ${12\over 12+3k\pm1}$  \\ \hline
			~& $6$ &$x_1^2+x_2^{3}+x_2x_3^2+z^{k}=0$ & ${6\over 6+k}$  \\ \hline
			$E_6/Z_2$& $18$ &$x_1^2+x_2^{3}+x_3^4+x_3z^{k+{1\over2}}=0$ & ${18\over 18+2k+1}$  \\ \hline
			~& $12$ &$x_1^2+x_2^{3}+x_3^4+z^{k}=0$ & ${12\over 12+k}$  \\ \hline
			~ & $8$ &$x_1^2+x_2^{3}+x_3^4+x_2z^{k}=0$ & ${8\over 12+k}$  \\ \hline
		\end{tabular}
		\caption{SW geometry of twisted theories at the SCFT point. Here we also list the scaling dimension of coordinate $z$. All $k$'s in this table are integer valued and the power of $z$ coordinate in singularity is equal to  $k_t$ used in equation \ref{eq:irregularPunctureTwisted}.}
		\label{table:SW:twisted}
	\end{center}
\end{table}

To get non-simply laced flavor groups, we need to consider the outer-automorphism twist of ADE Lie algebra and its Langlands dual. A systematic study of these AD theories 
was performed in \cite{Wang:2018gvb}. Denoting the twisted Lie algebra of $\mathfrak{j}$ as $\fg^{\vee}$ and its Langlands dual as $\fg$, outer-automorphisms and twisted algebras of $\mathfrak{j}$ are summarized in table \ref{table:outm}. The irregular singularity of regular semi-simple type is also classified as in table \ref{table:SW:twisted} with the following form,
\begin{equation}
\label{eq:irregularPunctureTwisted}
\Phi={T^t\over z^{2+{k_t\over b_t}}}+\ldots
\end{equation} 
Here $T^t$ is an element of Lie algebra $\fg^{\vee}$ or other parts of the decomposition of $\mathfrak{j}$ under the outer automorphism. $k_t>-b_t$, and
the novel thing  is that $k_t$ could take half-integer value or 
in  thirds ($\fg=G_2$). One can also represent those irregular singularities by 
3-fold singularities as in table \ref{table:SW:twisted}. We could again add 
a twisted regular puncture labeled also by a nilpotent orbit $f$ in $\fg$.

Some interesting physical properties about these theories are:
\begin{itemize}
\item If there is no mass parameter  in the irregular singularity (the constraint on $k$ will be listed in the next section),
the corresponding VOA is conjectured to be given by the following $W$ algebra \cite{Wang:2018gvb}:
\begin{equation}
\label{eq:ADtoWalgebra}
\boxed{W^{k_{2d}}(\fg,f),~~k_{2d}=-h^{\vee}+{1\over n} {b\over k^{'}+b}},
\end{equation}
where $h^{\vee}$ is the dual Coxeter number of $\fg$, $n$ is the number listed in table \ref{table:lie}, and $k'$ is restricted to 
the value such that no mass parameter is in the irregular singularity. $k'$ is related to the value $k$ in table \ref{table:sing:b} and \ref{table:SW:twisted} as follows: a)~$n=1,~~k'=k$; b)~$n=2$ or $n=4$,~~$k'=2k+1$;~~c)~$n=3,~~k'=3k\pm1$.
 
\begin{table}[h]
	\begin{center}
		\begin{tabular}{|c|c|c|c|c|}
			\hline
			~&dimension & $h$ & $h^{\vee}$&$n$ \\ \hline
			$A_{N-1}$&$N^2-1$& $N$ & $N$&1  \\ \hline
			$B_N$ & $(2N+1)N$ & $2N$& $2N-1$&2 \\ \hline
			$C_N^{(1)}$ &  $(2N+1)N$&  $2N$ &$N+1$&4 \\ \hline
			$C_N^{(2)}$ &  $(2N+1)N$&  $2N$ &$N+1$&2 \\ \hline
			$D_N$ & $N(2N-1)$ & $2N-2$ & $2N-2$&1 \\ \hline
			$E_6$& 78 & 12 & 12 &1\\ \hline
			$E_7$& 133 & 18 & 18 &1\\ \hline
			$E_8$& 248 & 30 & 30 &1\\ \hline
			$F_4$& 52 & 12 & 9 &2\\ \hline
			$G_2$& 14 & 6 & 4&3 \\ \hline
			
		\end{tabular}
	\end{center}
	\caption{Lie algebra data. $h$ is the Coxeter number and $h^{\vee}$ is the dual Coexter number. $n$ is the number which appears in the level of 2d W-algebra, see \ref{eq:ADtoWalgebra}.}
	\label{table:lie}
\end{table}

\item The SW geometry of these theories is  identified with the spectral curve 
of the corresponding Hitchin system \cite{hitchin1987stable}:
\begin{equation}
\det(x-\Phi)=0,
\end{equation}
and one can read off the Coulomb branch spectrum from an associated Newton polygon\cite{Xie:2012hs,Wang:2015mra,Wang:2018gvb}, which is also reviewed in the appendix B of \cite{Xie:2019yds}. 

 \end{itemize}

\textbf{Remark}: We could also consider theories constructed using regular singularities only, and such theories are studied extensively in \cite{Gaiotto:2009we}. 

\subsubsection{4d $\mathcal{N}=2$ SCFTs from 3-fold singularities}
One can also engineer a large class of 4d $\mathcal{N}=2$ SCFTs by putting type IIB string theory on a 3-fold canonical singularity with a $\mathbb{C}^*$ action \cite{Xie:2015rpa}.  We first review 
the hypersurface case: starting with an isolated singularity which is defined by a polynomial $f:(\mathbb{C}^4,0)\rightarrow (\mathbb{C},0)$, which  is required to have a $\bbC^*$ action satisfying following condition
\begin{equation}
f(\lambda^{q_i}z_i)=\lambda f(z_i),~~~\sum q_i>1.
\end{equation}   
The SW geometry of the theory defined by $f$ is identified with the mini-versal deformation of $f$, i.e.
\begin{equation}
F(z,\lambda)=f(z)+\sum_{\alpha=1}^\mu \lambda_{\alpha} \phi_{\alpha},
\end{equation}
with $\phi_{\alpha}$ being a monomial basis of the Jacobi algebra
\begin{equation}
J_f={\bbC[z_1,z_2,z_3,z_4]\over \{{\partial f\over \partial z_1}, {\partial f\over \partial z_2},{\partial f\over \partial z_3},{\partial f\over \partial z_4}\}},
\end{equation}
and $\lambda_{\alpha}$ is identified with the Coulomb branch parameters with scaling dimension 
\begin{equation}
[\lambda_{\alpha}]={1-Q_\alpha\over \sum q_i-1}.
\end{equation}
Here $Q_\alpha$ is the charge of  $\phi_\alpha$ under the $\bbC^*$ action. In particular, if $[\lambda_\alpha]=1$,  it is a mass parameter.  Theories constructed using complete intersection singularity are discussed in \cite{Chen:2016bzh,Wang:2016yha}.
Theories which do not have Coulomb branch are discussed in \cite{Chen:2017wkw}. Some further physical aspects of those theories are discussed in \cite{Xie:2015xva,Li:2018rdd}.

\subsection{The Higgs branch}
As reviewed in the last subsection, one can find the SW geometry for the Coulomb branch once the geometric data specifying the theory is given.
The Higgs branch, on the other hand, is more complicated and less studied. However, the basic ideas for studying the Higgs branch were actually already 
given in \cite{Xie:2014pua,Chen:2017wkw}, and we review those methods first. 

\subsubsection{4d SCFTs from $(2,0)$ theories}
Higgs branches of 4d SCFTs from 6d $(2,0)$ theories have been studied in \cite{Xie:2014pua}. Taking theories constructed from 6d $A_N$ $(2,0)$ theory as examples, the low energy effective
theory of $N$ M5 branes is a 6d $(2,0)$ theory with transverse deformations described by 5 real scalars. Then put the $(2,0)$ theory on 
a Riemann surface $\Sigma$ with certain partial topological twist on $\Sigma$ so that the 4d theory has $\mathcal{N}=2$ supersymmetry.  After the twist, the transverse deformations are described by
two complex scalars $\Phi_1$ and $\Phi_2$ defined on $\Sigma$. $\Phi_1$ is a section of the canonical bundle on $\Sigma$, while $\Phi_2$ is a section 
of the trivial bundle. The low energy effective theory is described by the following spectral curve
\begin{equation}
\begin{split}
&\det(v-\Phi_1)=0\rightarrow v^N+\sum_{i=2}^N \phi_i(z)x^{N-i}=0, \\
&\det(w-\Phi_2)=0\rightarrow \prod_i(w-c_i)=0,
\end{split}
\end{equation}
where $\phi_i$ is a section of $K_\Sigma^i$, and $c_i$ are constants. Above equations are independent. However, the two Higgs fields $\Phi_1$ and $\Phi_2$
 are actually commuting \cite{Xie:2013rsa}, which give a holomorphic factorization condition on the above curve.  Different branches are given by following curves:
\begin{equation}
\begin{split}
 & (v^{n_1}+\ldots)(v^{n_2}+\ldots)\ldots (v^{n_s}+\ldots)=0,  \\
 & (w-c_1)^{n_1} (w-c_2)^{n_2}\ldots (w-c_s)^{n_s}=0,
 \end{split}
\end{equation}
where $\sum n_i=N$ and $\sum c_i =0$.  Assuming $n_1\geq n_2\geq \ldots \geq n_s$, the branch is then labeled by a Young Tableaux $[n_1, \ldots, n_s]$, although not all Young Tableaux can appear. Here are some further remarks:
\begin{itemize}
\item It is always possible to take the Young Tableaux $Y=[N]$ (A single column of $N$ boxes), and all $c$s are equal to zero, so we have a Coulomb branch.
\item The Higgs branch appears if $Y=[1^N]$ (A single row of $N$ boxes),  and $v^N=0$. Notice that this is not always possible. 
\item If our theory is constructed using regular singularities only,  it is always possible to have a Higgs branch or a mixed branch, as we can always turn off all 
the deformations in $v$ direction, i.e. $v^N=0$, and it has at least a $N-1$ dimensional Higgs branch.
\end{itemize}
So to find a theory without Higgs branch, we have to look at the theory constructed using irregular singularities. Furthermore, the necessary condition is that 
\textbf{there is no holomorphic factorization of the Coulomb branch spectral curve!}

This can be easily checked for the $A_{N-1}$ type theory, as the SW curve at the SCFT point takes the form $x^N+z^k=0$, and it is easy to see that 
no holomorphic factorization implies that $N$ and $k$ are coprime. For other Lie algebras, this condition is not easy to check, but luckily it can be translated into a much simpler criteria which will be discussed in the next section. 
Things become more complicated 
if there is a regular singularity, as one could have local contribution to Higgs branch. The detailed  discussion on regular singularities will also be given in next section.

\textbf{Example 1}: Consider a theory engineered by putting 6d $A_1$  $(2,0)$ theory on a sphere with the following irregular singularity
\begin{equation}
\Phi={T\over z^{3+{1\over 2}}}+\ldots
\end{equation}
We have only one branch
\begin{equation}
\begin{split}
& \text{I}:~~v^2=z^3+u_1  z + u_2 , \nonumber\\ 
&~~~~~~w^2=0,
\end{split}
\end{equation}
which is the usual Coulomb branch. This theory has no Higgs branch.

\textbf{Example 2}: Consider a theory engineered by putting 6d $A_1$  $(2,0)$ theory  on a sphere with the following irregular singularity
\begin{equation}
\Phi={T\over z^4}+\ldots,
\end{equation}
and there are two branches
\begin{equation}
\begin{split}
\text{I}:~~&v^2=z^4+u_1 z^2+ m z + u_2 ,\\~~
&w^2=0,\\
\text{II}:~~&(v-(z^2+a z+b))(v+(z^2+a z+b)=0, \\
&(w-c)(w+c)=0.
\end{split}
\end{equation}
Branch I is a pure Coulomb branch, while branch II is a mixed branch.

\subsubsection{4d SCFTs from  3-fold singularities}
For a 4d theory engineered from a 3-fold singularity, its Coulomb branch is described by a mini-versal deformation of the singularity \cite{Xie:2015rpa}. The Higgs branch, on the other hand, is described 
by its crepant resolution \cite{Chen:2017wkw}. A crucial theorem by Kawamata \cite{kawamata1988crepant} is that every 3-fold canonical singularity has a partial crepant resolution $f:Y\rightarrow X$, where $Y$ is a Q-factorial 
terminal singularity. Gorenstein terminal singularity takes the following form, 
\begin{equation}
f_{ADE}(x,y,z)+tg(x,y,z,t)=0.
\end{equation}
Here $f_{ADE}(x,y,z)$ is the familiar 2d $ADE$ singularity.  $g(x,y,z,t)$ is a polynomial such that the singularity is isolated. Other terminal singularities are constructed from the quotient of  Gorenstein terminal singularities.

The crepant resolution of a three-fold singularity is not unique, but the number of crepant divisor $C(X)$ is constant. 
Given a partial resolution, the number of free hypermultiplets is given by  $b_2(Y)$ which is given by following formula \cite{caibuar2003divisor}:
\begin{equation}
b_2(Y)=\rho(X)+C(X).
\end{equation}
Here $\rho(X)$ is the rank of local class group of the singularity. In the hypersurface case, $\rho(x)$ is identified with the number of mass parameter in the SW geometry.  We are looking for 
the singularity $X$ with $b_2(Y)=0$, therefore 
\begin{itemize}
\item $C(X)$ has to be zero, which implies that $X$ has to be a terminal singularity.
\item $\rho(X)$ is zero, which further implies that $X$ is a Q-factorial terminal singularity.
\end{itemize}
 The conclusion is then \textbf{theories without Higgs branch corresponds to Q-factorial terminal singularity!}

In general $b_3(Y)$ is nonzero, and this means that the low energy theory in the Higgs branch contains free vector multiplets. Moreover, if $Y$ is not smooth, one can 
have an interacting theory in the IR which is described by the singularity of $Y$.

\textbf{Example}: Consider the singularity $X$: $x^3+y^3+z^3+t^{n}=0$, and it is possible to do a weighted blow up so that the resolution is a crepant resolution, the new variety 
has singularity of type $x^3+y^3+z^3+t^{n-3}=0$. So we have a chain of blow ups 
\begin{equation}
X_s\rightarrow X_{s-1}\rightarrow \ldots \rightarrow X_1\rightarrow X         
\end{equation} 
with $s=[{n\over 3}]$. The ending variety $X_s$ is either a non-singular variety if $n\equiv0~\mathrm{or}~1~\mathrm{mod}~3$, or a Q-factorial terminal singularity if $n\equiv2~\mathrm{mod}~3$. 
The topology of $X_s$ is then computed as follows. We have $b_2(X_s)=C(X)+\rho(X)$ with
\begin{equation}
C(X)=[{n\over 3}]
\end{equation}
and 
\begin{equation}
\rho(X)= \begin{cases}
6~~\mathrm{if}~n\equiv0~\mathrm{mod}~3 \\
0~~\mathrm{otherwise}
\end{cases}.
\end{equation}
The third Betti number $b_3(X_s)$ can be computed as 
\begin{equation}
b_3(X_s)=\begin{cases}
2([n/3]-1)~\mathrm{if}~n\equiv0~\mathrm{mod}~3\\
2([n/3])~~~~~\mathrm{otherwise}
\end{cases}.
\end{equation} 
All in all we have following interesting situations:
\begin{itemize}
\item [1] For $x^3+y^3+z^3+t^{2}=0$,~~this is a Q-factorial terminal singularity, and there is no Higgs branch!
\item [2] For $x^3+y^3+z^3+t^{3}=0$,~~$b_3(X_s)=0$ and $b_2(X_s)=7$. This theory has a pure Higgs branch. This theory actually has a quiver gauge theory description, and the quiver is of the affine $E_6$ shape \cite{DelZotto:2015rca}.   
\item [3] For $x^3+y^3+z^3+t^{n}=0$ with $n>3$, ~$b_3(X_s)\neq 0$ and $b_2(X_s)\neq 0$,  the theory has a Higgs branch and the low energy effective theory involves free hypermultiplets and free vector multiplets for $n\equiv1~\mathrm{mod}~3$, 
and the low energy effective theory has free hypermultiplets, vector multiplets, and an interacting theory described by the singularity $x^3+y^3+z^3+t^{2}=0$  for $n\equiv2~\mathrm{mod}~3$. 
\end{itemize}

\section{Absence of mass deformation: Classification}
\label{sec:nomass}

Now we would like to classify 4d SCFTs which do not have Higgs branch. A necessary condition is that the theory does not have a mass deformation, as the existence of mass deformation implies the existence of $\hat{{\cal B}}_1$ operators. 
We first consider 4d theories constructed from 6d $(2,0)$ theories, and classify irregular singularities which do not have mass parameters.  We then list those regular singularities which do not have a flavor symmetry and therefore do not have mass deformation. 

For theories constructed from three fold singularities, the Higgs branch deformation is identified with the crepant resolution of the singularity, and a very powerful theorem by Kawamata
reduces the classification of theory without Higgs branch to the classification of Q-factorial terminal singularity. The classification of quasi-homogeneous \textbf{Gorenstein Q-factorial terminal singularities} actually coincides with 
the classification of special irregular singularity which do not have mass parameter in class ${\cal S}$ construction.  

\subsection{Irregular singularity: absence of Higgs branch deformation}

\subsubsection{Untwisted theory and Q factorial Gorenstein terminal singularity}
 Let's consider class ${\cal S}$ theory constructed using 6d $(2,0)$ theory of type $\mathfrak{g}=ADE$, and $f$ to be a regular nilpotent orbit of $\fg$. The same theory can be then engineered from the three fold singularity listed in table \ref{table:sing:b}. They can be written in the form 
\begin{equation}
f_{ADE}(x,y,z)+tg(x,y,z,t)=0,
\end{equation}
with $f_{ADE}(x,y,z)$ being the famous 2d ADE singularity. 
The Higgs branch of these theories is conjectured to be described by crepant resolutions of these singularities. So we would like to classify above singularities which do not admit crepant resolutions. 
The singularity of above type is called Gorenstein terminal singularity, and those which do not admit crepant resolution is called Q-factorial terminal singularity \cite{kawamata1988crepant}, which is actually equivalent to the fact that the corresponding 4d 
$\mathcal{N}=2$ SCFT does not admit a mass parameter. Such singularities have already been classified in \cite{Xie:2016evu}, and the complete list is shown in table \ref{table:constraintADEirregular}. 
\begin{table}[h]
\begin{center}
\begin{tabular}{|c|l|c|l|}
\hline
  ${\cal T}$ &$no mass$&${\cal T}$&$no mass$  \\ \hline
     $A_{N-1}^N[k]$ &$(k,N)=1$& $A_{N-1}^{N-1}[k]$ &$\text{No solution}$\\ \hline
          $D_{N}^{2N-2}[k]$ &${2N-2\over \mathrm{gcd}(k,2N-2)}$ is even $\&$ $\mathrm{gcd}(k,2N-2)$ is odd& $D_{N}^{N}[k]$&${N\over \mathrm{gcd}(k,N)}$ is even\\ \hline
     $E_{6}^{12}[k]$ &$k\neq 3n$& $E_6^{9}[k]$ &$k\neq 9n$\\ \hline
     $E_{6}^8[k]$ &$\text{No solution}$& $E_{7}^{18}[k]$ &$k\neq 2n$\\ \hline
     $E_7^{14}[k]$ &$k\neq 2n,n>1$& $E_{8}^{30}[k]$ &$k\neq 30n$\\ \hline
     $E_{8}^{24}[k]$ &$k\neq 24n$& $E_{8}^{20}[k]$ &$k\neq 20 n$\\ \hline
\end{tabular}
\end{center}
\caption{Constraint on $k$ so that irregular singularity denoted by $J^b[k]$ has no mass deformation.}
  \label{table:constraintADEirregular}
\end{table}
\begin{table}[h]
	\begin{center}
		\begin{tabular}{|c|l|l|}
			\hline
			$j$ with twist & SW geometry at SCFT point & no mass \\ \hline
			$A_{2N}/Z_2$  &$x_1^2+x_2^2+x^{2N+1}+z^{k+{1\over2}}=0$ & ${4N+2\over gcd(4N+2,2k+1)}$  is  even  \\ \hline
			~ & $x_1^2+x_2^2+x^{2N+1}+xz^k=0$ & $2N\over gcd(2N,k) $ is even \\ \hline
			$A_{2N-1}/Z_2$ & $x_1^2+x_2^2+x^{2N}+xz^{k+{1\over2}}=0$ & ${4N-2\over gcd(4N-2,2k+1)}$  is  even  \\ \hline
			~ &$x_1^2+x_2^2+x^{2N}+z^{k}=0$ & $2N\over gcd(2N,k) $ is even  \\ \hline
			$D_{N}/Z_2$ & $x_1^2+x_2^{N-1}+x_2x_3^2+x_3z^{k+{1\over2}}=0$ & $2N\over gcd(2k+1,2N)$ is even \\ \hline
			~  &$x_1^2+x_2^{N-1}+x_2x_3^2+z^{k}=0$ & ${2N-2\over gcd(k,2N-2)}$ $\&$ $gcd(k,2N-2)$ are even   \\ \hline
			$D_4/Z_3$  &$x_1^2+x_2^{3}+x_2x_3^2+x_3z^{k\pm {1\over3}}=0$ & no constraint  \\ \hline
			~ &$x_1^2+x_2^{3}+x_2x_3^2+z^{k}=0$ & $k\neq 6n$  \\ \hline
			$E_6/Z_2$ &$x_1^2+x_2^{3}+x_3^4+x_3z^{k+{1\over2}}=0$ &  no constraint\\ \hline
			~ &$x_1^2+x_2^{3}+x_3^4+z^{k}=0$ & $k\neq 12n$  \\ \hline
			~  &$x_1^2+x_2^{3}+x_3^4+x_2z^{k}=0$ & $k\neq 8n~and~k~even$  \\ \hline
		\end{tabular}
		\caption{Constraint on  twisted irregular singularity which does not have mass deformation.}
		\label{table:constraintwist}
	\end{center}
\end{table}

\subsubsection{Twisted theory}
Now we would like to consider the classification of twisted irregular singularity which does not admit a Higgs branch. We do not know the geometric criteria for them yet, however, we 
would still like to conjecture that {\bf the absence mass parameters in the Coulomb branch implies that there is no higgs branch!} The complete list of these irregular singularities is shown in table \ref{table:constraintwist}.

\subsubsection{Classification using the positive grading}
As described in \cite{Wang:2015mra}, the irregular singularity described in section \ref{sec:2dVOAand4dSCFTs} is classified by the positive grading of regular semi-simple type \footnote{We take $g$ to be simply-laced here.}, i.e. 
\begin{equation}
g=\bigoplus_{0\leq j\leq b-1,\,j\in\mathbb{Z}} g_j.
\end{equation}
The coefficient of the $z^{-1}$ term in Higgs field takes value in $g_0$. The coefficient $T$ of the leading order term is chosen as a regular semi-simple element in $g_1$, 
and the number of mass parameter is determined by the dimension of $\{g_0^T\}$ which are all semi-simple elements in $g_0$ which also commute with $T$. 

Some positive gradings of regular semi-simple type are determined by a nilpotent element $e$, and the number of mass parameter has 
a simple description. Each nilpotent element gives a $\fsl_2$ triple $\phi$, and $g_0^T$ is given by $g^\phi$ (this is the sub-algebra which commutes with $\fsl_2$ triple).  In particular, if $b=h^\vee$ and $(k,h^\vee)=1$, 
the nilpotent element is given by the regular nilpotent element whose flavor symmetry is trivial, so the corresponding irregular singularity does not 
have any flavor symmetries. This agrees with results shown in table. \ref{table:constraintADEirregular}, which is derived using the explicit computation of Coulomb branch spectrum.

\subsection{Regular singularity: absence of mass deformation}
A necessary condition for the absence of Higgs branch is that there is no flavor symmetry in the regular singularity.
Such regular singularities can be found from the table listed in \cite{Chacaltana:2010ks, Chacaltana:2011ze, Chacaltana:2013oka, Chacaltana:2014jba, Chacaltana:2015bna, Chacaltana:2016shw, Chacaltana:2017boe, Chcaltana:2018zag}. Here we reproduce them in table \ref{table:rigidregular}.  

\begin{table}[h]
\begin{center}
\begin{tabular}{|c|c|}
\hline
  $\mathfrak{g}$ & No flavor symmetry \\ \hline
  $A_{N-1}$ & $Y=[N]$ \\ \hline 
  $B_N$ & $[n_1,\ldots, n_s]$,$n_i\neq n_j$ and $n_i$ are all odd  \\ \hline
    $C_N$ & $[n_1,\ldots, n_s]$,$n_i\neq n_j$ and $n_i$ are all even  \\ \hline
  $D_{N}$ & $[n_1,\ldots, n_s]$,$n_i\neq n_j$ and $n_i$ are all odd  \\ \hline
   $E_6$ & $E_6, E_6(a_1), E_6(a_3)$ \\ \hline
   $E_7$ & $E_7, E_7(a_1), E_7(a_2), E_7(a_3), E_7(a_4), E_7(a_4),E_7(a_5)$ \\ \hline
   $E_8$ & $ E_8, E_8(a_1), E_8(a_2), E_8(a_3), E_8(a_4), E_8(a_5),E_8(a_6), E_8(a_7),E_8(b_4),E_8(b_5),E_8(b_6)$ \\ \hline
   $G_2$ &  $G_2, G_2(a_1)$ \\  \hline
   $F_4$ & $F_4, F_4(a_1), F_4(a_2), F_4(a_3)$  \\ \hline
\end{tabular}
\end{center}
\caption{Nilpotent orbits which do not have flavor symmetry. A nilpotent orbit is labeled by a partition for  classical groups, and by a Bala-Carter label for exceptional groups.}
  \label{table:rigidregular}
\end{table}

\newpage
\section{Admissible lisse W-algebras}
\label{sec:admissble}

Considering 4d $\mathcal{N}=2$ SCFTs constructed from 6d $(2,0)$ theory on a sphere with an irregular singularity and a regular singularity discussed in section \ref{sec:2dVOAand4dSCFTs}, 
the necessary condition for a theory in this class to have no Higgs branch is that the irregular singularity does not admit mass deformation, which provides a constraint on the irregular singularity (such singularities are summarized in table \ref{table:constraintADEirregular} and \ref{table:constraintwist}).

The choice of the regular singularity is  subtle on the other hand. Naively, we would think that we have to choose regular singularity which do not have a flavor symmetry, however this is not that simple for following reasons.
 First of all, there is still possible contribution to Higgs branch from regular singularity even if 
there is no mass deformation in it, and we have to choose very special regular singularity (whose choice depends on the choice of irregular singularity) to make sure there is no Higgs branch. 
The main tool to select such special regular singularities is the associated variety of the AKM algebra which corresponds to the theory defined by the same irregular singularity and a regular singularity with trivial nilpotent orbit. Second of all, even if a regular puncture 
carries a flavor symmetry, it is possible that the 4d theory actually does not have such flavor symmetry which is related to the so-called collapsing levels of 2d W-algebra \cite{Xie:2019yds}.

In this section, we focus on the irregular singularity such that the level of the corresponding W-algebra is \textbf{admissible}, namely, the level takes the following form
\begin{equation}
k_{2d}=-h^\vee+{h^\vee \over b_{max}+k^{'}}.
\label{eq:level1}
\end{equation}
Here $(k^{'},h^\vee)=1$ \footnote{The W-algebra of $A_{2N}/Z_2$ class theory given in \cite{Wang:2018gvb} is not principal admissible, however, using the result of \cite{Xie:2019yds}, the previous  W-algebra is actually 
isomorphic to an admissible W-algebra.}, see equation \ref{eq:ADtoWalgebra} for the convention. The associated variety of the corresponding admissible W-algebra is (defining $u\equiv b_{max}+k^{'}$)
\begin{equation}
S_f\cap \overline{\CO}_{u},
\end{equation}
where $S_f$ is the Slowdoy slice associated with the nilpotent orbit $f$, $\CO_{u}$ is the nilpotent orbit depending on $u$, see tables in \cite{MR3456698}, and $\overline{\CO}_u$ is the closure of $\CO_u$.  $\overline{\CO}_u$ is actually the Higgs branch of the theory defined by an irregular singularity determined by the pair $(h^\vee, u)$ and
a regular puncture labeled by trivial nilpotent orbit (the corresponding VOA is the AKM with  level \ref{eq:level1}). So the regular singularity $f$ should be taken as an element in ${\cal O}_u$
\begin{equation}
f\in \CO_{u}.
\end{equation}
and the corresponding 4d theory would have no Higgs branch!

Since $ \CO_{u}$ is determined by the data of irregular singularities, the above condition further restrict the choice of $f$ given the choice of irregular singularity. We first consider all $f$'s which do not have flavor symmetries which are summarized
 in \ref{table:admissibleregular}. More data of those 4d theories defined using exceptional Lie algebra are listed in table \ref{table:admissibledata}. Other cases where the regular singularity $f$ has non-trivial flavor symmetry but the whole theory has no Higgs branch are summarized in table \ref{table:admissibledata2} and \ref{table:charactersClassical}.
\begin{table}[!hb]
\begin{center}
\resizebox{\textwidth}{!}{
\begin{tabular}{|c|c|}
\hline
  $\mathfrak{g}$ & Exceptional Pair $(f, u)$, and $\mathrm{gcd}(u,h^\vee)=1$ \\ \hline
  $A_{N-1}$ & $([N],u>N)$\\ \hline 
    $B_N$ & a: $([2N+1],~u>2N)$; b: $([u,s,1],u)$,~u~odd  \\ \hline
            $C_N^{(2)}$ & a:~$([2N],~u>2N)$;~~b:~$([u-1,s],u)$,~~u~odd \\ \hline
  $D_{N}$ & a:~$([2N-1,1],u>2N-2)$;~b:~$([u,s],q)$,~u~odd~\\ \hline
   $E_6$ & $(E_6, u>12)$,~$(E_6(a_1),11)$ \\ \hline
   $E_7$ & $(E_7,u>18)$, $(E_7(a_1),17)$, $(E_7(a_2),13)$, $(E_7(a_3),11)$ \\ \hline
   $E_8$ & $ (E_8,u>30), (E_8(a_1),29), (E_8(a_2),23), (E_8(a_3),19), (E_8(a_4),17), (E_8(a_5),13),(E_8(a_6),11)$ \\ \hline
      $G_2$ &  $(G_2,u>6), (G_2(a_1),5)$ \\  \hline
   $F_4$ & $(F_4,u>12), (F_4(a_1),11), (F_4(a_2),7), (F_4(a_3),5)$  \\ \hline
\end{tabular}
}
\end{center}
\caption{$f$ is a nilpotent element labelling a regular singularity. $u$ specifies the irregular singularity. The level $k$ of the corresponding 2d VOA is $k=-h^\vee+ {h^\vee \over u}$. These pairs give rise to 4d $\mathcal{N}=2$ SCFTs with no Higgs branch. Here we list only those regular singularities without mass deformation. }
  \label{table:admissibleregular}
\end{table}

\begin{table}[h]
\begin{center}
\resizebox{\textwidth}{!}{
\begin{tabular}{|l|l|l|l|l|}
\hline
 $(f,u)$ & Coulomb branch & Schur index & 4d Isomorphism & W-algebra \\ \hline
  $(E_6(a_1),11)$& $ ({24\over11},{18\over 11},{16\over11},{15\over11},{12\over11})$ & $\pe{{q^2+q^5-(q^7+q^{10})\over (1-q)(1-q^{11})}}$ & $(A_2, D_5)$ & $\mathrm{Vir}_{3,22}\oplus L(21,1)$  \\ \hline
    $(E_6(a_3),7)$&  $({12\over 7},{9\over 7},{8\over7})$& $\pe{{q^2+q^3-(q^5+q^{6})\over (1-q)(1-q^{7})}}$ & $(A_2, A_3)$  & $\mathrm{Vir}_{3,14}\oplus L(13,1)$ \\ \hline
        $(E_7(a_1),17)$ & \parbox[c]{2.9cm}{ $({54\over17},{42\over 17},{40\over17},{36\over17},{30\over17})$ \\ $({28\over17},{26\over 17},{24\over17},{22\over17},{18\over17})$ } & $\pe{ {q^2+q^6-(q^{12}+q^{16})\over (1-q)(1-q^{17})} }$ & $(D_4/Z_3, k-{1\over3}={5\over3})$ & \\ \hline 
    $(E_7(a_2),13)$& $({22\over13},{20\over13},{18\over13},{16\over13},{14\over13})$ & $\pe{{q^2-q^{12}\over (1-q)(1-q^{13})}}$ & $(A_1,A_{10})$ & $\mathrm{Vir}_{2,13}$ \\ \hline
    $(E_7(a_3),11)$& $({24\over11},{18\over 11},{16\over11},{15\over11},{12\over11})$  & $\pe{{q^2 + q^5 - (q^7 + q^{10}))\over (1-q)(1-q^{11})}}$ & $(A_2, D_5)$  & $\mathrm{Vir}_{3,22}\oplus L(21,1)$ \\ \hline
      $(E_8(a_1),29)$& 
      \parbox[c]{4.7cm}{$(\frac{30}{29},\frac{36}{29},\frac{40}{29},\frac{42}{29},\frac{46}{29},\frac{48}{29}, \frac{52}{29} )$ \\ $(\frac{54}{29},\frac{60}{29},\frac{66}{29},\frac{70}{29},\frac{72}{29}, \frac{76}{29},\frac{78}{29})$ \\  $(\frac{90}{29},\frac{96}{29},\frac{100}{29},\frac{102}{29},\frac{120}{29},\frac{126}{29},\frac{150}{29})$ } & $\pe{ \frac{q^2+q^8+q^{14} - (q^{16}+q^{22}+q^{28})}{(1-q)(1-q^{29})} }$ & $x^2+y^3+z^5+zw^5=0$ & \\ \hline
            $(E_8(a_2),23)$&  \parbox[c]{4cm}{ $(\frac{50}{23},\frac{52}{23},\frac{54}{23},\frac{60}{23},\frac{70}{23},\frac{72}{23},\frac{90}{23} )$ \\ $(\frac{24}{23},\frac{30}{23},\frac{32}{23},\frac{34}{23},\frac{36}{23},\frac{40}{23},\frac{42}{23})$  } & $\pe{ \frac{q^2+q^8 - (q^{16}+q^{22})}{(1-q)(1-q^{23})} }$ & $(A_4, E_7)$ &~ \\   \hline
      $(E_8(a_3),19)$& \parbox[c]{3cm}{ $({48\over19},{42\over19},{36\over19},{32\over19},{30\over19})$ \\ $({27\over14},{26\over14},{24\over19},{20\over19})$ } & $\pe{ \frac{q^2+q^9 - (q^{11}+q^{18})}{(1-q)(1-q^{19})} }$ & $(A_2,D_9)$ & $\mathrm{Vir}_{3,38}\oplus L(37,1)$  \\ \hline
      $(E_8(a_4),17)$& \parbox[c]{3.5cm}{ $(\frac{18}{17},\frac{21}{17},\frac{24}{17},\frac{25}{17},\frac{28}{17}, \frac{30}{17})$ \\  $(\frac{33}{17},\frac{36}{17},\frac{40}{17},\frac{45}{17},\frac{48}{17},\frac{60}{17})$}  & $\pe{ \frac{q^2+q^5+q^{8} - (q^{10}+q^{13}+q^{16})}{(1-q)(1-q^{17})} }$&  $(E_6, A_4)$ &  \\  \hline
        $(E_8(a_5),13)$& $({30\over13},{24\over13},{20\over13},({18\over13})^2,{14\over13})$ & $\pe{ \frac{q^2 + q^6 - (q^8 + q^{12})}{(1-q)(1-q^{13})} }$ & $(D_4/Z_3, k+{1\over3}={1\over3})$& $\mathrm{Vir}_{3,26}\oplus L(25,1)$ \\  \hline
                $(E_8(a_6),11)$& $({12\over11},{18\over11},{24\over11},{30\over11},({14\over 11})^2,({20\over11})^2)$& $\pe{ \frac{q^2 + 2q^4 - (2q^8 + q^{10})}{(1-q)(1-q^{11})} }$ & $(D_4,A_4)$ &\\ \hline
$(G_2(a_1),5)$& $({6\over5}^3)$ & $\pe{ \frac{3 q^2 - 3q^4}{(1-q)(1-q^{5})} }$ & $3\times (A_1, A_2)$& $\mathrm{Vir}_{2,5}^{\otimes 3}$ \\ \hline
$(F_4(a_1),11)$& $({24\over11},{21\over11},{18\over11},{15\over11},{12\over11},{16\over 11},{13\over11})$ & $\pe{ \frac{q^2 + q^3 - (q^9 + q^{10})}{(1-q)(1-q^{11})} }$ & $(A_2,A_7)$ & dd \\ \hline
$(F_4(a_2),7)$& $({10\over 7}, {8\over 7})$ & $\pe{ \frac{q^2 - q^6 }{(1-q)(1-q^{7})} }$ &$(A_1, A_4)$& $\mathrm{Vir}_{2,7}$ \\ \hline
$(F_4(a_3),5)$& $({6\over5}^4)$ & $\pe{ \frac{4q^2 - 4q^4}{(1-q)(1-q^{5})} }$ & $4\times(A_1, A_2)$ &  $\mathrm{Vir}_{2,5}^{\otimes 4}$  \\ \hline
 \end{tabular} 
 }               
\end{center}
\caption{$f$ denotes the regular singularity, and $u$ specifies the irregular singularity.  The corresponding W algebra is $W^{-h^\vee+{h^\vee \over u}}(\mathfrak{g},f)$. W-algebra ismorphisms and other properties for cases with $f$ subregular were studied in \cite{2019arXiv190511473A}. $\mathrm{Vir}_{p,q}$ is the vacuum module of Virasoro minimal model with label $(p,q)$, and $L(a,b)$ is a module of the previous minimal model with label $(a,b)$.}
  \label{table:admissibledata}
\end{table}

\begin{table}[h]
\begin{center}
\resizebox{\textwidth}{!}{
\begin{tabular}{|l|l|l|l|l|}
\hline
 $(\mathfrak{g},f,u)$ & Coulomb branch & Schur index & 4d Isomorphism & W-algebra \\ \hline
  $(\mathfrak{e}_6,A_4+A_1,5)$& $ {6\over 5}$ & $\pe{{q^2-q^4\over (1-q)(1-q^{5})}}$ & $(A_1, A_2)$ & $\mathrm{Vir}_{2,5}$  \\ \hline
  $(\mathfrak{e}_7,A_4+A_2,5)$& $ {6\over 5}$ & $\pe{{q^2-q^4\over (1-q)(1-q^{5})}}$ & $(A_1, A_2)$ & $\mathrm{Vir}_{2,5}$  \\ \hline
  $(\mathfrak{e}_8,A_6+A_1,7)$& $ {10\over 7},{8\over 7}$ & $\pe{{q^2-q^6\over (1-q)(1-q^{7})}}$ & $(A_1, A_4)$ & $\mathrm{Vir}_{2,7}$  \\ \hline
 \end{tabular} 
 }               
\end{center}
\caption{Theories without the Higgs branch but whose regular singularity has a flavor symmetry (mass deformation). $f$ denotes the nilpotent orbit of the regular singularity, and $u$ specifies the irregular singularity. The corresponding W algebra is $W^{-h^\vee+{h^\vee \over u}}(\mathfrak{g},f)$.}
  \label{table:admissibledata2}
\end{table}

For theories whose VOA is  admissible lisse W-algebra, we can compute following things:
\begin{itemize}
\item \textbf{Coulomb branch spectrum}: One can compute the Coulomb branch spectrum of these theories using the spectral curve of the corresponding Hitchin system. The irregular singularity part can be found from 
the Newton polygon, and one also need to use the pole structure of regular singularities listed in \cite{Chacaltana:2010ks, Chacaltana:2011ze, Chacaltana:2013oka, Chacaltana:2014jba, Chacaltana:2015bna, Chacaltana:2016shw, Chacaltana:2017boe, Chcaltana:2018zag}.

\item \textbf{Schur index}: The Schur index (hence the vacuum character of the corresponding W-algebra) takes the following form \cite{Xie:2019zlb}:
\begin{equation}
\ch=\pe{{\sum q^{1+j} \chi_{R_j}-q^u\sum q^{-j}\chi_{R_j}\over (1-q)(1-q^u)}}.
\end{equation}
Here $R_j$ is determined by the nilpotent orbit $f$, which in turn gives a decomposition of the adjoint representation of $\mathfrak{g}$, $adj_{\mathfrak{g}}\rightarrow \oplus V_j \otimes R_j$. Here $V_j$ is spin $j$ representation of $SU(2)$ group.  In our case $u=q+k^{'}$, and $(u,h^\vee)=1,~~(u,r)=1$. Here $h^\vee$ is the dual Coxeter number, and $r$ is the lacety of lie algebra.

\item \textbf{Zhu's $C_2$ algebra}: Zhu's $C_2$ algebra can be found from Jacobi algebra of an isolated hypersurface singularity \cite{Xie:2019zlb}.

\end{itemize}

Here are some further remarks about those 4d theories whose VOA is admissible lisse W-algebras:
\begin{itemize}
\item[1] If $u>h^\vee$, the nilpotent orbit $O_{u}$ is just 
the regular nilpotent orbit, so $f$ has to be taken to regular nilpotent orbit. The corresponding W-algebra is just the minimal model of W-algebra listed in table \ref{table:admissibleprinciple}. 
Some of the physical information such as the Coulomb branch, the Schur index, central charge, etc, can be found in \cite{Xie:2019zlb}.  These might be called {\bf admissible principal lisse W-algebra}.  These theories (for $\mathfrak{g}=ADE$ ) can be engineered by following singularity 
\begin{equation}
f_{ADE}(x,y,z)+w^k=0.
\end{equation}
and it is also labeled as $(G,A_{k-1})$ (or $(A_{k-1}, G)$ theory) theory using the notation of \cite{Xie:2019yds}. Here $G$ is the ADE type appearing in above singularity.

\item[2] If $u<h^\vee$ and for classical lie algebra, we can actually see from the index that most of them are isomorphic to an admissible principal lisse W-algebra. For example, consider D type case with the regular singularity of the form $[u,s]$ (here $u$ iand $s$ are odd integers), and the corresponding W-algebra is 
\begin{equation}
W^{-(2N-2)+{2N-2\over u}}(\mathfrak{so}_{2N},[u,s])
\end{equation} 
Using the index formula in \cite{Xie:2019zlb}, we have 
\begin{equation}
\ch=\pe{{q^2+q^4+\ldots+ q^{s-1}-q^u(q^{-1}+q^{-3}+\ldots+q^{-(s-2)})\over (1-q)(1-q^u)}}.
\end{equation}
We now recognize that this is actually the vacuum character of $W^{-(s-2)+{s-2\over u}}(\mathfrak{so}_{s},f_{prin})$!  In fact, one can check that 4d Coulomb branch spectrum agrees with each other.

\item[3]  The data for exceptional case is listed in table \ref{table:admissibledata} and \ref{table:admissibledata2}. Almost all of them is actually isomorphic to principal W-algebras (All of them are admissible principal W-algebra except $E_8$ theory with $u=29$.).

\end{itemize}

\begin{table}[!htb]
\begin{center}
	\begin{tabular}{ |c|l|l|}
		\hline
		~  & SW geometry &VOA  \\ \hline
		$A_{N-1}$ &$x_1^2+x_2^2+x_3^N+z^k=0$ & $W^{-N+{N\over N+k}}(\mathfrak{sl}_N,f_{prin})$\\ \hline
		$D_N$   & $x_1^2+x_2^{N-1}+x_2x_3^2+z^k=0$ &$W^{-(2N-2)+{2N-2\over 2N-2+k}}(\mathfrak{so}_{2N},f_{prin})$ \\     \hline		
		$E_6$  & $x_1^2+x_2^3+x_3^4+z^k=0$ &$W^{-12+{12\over 12+k}}(\mathfrak{e}_6,f_{prin})$  \\     \hline	
		$E_7$   & $x_1^2+x_2^3+x_2x_3^3+z^k=0$ &$W^{-18+{18\over 18+k}}(\mathfrak{e}_7,f_{prin})$  \\     \hline	
		$E_8$    & $x_1^2+x_2^3+x_3^5+z^k=0$ &$W^{-30+{30\over 30+k}}(\mathfrak{e}_8,f_{prin})$ \\     \hline
		$A_{2N-1}/Z_2$& $x_1^2+x_2^2+x^{2N}+xz^{k+{1\over2}}=0$ & $W^{-(2N-1)+{2N-1\over 4N+2k-1}}(\mathfrak{so}_{2N+1},f_{prin})$ \\ \hline
			$D_{N+1}/Z_2$ & $x_1^2+x_2^{N}+x_2x_3^2+x_3z^{k+{1\over2}}=0$ & $W^{-N+{N\over 2N+2k+1}}(\mathfrak{sp}_{2N-2},f_{prin})$ \\ \hline
			$D_4/Z_3$ &$x_1^2+x_2^{3}+x_2x_3^2+x_3z^{k\pm {1\over3}}=0$ & $W^{-4+{4\over 12+3k\pm1}}(\mathfrak{g}_2,f_{prin})$ \\ \hline
			$E_6/Z_2$ &$x_1^2+x_2^{3}+x_3^4+x_3z^{k+{1\over2}}=0$ & $W^{-9+{9\over 18+2k+1}}(\mathfrak{f}_4,f_{prin})$  \\ \hline
	\end{tabular}
\end{center}
\caption{4d SCFT whose VOA is admissible lisse W-algebra. Here $k$ is constrained so that 4d theory has no mass deformation and the level of W-algebra is admissible. Namely, $(k,h^\vee)=1$ for ADE case, $(2k+1, h^\vee)=1$ for BCF case, and finally $(3k\pm 1, h^\vee)=1$ for $G_2$  case! The 
corresponding field theory has no exact marginal deformation.}
\label{table:admissibleprinciple}
\end{table}

\begin{table}[h]
\begin{center}
\resizebox{\textwidth}{!}{
\begin{tabular}{|c|c|c|c|c|}
\hline
$\fg$ & Nilpotent orbit  $f$& Vacuum character &VOA &Isomorphism\\
\hline
$A_{N-1}$ & $[u,\cdots,u,s]$, $0\leq s\leq u-2$ & $\pe{\frac{\sum_{j=1}^{s-1}(q^{j+1}-q^{u-j})}{(1-q)(1-q^u)}}$ & $W^{-N+{N\over u}}(\mathfrak{sl}_N,f)$& $W^{-s+{s\over u}}(\mathfrak{sl}_s,f_{prin})$ \\ \hline
$B_N$ & $[\underbrace{u,\cdots,u}_{even},s]$, $s$ odd, $0\leq s\leq u$ & $\pe{\frac{\sum_{j=1}^{\frac{s-1}{2}}(q^{2j}-q^{u-2j+1})}{(1-q)(1-q^u)}} $ & $W^{-(2N-1)+{2N-1\over u}}(\mathfrak{so}_{2N+1},f)$ &$W^{-(s-2)+{s-2\over u}}(\mathfrak{so}_s,f_{prin})$   \\
\hline 
& $[\underbrace{u,\cdots,u}_{odd},s,1]$, $s$ odd, $1\leq s\leq u-1$ & $\pe{\frac{\sum_{j=1}^{\frac{s-1}{2}}(q^{2j}-q^{u-2j+1})+q^{\frac{s+1}{2}}-q^{u-\frac{s}{2}+\frac{1}{2}}}{(1-q)(1-q^u)}}$&$W^{-(2N-1)+{2N-1\over u}}(\mathfrak{so}_{2N+1},f)$ &$W^{-(s-1)+{s-1\over u}}(\mathfrak{so}_{s+1},f_{prin})$ \\
\hline
$C_N$ & $[\underbrace{u,\cdots,u}_{even},s]$, $s$ even, $0\leq s\leq u-1$ & $\pe{\frac{\sum_{j=1}^{s/2}(q^{2j}-q^{u-2j+1})}{(1-q)(1-q^u)}}$  & $W^{-(N+1)+{N+1\over u}}(\mathfrak{sp}_{2N},f)$ &$W^{-({s+2\over2})+{{s+2\over2}\over u}}(\mathfrak{sp}_s,f_{prin})$ \\
\hline
& $[\underbrace{u,\cdots,u}_{odd},u-1,s]$, $s$ even, $1\leq s\leq u-1$ & $\pe{\frac{\sum_{j=1}^{s/2}(q^{2j}-q^{u-2j+1})+q^{u-s+\frac{1}{2}}-q^{u+s+\frac{1}{2}}}{(1-q)(1-q^u)}}$ &$W^{-(N+1)+{N+1\over u}}(\mathfrak{sp}_{2N},f)$ &~  \\
\hline
$D_N$ & $[\underbrace{u,\cdots,u}_{odd},s]$, $s$ odd, $0\leq s\leq u$ & $\pe{\frac{\sum_{j=1}^{\frac{s-1}{2}}(q^{2j}-q^{u-2j+1})}{(1-q)(1-q^u)} } $ & $W^{-(2N-2)+{2N-2\over u}}(\mathfrak{so}_{2N},f)$ &$W^{-(s-2)+{s-2\over u}}(\mathfrak{so}_s,f_{prin})$  \\
\hline
& $[\underbrace{u,\cdots,u}_{even},s,1]$, $s$ odd, $0\leq s\leq u-1$ & $\pe{\frac{ \sum_{j=1}^{\frac{s-1}{2}}(q^{2j}-q^{u-2j+1}) + q^{\frac{s}{2}+\frac{1}{2}} - q^{u-\frac{s}{2}+\frac{1}{2}} }{(1-q)(1-q^u)} } $ & $W^{-(2N-2)+{2N-2\over u}}(\mathfrak{so}_{2N},f)$&$W^{-(s-1)+{s-1\over u}}(\mathfrak{so}_{s+1},f_{prin})$\\
\hline
\end{tabular} 
}
\end{center}
\caption{Vacuum characters for W-algebra $W^{-h^\vee+{h^\vee \over u}}(\mathfrak{g},f)$, where $\mathfrak{g}$ is classical lie algebra. $u$ is an odd integer and $(u,h^\vee)=1$.}
\label{table:charactersClassical}
\end{table}

\newpage
\section{Non-admissible lisse W-algebra: Classical Lie algebra}
\label{sec:nonadm:classical}

Now consider 4d $\mathcal{N}=2$ SCFT whose associated VOA is non-admissible lisse W-algebra with classical Lie algebra. The 4d theories 
are constructed from 6d $(2,0)$ theory of $A$ and $D$ type (We also consider outer automorphism twist so that VOAs  also include W-algebra with non-simply-laced classical Lie algebra). 
 The irregular singularity is chosen so that there is no mass parameter in it, as in table \ref{table:constraintADEirregular} and \ref{table:constraintwist}, and if the regular singularity is chosen to be the principal one, the corresponding W-algebra 
 is conjectured to be lisse. Such theories are summarized in table \ref{table:lisseprinciple1}. 
 
 \begin{table}[h]
\begin{center}
\resizebox{\textwidth}{!}{
		\begin{tabular}{|c|l|l|l|l|}
			\hline
			$j$ with twist & SW geometry at SCFT point & VOA & Constraint &$[z]$ \\ \hline
			$A_{2N}/Z_2$  &$x_1^2+x_2^2+x^{2N+1}+z^{k+{1\over2}}=0$ & $W^{-(N+1)+{n\over  2u}}(\mathfrak{sp}_{2N},f_{prin})$ & $n$ odd and $g={2N+1\over n}$ odd & ${2n\over u}$ \\ \hline
			~ & $x_1^2+x_2^2+x^{2N+1}+xz^k=0$ & $W^{-(N+1)+{n\over  2u}}(\mathfrak{sp}_{2N},f_{prin})$ & $n$ odd and $g={2N\over n}$ even & ${2n\over u}$\\ \hline
			$A_{2N-1}/Z_2$ & $x_1^2+x_2^2+x^{2N}+xz^{k+{1\over2}}=0$ & $W^{-(2N-1)+{n\over u}}(\mathfrak{so}_{2N+1},f_{prin})$ & $n$ odd and $g={2N-1\over n}$ odd & ${2n\over u}$  \\ \hline
			~ &$x_1^2+x_2^2+x^{2N}+z^{k}=0$ & $W^{-(2N-1)+{n\over u}}(\mathfrak{so}_{2N+1},f_{prin})$ & $n$ odd and $g={2N\over n}$ even & ${2n\over u}$  \\ \hline
			$D_{N}/Z_2$ & $x_1^2+x_2^{N-1}+x_2x_3^2+x_3z^{k+{1\over2}}=0$ &$W^{-N+{n\over 2u}}(\mathfrak{sp}_{2N-2},f_{prin})$ & $n$ even and $g={2N\over n}$ odd & ${n\over u}$\\ \hline
			~  &$x_1^2+x_2^{N-1}+x_2x_3^2+z^{k}=0$ & $W^{-N+{n\over 2u}}(\mathfrak{sp}_{2N-2},f_{prin})$ &$n$ even and $g={2N-2\over n}$ even & ${n\over u}$ \\ \hline
				$D_N$   & $x_1^2+x_2^{N-1}+x_2x_3^2+z^k=0$ &$W^{-(2N-2)+{n\over u}}(\mathfrak{so}_{2N},f_{prin})$ & $n$ even and $g={2N-2\over n}$ odd & ${n\over u}$  \\     \hline
		       $~$   &$x_1^2+x_2^{N-1}+x_2x_3^2+z^k x_3=0$ & $W^{-(2N-2)+{n\over u}}(\mathfrak{so}_{2N},f_{prin})$ & $n$ even and $g={2N\over n}$ even &${n\over u}$\\    \hline
		\end{tabular}
		}
		\caption{4d $\mathcal{N}=2$ SCFT whose 2d VOA is non-admissible lisse W-algebra defined using classical Lie algebra!  $u$ is coprime with $n$, and $u$ is always odd. 
		}
				\label{table:lisseprinciple1}
	\end{center}
\end{table}

\subsection{A physical derivation of associated variety}

Now we consider adding a regular singularity of other type instead of the principal one, and try to find the criteria on $f$ such that the resulting theory has no Higgs branch. As discussed in last section, the crucial input is 
the associated variety of W-algebra with trivial nilpotent orbit, and the corresponding W-algebra is actually the AKM
\begin{equation}
V^{k_{2d}}(\mathfrak{g}),~~~k_{2d}=-h^{\vee}+{n\over r u}.
\label{akm}
\end{equation}
For classical groups,  proper values of $k_{2d}$ are listed in the VOA column of table \ref{table:lisseprinciple1}.  
The associated variety of the above AKM is the closure of a nilpotent orbit which is denoted as $\overline{{\cal O}}_{\mathfrak{g},n,u}$. The growth function of the Schur index (equivalently the character of 
vacuum module of the above AKM algebra) is proportional to the difference of 4d central charge which can be computed using the Coulomb branch data \cite{Xie:2019zlb}
\begin{equation}
{\cal G}=-48(a_{4d}-c_{4d}).
\end{equation}
In our case the growth function of VOA (\ref{akm}) can  be summarized by the following formula
\begin{equation}
{\cal G}_{\mathfrak{g},n,u}=\dim(\mathfrak{g})-{d_{\mathfrak{g}}(n)\over u}
\end{equation}
with $d_{\mathfrak{g}}(n)$ summarized in table \ref{Growth:classical}.  We found this formula by explicitly computing the 4d central charges.

\begin{table}[h]
\begin{center}
\begin{tabular}{|c|c|c|c|}
\hline
 $j$ with twist &  $\mathfrak{g}$ & Constraint on $n$ &$d_{\mathfrak{g}}(n)$ \\ \hline
$A_{2N}/Z_2$  & $\mathfrak{sp}_{2N}$& $n$ odd and $g={2N+1\over n}$ odd  &${g\over 2}n^2-\frac{3n}{2}+g$ \\ \hline
			~ & $\mathfrak{sp}_{2N}$&$n$ odd and $g={2N\over n}$ even &${g\over 2} n^2+g$ \\ \hline
			$A_{2N-1}/Z_2$ & $\mathfrak{so}_{2N+1}$& $n$ odd and $g={2N-1\over n}$ odd&  ${g\over 2}n^2+\frac{3n}{2}+g$  \\ \hline
			~  &  $\mathfrak{so}_{2N+1}$&$n$ odd and $g={2N\over n}$ even  & ${g\over 2} n^2+g$  \\ \hline
			$D_{N}/Z_2$ &$\mathfrak{sp}_{2N-2}$&  $n$ even and $g={2N\over n}$ odd&${g\over 2}n^2-\frac{3n}{2}+g$   \\ \hline
			~  &$\mathfrak{sp}_{2N-2}$&$n$ even and $g={2N-2\over n}$ even & ${g\over 2} n^2+g$  \\ \hline
				$D_N$    &$\mathfrak{so}_{2N}$& $n$ even and $g={2N-2\over n}$ odd & ${g\over 2}n^2+\frac{3n}{2}+g$  \\     \hline
		       $~$   &$\mathfrak{so}_{2N}$& $n$ even and $g={2N\over n}$ even & ${g\over 2} n^2+g$  \\    \hline
\end{tabular}
\end{center}
\caption{Data required to compute the growth function ${\cal G}$ of VOA $V^{-h^\vee+{n\over r u}}(\mathfrak{g})$. ${\cal G}=\dim(\mathfrak{g})-{d_{\mathfrak{g}}(n)\over u}$.}
  \label{Growth:classical}
\end{table}

The lisse condition on 4d theory is equivalent to the following choices of irregular and regular singularity:
\begin{itemize}
\item The irregular singularity is restricted to those listed in table \ref{table:lisseprinciple1}.
\item  The regular singularity is chosen such that $f\in {\cal O}_{\mathfrak{g},n,u}$, therefore $S_f\cap   \overline{\cal O}_{\fg, n,u}$ is trivial.
\end{itemize}
So the choice of regular singularity is determined once we find the associated variety $\overline{\cal O}_{\mathfrak{g},n,u}$. 

\begin{figure}
\begin{center}
\tikzset{every picture/.style={line width=0.75pt}} 

\begin{tikzpicture}[x=0.75pt,y=0.75pt,yscale=-1,xscale=1]

\draw   (71.08,123.46) .. controls (71.08,87.86) and (99.94,59) .. (135.54,59) .. controls (171.14,59) and (200,87.86) .. (200,123.46) .. controls (200,159.06) and (171.14,187.92) .. (135.54,187.92) .. controls (99.94,187.92) and (71.08,159.06) .. (71.08,123.46) -- cycle ;
\draw   (131,81.92) -- (141.08,81.92) -- (141.08,92) -- (131,92) -- cycle ;
\draw    (130,149.92) -- (140,160.92) ;

\draw    (129,160.92) -- (140,149.92) ;

\draw   (365.08,122.46) .. controls (365.08,86.86) and (393.94,58) .. (429.54,58) .. controls (465.14,58) and (494,86.86) .. (494,122.46) .. controls (494,158.06) and (465.14,186.92) .. (429.54,186.92) .. controls (393.94,186.92) and (365.08,158.06) .. (365.08,122.46) -- cycle ;
\draw   (425,80.92) -- (435.08,80.92) -- (435.08,91) -- (425,91) -- cycle ;
\draw    (424,148.92) -- (434,159.92) ;

\draw    (423,159.92) -- (434,148.92) ;

\draw    (219,129.92) -- (349,130.91) ;
\draw [shift={(351,130.92)}, rotate = 180.43] [color={rgb, 255:red, 0; green, 0; blue, 0 }  ][line width=0.75]    (10.93,-3.29) .. controls (6.95,-1.4) and (3.31,-0.3) .. (0,0) .. controls (3.31,0.3) and (6.95,1.4) .. (10.93,3.29)   ;

\draw (162,153) node   {$full$};
\draw (448,151) node   {$f$};
\draw (285,114) node  [align=left] {Going to Higgs branch};
\draw (284,145) node  [align=left] {Closing puncture};

\end{tikzpicture}
\caption{Left: a 4d theory defined by an irregular singularity and a full regular singularity (the nilpotent orbit $f$ is trivial). Right: a 4d theory defined by an irregular singularity and a  regular singularity labeled by $f$. Going to Higgs branch of this theory is equivalent to closing off the regular puncture, i.e. change the type of regular singularity.}
\label{closing}
\end{center}
\end{figure}
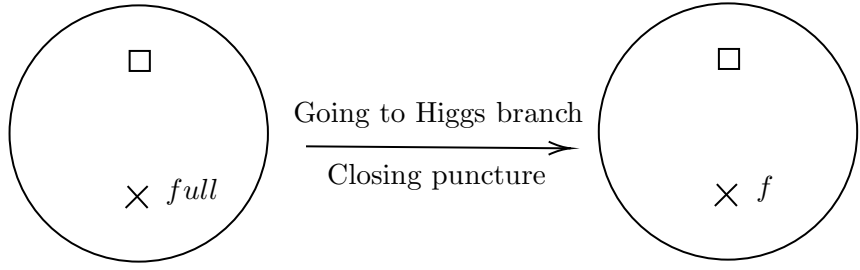

Unlike the admissible case, associated varieties of VOA (\ref{akm}) are not known in the mathematics literature. Here we will use a simple physical method to compute it. 
The idea is as the following: our 4d theory is engineered by a 6d $(2,0)$ theory on a sphere with an irregular and a regular singularity labeled by a trivial nilpotent orbit. Going to the Higgs branch of this theory is 
equivalent to closing the regular puncture, in other words, changing the type of regular puncture (equivalent to choice a different nilpotent orbit) as sketched in figure \ref{closing}. The IR theory consists of free hypers and 
an interacting theory which is described by 6d $(2,0)$ theory on a sphere with one irregular and one regular puncture labeled by the nilpotent element $f$. We would like to choose the right  $f$ such that $\overline{O}_f$ gives the Higgs branch of original theory (the associated variety of the non-admissible AKM). To accomplish that, $f$ has to be chosen as the following:
\begin{enumerate}
\item \label{higgscondition1}The number of Coulomb branch operators in the leading order differential of the corresponding SW curve defined by $f$ should be non-negative. Since otherwise, the right configuration in figure \ref{closing} consists of free hypermultiplets, and we can not simply count the dimension of $\CO_f$ as the dimension of free hypermultiplets.  We also imposing the constraint that the Coulomb branch parameters in the leading order differential is as small as possible. 
\item Once we use above condition to constrain possible $f$, we choose the maximal among them, namely, there is an order among nilpotent orbit, and we choose the maximal one among the nilpotent orbits satisfying condition 1. 
\label{higgscondition2}
\end{enumerate}
Once such a $f$ is found, we conjecture that the Higgs branch (associated variety) of the left theory (non-admissible AKM) of figure \ref{closing} is the closure of $\CO_f$.

\textbf{Example 1}: First we illustrate the idea for the simple case $u>n$, and in this case, we can completely close off the regular puncture, i.e. we can take the regular puncture to be 
labeled by the principal nilpotent orbit (which is the maximal regular singularity we can choose), and the remaining theory is defined by an irregular singularity which has no Higgs branch. So we claim that the associated variety in this case
is given by the closure of principal nilpotent orbit.

\textbf{Example 2}: Now consider 4d theory whose VOA is $V^{-N+{N\over u}}(\mathfrak{sl}_N)$, here $u$ is coprime with $N$ so that there is no mass parameter in irregular singularity of corresponding $(2,0)$ construction. We consider the case $u<N$. The SW curve takes the form 
\begin{equation}
x^N+\sum_{i=2}^N\phi_i(z)x^{N-i}=0.
\end{equation}
Given a general regular puncture labeled by a  Young Tableaux $[h_1^{r_1},h_2^{r_2},\ldots,h_s^{r_s}]$, the maximal scaling dimension of Coulomb branch operator in differential $\phi_N$ is
 \begin{equation}
N-{N\over u}s_N,
 \end{equation}
where $s_N$ is the height of $N$th box of the Young Tableaux (we label the Young Tableaux row by row starting with the bottom-left corner), which is equal to $h_1$. The maximal one we 
can choose so that it satisfies condition \ref{higgscondition1} is $h_1=u$, and the maximal scaling dimension in $\phi_N$ is then $0$.  Using the partial ordering of nilpotent orbit of $\mathfrak{sl}_N$ Lie algebra described in \cite{Collingwood:1993rr}, we find that the regular puncture $f$ which satisfies condition \ref{higgscondition2} is $[u,\ldots, u, s],~~s\leq u$, which agrees with results found by Arakawa \cite{MR3456698}. 

Using the condition \ref{higgscondition1} and \ref{higgscondition2}, we are able to find associated varieties for non-admissible VOAs listed in table \ref{table:associated}. 

Alternatively, we notice that we can find the associated variety using the isomorphism found in \cite{Xie:2019yds}:  our theory can be denoted as $((\mathfrak{g},n,u),f)$, and if $f$ can be chosen so that the above theory is isomorphic to another configuration $((\mathfrak{g}^{'},n^{'},u^{'}),f_{prin})$ which is conjectured to have no Higgs branch, then 
the associated variety of original theory is equal to the closure of $f$. This method can be used to confirm the result from using previous result, and furthermore we can find the resulting lisse W algebra using this method. 

\textbf{Example}: Let's consider following example: consider a four dimensional theory whose 2d VOA is (here $n$ is even, and $g={2N-2\over n}$ is odd)
\begin{equation}
W^{-(2N-2)+{n\over u}}(\mathfrak{so}_{2N},[\underbrace{gu,\ldots, gu}_{even},s,1)]),
\end{equation}
and this is the $D_N$ type theory, see  table. \ref{table:associated}. Using the similar argument of \cite{Xie:2019yds}, we have following equivalence
\begin{equation}
W^{-(2N-2)+{n\over u}}(\mathfrak{so}_{2N},[\underbrace{gu,\ldots, gu}_{even},s,1)])=W^{-(s-1)+{s-1\over u}}(\mathfrak{so}_{s+1},f_{prin})
\end{equation}
The W algebra $W^{-(s-1)+{s-1\over u}}(\mathfrak{so}_{s+1},f_{prin})$ is lisse, which confirms our conjecture of the associated variety of AKM VOA listed in table. \ref{table:associated}. 

 \begin{table}[h]
\begin{center}
\resizebox{\textwidth}{!}{
		\begin{tabular}{|c|c|c|c|}
			\hline
			$j$ with twist & VOA & constraint &associated variety\\ \hline
			$A_{2N}/Z_2$   & $V^{-(N+1)+{n\over  2u}}(\mathfrak{sp}_{2N})$ & $n$ odd and $g={2N+1\over n}$ odd &  \begin{tabular}{c}$[gu+1,\underbrace{gu\ldots,gu}_{\mathrm{even}},s],~0\leq s \leq gu-1, s~\mathrm{even}$ \\  $[gu+1,\underbrace{gu\ldots,gu}_{\mathrm{even}},gu-1,s],~2\leq s \leq gu-1, s~\mathrm{even}$ \end{tabular} \\ \hline
			$~$ &  $V^{-(N+1)+{n\over  2u}}(\mathfrak{sp}_{2N})$ & $n$ odd and $g={2N\over n}$ even & $[gu,\cdots, gu, s]$, $0\leq s\leq gu-1$, $s$ even  \\ \hline
			$A_{2N-1}/Z_2$  & $V^{-(2N-1)+{n\over u}}(\mathfrak{so}_{2N+1})$ & $n$ odd and $g={2N-1\over n}$ odd   & \begin{tabular}{c}$[\underbrace{gu,\cdots, gu}_{\mathrm{even}},s]$, $0\leq s\leq gu$, $s$ odd \\  $[\underbrace{gu,\cdots, gu}_{\mathrm{odd}}, s,1]$, $0\leq s\leq gu-1$, $s$ odd\end{tabular}\\ \hline
			$~$  & $V^{-(2N-1)+{n\over u}}(\mathfrak{so}_{2N+1})$ & $n$ odd and $g={2N\over n}$ even  & \begin{tabular}{c} $[gu+1,\underbrace{gu,\cdots, gu}_{\mathrm{even}},s,1]$, $0\leq s\leq gu-1$, $s$ odd \\$[gu+1,\underbrace{gu,\cdots, gu}_{\mathrm{even}},gu-1,s]$, $0\leq s\leq gu-1$, $s$ odd \\$[gu+1,\underbrace{gu,\cdots, gu}_{\mathrm{even}},gu-1,gu-1,1,1]$   \end{tabular} \\ \hline
			$D_{N}/Z_2$  &$V^{-N+{n\over 2u}}(\mathfrak{sp}_{2N-2})$ & $n$ even and $g={2N\over n}$ odd & \begin{tabular}{c} $[\underbrace{gu,\cdots, gu}_{\mathrm{even}},s]$, $0\leq s\leq gu-1$, $s$ even \\ $[\underbrace{gu,\cdots, gu}_{\mathrm{even}},gu-1, s]$, $0\leq s\leq gu-1$, $s$ even \end{tabular} \\ \hline
			$~$  & $V^{-N+{n\over 2u}}(\mathfrak{sp}_{2N-2})$ &$n$ even and $g={2N-2\over n}$ even & $[gu,\cdots, gu, s]$, $0\leq s\leq gu-1$, $s$ even \\ \hline
				$D_N$   &$V^{-(2N-2)+{n\over u}}(\mathfrak{so}_{2N})$ & $n$ even and $g={2N-2\over n}$ odd & \begin{tabular}{c}$[\underbrace{gu,\cdots, gu}_{\mathrm{odd}},s]$, $0\leq s\leq gu$, $s$ odd \\  $[\underbrace{gu,\cdots, gu}_{\mathrm{even}}, s,1]$, $0\leq s\leq gu-1$, $s$ odd\end{tabular} \\     \hline
		      $~$  & $V^{-(2N-2)+{n\over u}}(\mathfrak{so}_{2N})$ & $n$ even and $g={2N\over n}$ even  & \begin{tabular}{c} $[\underbrace{gu,\cdots, gu}_{\mathrm{even}},s,1]$, $0\leq s\leq gu-1$, $s$ odd \\$[\underbrace{gu,\cdots, gu}_{\mathrm{even}},gu-1,s]$, $0\leq s< gu-1$, $s$ odd  \\ $[\underbrace{gu,\cdots, gu}_{\mathrm{even}},gu-1,gu-1,1,1]$ \end{tabular}\\ \hline
		\end{tabular}
		}
		\caption{The associated variety of  non-admissible lisse W-algebra defined using classical Lie algebra. $u$ is coprime with $n$, and $u$ is always odd. The associated variety is the closure of the nilpotent orbit of Lie algebra used to define W-algebra.
		}
				\label{table:associated}
	\end{center}
\end{table}

\subsection{Some details of D  type theory}
\begin{figure}

\tikzset{every picture/.style={line width=0.75pt}} 
\resizebox{\textwidth}{!}{
\begin{tikzpicture}[x=0.75pt,y=0.75pt,yscale=-1,xscale=1]

\draw    (82.5,58.92) -- (112.5,58.92) ;

\draw   (112.5,58.92) .. controls (112.5,41.8) and (126.38,27.92) .. (143.5,27.92) .. controls (160.62,27.92) and (174.5,41.8) .. (174.5,58.92) .. controls (174.5,76.04) and (160.62,89.92) .. (143.5,89.92) .. controls (126.38,89.92) and (112.5,76.04) .. (112.5,58.92) -- cycle ;
\draw    (174.5,58.92) -- (204.5,58.92) ;

\draw    (237.5,58.92) -- (267.5,58.92) ;

\draw   (267.5,58.92) .. controls (267.5,41.8) and (281.38,27.92) .. (298.5,27.92) .. controls (315.62,27.92) and (329.5,41.8) .. (329.5,58.92) .. controls (329.5,76.04) and (315.62,89.92) .. (298.5,89.92) .. controls (281.38,89.92) and (267.5,76.04) .. (267.5,58.92) -- cycle ;
\draw  [dash pattern={on 4.5pt off 4.5pt}]  (329.5,58.92) -- (383.5,58.92) ;

\draw    (415.5,58.92) -- (445.5,58.92) ;

\draw   (445.5,58.92) .. controls (445.5,41.8) and (459.38,27.92) .. (476.5,27.92) .. controls (493.62,27.92) and (507.5,41.8) .. (507.5,58.92) .. controls (507.5,76.04) and (493.62,89.92) .. (476.5,89.92) .. controls (459.38,89.92) and (445.5,76.04) .. (445.5,58.92) -- cycle ;
\draw    (507.5,58.92) -- (537.5,58.92) ;

\draw    (570.5,58.92) -- (600.5,58.92) ;

\draw   (601,33.92) -- (651,33.92) -- (651,83.92) -- (601,83.92) -- cycle ;
\draw    (82.5,148.92) -- (112.5,148.92) ;

\draw   (112.5,148.92) .. controls (112.5,131.8) and (126.38,117.92) .. (143.5,117.92) .. controls (160.62,117.92) and (174.5,131.8) .. (174.5,148.92) .. controls (174.5,166.04) and (160.62,179.92) .. (143.5,179.92) .. controls (126.38,179.92) and (112.5,166.04) .. (112.5,148.92) -- cycle ;
\draw    (174.5,148.92) -- (204.5,148.92) ;

\draw    (237.5,148.92) -- (267.5,148.92) ;

\draw   (267.5,148.92) .. controls (267.5,131.8) and (281.38,117.92) .. (298.5,117.92) .. controls (315.62,117.92) and (329.5,131.8) .. (329.5,148.92) .. controls (329.5,166.04) and (315.62,179.92) .. (298.5,179.92) .. controls (281.38,179.92) and (267.5,166.04) .. (267.5,148.92) -- cycle ;
\draw  [dash pattern={on 4.5pt off 4.5pt}]  (329.5,148.92) -- (383.5,148.92) ;

\draw    (415.5,148.92) -- (445.5,148.92) ;

\draw   (445.5,148.92) .. controls (445.5,131.8) and (459.38,117.92) .. (476.5,117.92) .. controls (493.62,117.92) and (507.5,131.8) .. (507.5,148.92) .. controls (507.5,166.04) and (493.62,179.92) .. (476.5,179.92) .. controls (459.38,179.92) and (445.5,166.04) .. (445.5,148.92) -- cycle ;
\draw    (507.5,148.92) -- (537.5,148.92) ;

\draw    (570.5,148.92) -- (600.5,148.92) ;

\draw   (601,123.92) -- (651,123.92) -- (651,173.92) -- (601,173.92) -- cycle ;

\draw (66,61) node [scale=1.2]  {$T^{u}_{1}$};
\draw (143.5,58.92) node [scale=0.7]  {$Sp( n-2)$};
\draw (221,61) node [scale=1.2]  {$T^{u}_{2}$};
\draw (298.5,58.92) node [scale=0.7]  {$SO( 2n)$};
\draw (401,60) node [scale=1.2]  {$T^{u}_{l}$};
\draw (476.5,58.92) node [scale=0.7]  {$Sp( ln-2)$};
\draw (556,60) node [scale=1.2]  {$T^{u}_{l+1}$};
\draw (626,58.92) node [scale=0.7]  {$SO( 2N)$};
\draw (27,52.92) node  [align=left] {a)};
\draw (66,151) node [scale=1.2]  {$T^{u}_{1}$};
\draw (143.5,148.92) node [scale=0.7]  {$SO( n+2)$};
\draw (221,151) node [scale=1.2]  {$T^{u}_{2}$};
\draw (298.5,148.92) node [scale=0.7]  {$Sp( 2n)$};
\draw (401,150) node [scale=1.2]  {$T^{u}_{l}$};
\draw (478.5,147.92) node [scale=0.7]  {$Sp( ln)$};
\draw (556,150) node [scale=1.2]  {$T^{u}_{l+1}$};
\draw (626,148.92) node [scale=0.7]  {$SO( 2N)$};
\draw (27,142.92) node  [align=left] {b)};
\draw (27,78.92) node   {$l\ \mathrm{odd}, ~n\ \mathrm{even}$};
\draw (29,164.92) node   {$l\ \mathrm{even},~~n\ \mathrm{even}$};

\end{tikzpicture}
}

\caption{Quiver for untwisted $D_N$ type theory defined using irregular singularity listed in \ref{Dirregular}, and a full D type regular singularity. Here we have a) $l=\frac{2N}{n}-1$ b) $l=\frac{2N-2}{n}-1$; and  $u=n+k$.}
\label{Dtypequiver}
\end{figure}
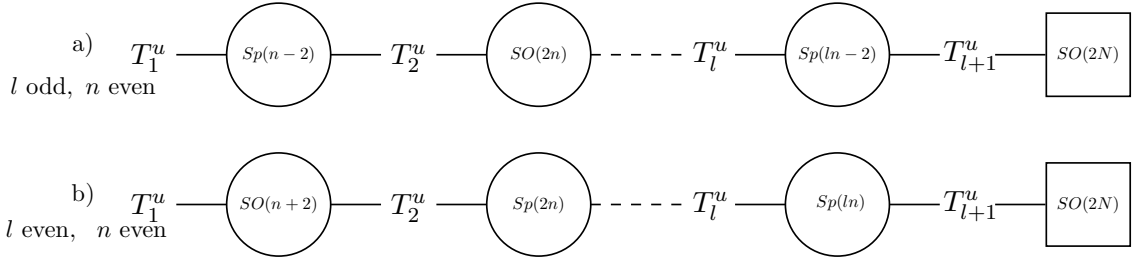

For the untwisted  $D_N$ type theory, the irregular singularity takes following form
\begin{equation}
\Phi={T\over z^{2+{k\over n}}}+\ldots
\label{Dirregular}
\end{equation}
with $(n,k)=1$,  and $n$ is chosen such that there is no mass parameter in the irregular singularity: 
either \textbf{a}) n is  an even divisor of $2N-2$ and ${2N-2\over n}$ is odd, or \textbf{b}) n is an even divisor of $N$. 
$T$ is chosen to be a regular semi-simple element of $D$ type. 
We first consider adding a full regular singularity of $D_N$ type besides the irregular singularity.  These theories have a weakly coupled 
gauge theory descriptions \cite{Xie:2017aqx} with its quiver shown in figure \ref{Dtypequiver}.  The corresponding 2d VOA is AKM
\begin{equation}
V^{-(2N-2)+{n\over n+k}}(\mathfrak{so}_{2N}).
\end{equation}

The Schur index can be computed from the index of each individual matter system.  We illustrate this by using the 
quiver listed in top row of figure \ref{Dtypequiver}. Define $u=n+k$. 
Each AD matter $T_i^u$ carries flavor symmetry $C_{i}$ and $D_{i}$ (Here $C$ means the symplectic group, and $D$ means orthogonal group).
The character of the AD matter is given by the following formula \cite{Xie:2019yds}
\begin{equation}
\ch(T_i^u)=\pe{{(q-q^u)\chi_{C_i}^{adj}+(q-q^u)\chi_{D_i}^{adj}+(q^{u\over2}-q^{{u\over2}+1})(\chi_{C_i}^{f}\chi_{D_i}^{\bar{f}}+\chi_{C_i}^{\bar{f}}\chi_{D_i}^{f})\over (1-q)(1-q^u)}}.
\end{equation}
Here $\chi_{C_i}^{adj}$ is the character of the adjoint representation of $C_i$ Lie algebra, and $\chi_{C_i}^{f}$ is the character of the fundamental representation of $C_i$, and etc.  It is interesting to 
note that when $u=1$, the above index reduces to the index of the free bi-fundamental hypermultiplets in $C_i$ and $D_i$ group. So our matter $T_i^u$ can be thought of as an infinite sequence of 
matter generalizing the usual bi-fundamental matter.  Now the vector multiplet (the gauge group is $\mathfrak{g}$) would contribute to the index by
\begin{equation}
\ch(C_i)=\pe{{-2q\chi_{\mathfrak{g}}^{adj}\over 1-q}}.
\end{equation}
Combine the contribution from matters and vector multiplets, we have the following index formula
\begin{equation}
\begin{split}
&\ch=\pe{{(q-q^u)\chi_{so(2N)}^{adj}\over (1-q)(1-q^u)}}\times \nonumber\\
&\int \prod_{i=1}^{{l+1\over2}} dC_i \prod_{i=1}^{{l-1\over2}}d D_i \pe{{-2 q^u (\sum_{i=1}^{{l+1\over2}} (\chi_{C_i}^{adj} +\chi_{D_i}^{adj}))+2q^{u\over2}(\sum_{i=1}^{{l+1\over2}} \chi_{C_i}^f\chi_{D_i}^f+\sum_{i=1}^{{l-1\over2}}\chi_{D_i}^f\chi_{C_{i+1}}^f) \over (1-q^u)} }.
\end{split}
\end{equation}
Notice that the integral does not involve the flavor group $SO(2N)$ and the integral actually does depend on fugacities of $SO(2N)$ flavor group.
This formula can be generalized to the bottom quiver in figure \ref{Dtypequiver}. The rule is quite simple: $u=1$ is just the quiver gauge theory with free bi-fundamental hypermultiplets as matter, and we write down the index using the free field realization, 
and to get the index for general case one simply replaces $q$ by $q^u$ in the full index (a prefactor which is zero for $u=1$ is also needed).

The growth function can be computed using the following two formulas
\begin{equation}
2a_{4d}-c_{4d}={1\over 4} \sum_{[u]>1} (2[u] -1),~~~~~c_{4d}=-{1\over12}{k_{2d}\dim(\mathfrak{so}_{2N}) \over h^\vee +k_{2d}}.
\end{equation}
Here we use the fact that the 2d VOA of above theories are AKM  with level $k_{2d}= -h^\vee+{n\over u}$, and the relation $c_{4d}=-{1\over12} c_{2d}$.  Since we know the Coulomb branch spectrum, 
we can compute the central charge $a_{4d}$ and therefore compute the growth function of the character ${\cal G}=-48(a_{4d}-c_{4d})$.  We find that 
the growth function takes the following form
\begin{equation}
{\cal G}=\dim \mathfrak{so}_{2N}-{d_{\mathfrak{so}_{2N}}(n)\over u}.
\end{equation}
with the function ${d_{\mathfrak{so}_{2N}}(n)}$ given by:
\begin{equation}
\begin{split}
&l={2N\over n}-1,~l~\mathrm{odd}:~~~~{d_{\mathfrak{so}_{2N}}(n)}={l+1\over 2}n^2+(l+1), \nonumber\\
&l={2N-2\over n}-1,~~l~\mathrm{even}:~~~{d_{\mathfrak{so}_{2N}}(n)}={l+1\over 2}n^2+\frac{3n}{2}+(l+1).
\end{split}
\end{equation}

Now we discuss the associated variety (the Higgs branch) of out theory. 
Consider $u=1$ case where the theory is just a linear quiver with Lagrangian description and the Higgs branch is just the closure of a nilpotent orbit, which can be read from the quiver shown in figure \ref{Dtypequiver}.  

First we recall how to associate a quiver tail to a D type nilpotent orbit \cite{Benini:2010uu}. 
For a nilpotent orbit with partition $Y=[h_1, h_2,\ldots, h_J]$ (where even $h_i$ appears even times), we attach a quiver tail (See appendix for brane construction):
\begin{equation}
[SO(2N)]-USp(r_1)-O(r_2)-\ldots-USp(r_{J-1})
\end{equation}
Here $J$ is always even (The end USp gauge group might be rank zero.). The sizes are
\begin{equation}
r_a=[\sum_{b=a+1}^Jh_b]_{+,-},~~~+:O,~~-:USp
\end{equation}
and $[n]_{+(-)}$ means the smallest (largest) even integer $\geq n$ ($\leq n$).  The Higgs branch of the above quiver is the closure of the dual orbit $Y^d$ of the original orbit $Y$. For our quiver shown in figure. \ref{Dtypequiver},  we find:
\begin{equation}
\begin{split}
&l={2N\over n}-1,~l~\mathrm{odd}:~~~~Y=[n+1,\underbrace{n,\ldots,n}_{l-1},n-1],~~Y^d=[\underbrace{l+1,\ldots,l+1}_{n-2},l,l,1,1], \\
&l={2N-2\over n}-1,~~l~\mathrm{even}:~~~Y=[n+1,\underbrace{n,\ldots,n}_{l},1],~~Y^d=[\underbrace{l+1,\ldots,l+1}_{n},1,1]. 
\end{split}
\end{equation}
The results are consistent with associated varieties listed in table \ref{table:associated}. The dimension of the Higgs branch is the dimension of the nilpotent orbit $\CO_{Y^d}$ which can be computed from $Y^d$
\begin{equation}
\begin{split}
&l={2N\over n}-1,~l~\mathrm{odd}:~~~~\dim{\CO_{Y^d}}=2N^2-N-\half((l+1)n^2+4), \\
&l={2N-2\over n}-1,~~l~\mathrm{even}:~~~\dim{\CO_{Y^d}}=2N^2-N-\half((l+1)n^2+3n+2). 
\end{split}
\end{equation}
To match the UV anomaly $-48(a_{4d}-c_{4d})$ (growth function), the gauge group is not completely Higgsed, and the there are ${l-1}\over 2$ free vector multiplet for $l$ odd, 
and ${l\over2}$ free vector multiplet for $l$ even.

We can compute the Schur index or the vacuum character of non-admissible W-algebra using the gauge theory description too. 
We illustrate for the theory with only irregular singularity, which is in the form of \ref{Dirregular} with the following constraint:
\begin{equation}
n~\mathrm{even}, ~l={2N-2\over n}-1~\mathrm{even},~~u=n+k~\mathrm{odd}.
\end{equation}
We take $u>n$ for simplicity.  Now define an integer number $a=[{2N\over k}]$ and $b=l-a$. If $a=0$, the quiver takes the same form as in figure \ref{Dtypequiver}, and the only difference is that 
  $T_{l+1}^u$ has no $SO(2N)$ flavor symmetry (The matter systems and the gauge groups are different though). The index takes following form   
  \begin{equation}
\begin{split}
&\ch=\pe{{\sum q^{d_i}-q^{u+1}(\sum q^{-d_i})\over (1-q)(1-q^u)}}\times \nonumber\\
&\int \prod_{i=1}^{{l+1\over2}} dC_i \prod_{i=1}^{{l-1\over2}}d D_i \pe{{-2 q^u (\sum_{i=1}^{{l+1\over2}} (\chi_{C_i}^{adj} +\chi_{D_i}^{adj}))+2q^{u\over2}(\sum_{i=1}^{{l-1\over2}} \chi_{C_i}^f\chi_{D_i}^f+\sum_{i=1}^{{l-1\over2}}\chi_{D_i}^f\chi_{C_{i+1}}^f) \over (1-q^u)} }.
\end{split}
\end{equation}
Here $d_i$ are Casmiers of $SO(2N)$ group. If $a\neq 0$,  the quiver takes a more complicated form:
\begin{equation}
T_1^u-Sp(n-2)-T_2^u-SO(2n)-\ldots-G_0(bn)-T_{b+1}^u-G_1(ak)-\ldots-SO(2k+1)-T_l^u-Sp(k-1)-T_{l+1}^u-O(1)
\end{equation}
if $a$ is even, then $G_0(bn)=Sp(bn-2),~G_1(ak)=SO(ak+1)$; and if $a$ is odd, then $G_0(bn)=SO(bn),~G_1(ak)=Sp(ak-1)$. Notice that there is half-hypermultiplet coupled with $Sp(k-1)$ gauge group. 
The Schur index takes the following form:
\begin{equation}
\begin{split}
&\ch=\pe{{\sum q^{1+j}-q^{u}(\sum q^{-j})\over (1-q)(1-q^u)}}\times \nonumber\\
&\int \prod_{i=1}^{{l+1\over2}} dC_i \prod_{i=1}^{{l-1\over2}}d D_i \pe{{-2 q^u (\sum_{i=1}^{{l+1\over2}} (\chi_{C_i}^{adj} +\chi_{D_i}^{adj}))+2q^{u\over2}(\sum_{i=1}^{{l+1\over2}} \chi_{C_i}^f\chi_{D_i}^f+\sum_{i=1}^{{l-1\over2}}\chi_{D_i}^f\chi_{C_{i+1}}^f) \over (1-q^u)} }.
\end{split}
\end{equation}
Here the set of $\{j\}=\{1,3,5,\ldots,(b+1)n-ak-3\}$ for $a$ even, and  $\{j\}=\{1,3,5,\ldots,(b+1)n-ak-2\}$ for $a$ odd.

\textbf{Isomorphism}: if $n=N$ with $N$ even, then $l=1$, the theory is actually defined by following singularity: $x^2+y^{N-1}+y z^2+z w=0$, which is actually equivalent to $(A_1, A_{2N})$ theory. 

\subsection{Other cases}
\textbf{Twisted $D_{N}$ type theory}
For the twisted $D_N$ type theory, the irregular singularity takes the following form
\begin{equation}
\Phi={T^t\over z^{2+{k\over n}}}+\ldots
\label{twistedDirr}
\end{equation}
with $(n,k)=1$. Here $n$ is chosen such that there is no mass parameter in the irregular singularity: either \textbf{a}) n is 
 an even divisor of $2N-2$ and ${2N-2\over n}$ is even, or
\textbf{b}) n is a even divisor of $2N$, and ${2N\over n}$ is odd. 
$T^t$ is chosen to be a regular semi-simple element of $C$ type Lie algebra.  
These theories also have a weakly coupled gauge theory description, see figure \ref{twistedD} for the theory defined by the above irregular singularity plus a  regular singularity labeled by trivial nilpotent orbit.

The Schur index for the quiver gauge theory shown in figure \ref{twistedD} can be computed using the index for each matter content, and the result is
\begin{equation}
\ch=\pe{{(q-q^u)\chi_{sp(2N-2)}^{adj}\over (1-q)(1-q^u)}}\ch_{n,u=1}(q^u),
\end{equation}
here $\ch_{n,u=1}(q)$ is the Schur index for the theory defined with $u=1$, which is a Lagrangian theory. For other theories defined by the same irregular singularity and different regular singularity, we could also write down its Schur index using the weakly coupled gauge theory description, and we leave it for the interested reader.

Let us now discuss a little bit of the Higgs branch of the theory shown in figure \ref{twistedD}.
If $u=1$, the theory is just a Lagrangian theory. The idea is: 
 We start from a
$B_{N-1}$ partition $Y_B$ and find its associated quiver which has $C_{N-1}$ type flavor symmetry (See appendix for the explanation), we get the higgs branch of the associated quiver by finding its dual partition $Y_C^d$ in its Langlands dual $C_{N-1}$ algebra. For the quiver shown in   figure \ref{twistedD} with $u=1$, we have:
\begin{equation}
\begin{split}
&l={2N\over n}-1,~l~\mathrm{even}:~~~~Y_B=[\underbrace{n,\ldots,n}_{l},n-1],~~Y^d_C=[\underbrace{l+1,\ldots,l+1}_{n-2},l,l], \\
&l={2N-2\over n}-1,~~l~\mathrm{odd}:~~~Y_B=[\underbrace{n,\ldots,n}_{l+1},1],~~Y^d_C=[\underbrace{l+1,\ldots,l+1}_{n}]. \\
\end{split}
\end{equation}
This agrees with the result shown in table \ref{table:associated}. The dimension of the Higgs branch of the quiver in \ref{twistedD} with $u=1$  is
\begin{equation}
\begin{split}
&l={2N\over n}-1,~l~\mathrm{even}:~~~~\dim\CO_{Y^d_C}= 2N^2+N-\half((l+1)n^2-3n+2), \\
&l={2N-2\over n}-1,~~l~\mathrm{odd}:~~~\dim\CO_{Y^d_C}= 2N^2+N-\half(l+1)n^2.
\end{split}
\end{equation}
Comparing with the UV growth, we find that in the Higgs branch there are $\frac{l}{2}$ free vector multiplets for $l$ even, and $\frac{l+1}{2}$ free vector multiplets for $l$ odd.

\textbf{Twisted  $\fsl_{2N}$ type theory}:
For the twisted $\fsl_{2N}$ theory, the Higgs field takes following form
\begin{equation}
\Phi={T^t\over z^{2+{k+1/2\over n}}}+\ldots.
\label{Btwistedirr}
\end{equation}
Here $n$ can be chosen as follows: either \textbf{a}) $n$ is a odd divisor of $2N$, or \textbf{b}) $n$ is a divisor of $2N-1$.  The corresponding quiver description is shown in figure \ref{Btype}. $T^t$ is a regular semi-simple element of $C$ type Lie algebra.

To associate a quiver tail  whose Higgs branch is $B_N$ type nilpotent orbit (so the Higgs branch has $B_N$ type flavor symmetry),  
we actually need to start with a nilpotent orbit of $C_N$ type with partition $Y=[h_1, h_2,\ldots, h_J]$ (where odd $h_i$ appears even times), then attach the following quiver tail
\begin{equation}
[SO(2N+1)]-USp(r_1)-O(r_2)-\ldots-USp(r_{J-1})
\end{equation}
Here $J$ is always even (The end USp gauge group might be rank zero). The sizes are
\begin{equation}
r_a=[1+\sum_{b=a+1}^Jh_b]_{+,-},~~~+:O,~~-:USp
\end{equation}
and $[n]_{+}$ means the smallest  \textbf{odd} integer $\geq n$, and $[n]_{-}$ is the largest \textbf{even} integer $\leq n$.  The Higgs branch of above quiver is the closure of Langlands dual $Y_B^d$ of nilpotent orbit $Y_C$. For the quiver shown in figure. \ref{Btype}, we have:
\begin{equation}
\begin{split}
&l={2N\over n}-1,~l~\mathrm{odd}:~~~~Y_{C_N}=[n+1,\underbrace{n,\ldots,n}_{l-1},n-1],~~Y_{B_N}^d=[l+2,\underbrace{l+1,\ldots,l+1}_{n-3},l,l,1,1], \\
&l={2N-1\over n}-1,~l~\mathrm{even}:~~~Y_{C_N}=[n+1,\underbrace{n,\ldots,n}_{l}],~~Y_{B_N}^d=[\underbrace{l+1,\ldots,l+1}_{n},1,1]. \\
\end{split}
\end{equation}
The dimension is
\begin{equation}
\begin{split}
&l={2N\over n}-1,~l~\mathrm{odd}:~~~~\dim{\CO_{Y_{B}^d}}=2N^2+N-\half((l+1)n^2+4), \\
&l={2N-1\over n}-1,~l~\mathrm{even}:~~~\dim{\CO_{Y_{B}^d}}=2N^2+N-\half((l+1)n^2+3n+2). \\
\end{split}
\end{equation}
To match the UV growth, the gauge group is not completely Higgsed, and the there are ${l+1}\over 2$ free vector multiplet for $l$ odd, 
and ${l\over2}$ free vector multiplet for $l$ even.

\begin{figure}

\tikzset{every picture/.style={line width=0.75pt}} 
\resizebox{\textwidth}{!}{
\begin{tikzpicture}[x=0.75pt,y=0.75pt,yscale=-1,xscale=1]

\draw    (82.5,58.92) -- (112.5,58.92) ;

\draw   (112.5,58.92) .. controls (112.5,41.8) and (126.38,27.92) .. (143.5,27.92) .. controls (160.62,27.92) and (174.5,41.8) .. (174.5,58.92) .. controls (174.5,76.04) and (160.62,89.92) .. (143.5,89.92) .. controls (126.38,89.92) and (112.5,76.04) .. (112.5,58.92) -- cycle ;
\draw    (174.5,58.92) -- (204.5,58.92) ;

\draw    (237.5,58.92) -- (267.5,58.92) ;

\draw   (267.5,58.92) .. controls (267.5,41.8) and (281.38,27.92) .. (298.5,27.92) .. controls (315.62,27.92) and (329.5,41.8) .. (329.5,58.92) .. controls (329.5,76.04) and (315.62,89.92) .. (298.5,89.92) .. controls (281.38,89.92) and (267.5,76.04) .. (267.5,58.92) -- cycle ;
\draw  [dash pattern={on 4.5pt off 4.5pt}]  (329.5,58.92) -- (383.5,58.92) ;

\draw    (415.5,58.92) -- (445.5,58.92) ;

\draw   (445.5,58.92) .. controls (445.5,41.8) and (459.38,27.92) .. (476.5,27.92) .. controls (493.62,27.92) and (507.5,41.8) .. (507.5,58.92) .. controls (507.5,76.04) and (493.62,89.92) .. (476.5,89.92) .. controls (459.38,89.92) and (445.5,76.04) .. (445.5,58.92) -- cycle ;
\draw    (507.5,58.92) -- (537.5,58.92) ;

\draw    (570.5,58.92) -- (600.5,58.92) ;

\draw   (601,33.92) -- (651,33.92) -- (651,83.92) -- (601,83.92) -- cycle ;
\draw    (82.5,148.92) -- (112.5,148.92) ;

\draw   (112.5,148.92) .. controls (112.5,131.8) and (126.38,117.92) .. (143.5,117.92) .. controls (160.62,117.92) and (174.5,131.8) .. (174.5,148.92) .. controls (174.5,166.04) and (160.62,179.92) .. (143.5,179.92) .. controls (126.38,179.92) and (112.5,166.04) .. (112.5,148.92) -- cycle ;
\draw    (174.5,148.92) -- (204.5,148.92) ;

\draw    (237.5,148.92) -- (267.5,148.92) ;

\draw   (267.5,148.92) .. controls (267.5,131.8) and (281.38,117.92) .. (298.5,117.92) .. controls (315.62,117.92) and (329.5,131.8) .. (329.5,148.92) .. controls (329.5,166.04) and (315.62,179.92) .. (298.5,179.92) .. controls (281.38,179.92) and (267.5,166.04) .. (267.5,148.92) -- cycle ;
\draw  [dash pattern={on 4.5pt off 4.5pt}]  (329.5,148.92) -- (383.5,148.92) ;

\draw    (415.5,148.92) -- (445.5,148.92) ;

\draw   (445.5,148.92) .. controls (445.5,131.8) and (459.38,117.92) .. (476.5,117.92) .. controls (493.62,117.92) and (507.5,131.8) .. (507.5,148.92) .. controls (507.5,166.04) and (493.62,179.92) .. (476.5,179.92) .. controls (459.38,179.92) and (445.5,166.04) .. (445.5,148.92) -- cycle ;
\draw    (507.5,148.92) -- (537.5,148.92) ;

\draw    (570.5,148.92) -- (600.5,148.92) ;

\draw   (601,123.92) -- (651,123.92) -- (651,173.92) -- (601,173.92) -- cycle ;

\draw (66,61) node [scale=1.2]  {$T^{u}_{1}$};
\draw (143.5,58.92) node [scale=0.7]  {$SO( n+2)$};
\draw (221,61) node [scale=1.2]  {$T^{u}_{2}$};
\draw (298.5,58.92) node [scale=0.7]  {$Sp( 2n)$};
\draw (401,60) node [scale=1.2]  {$T^{u}_{l}$};
\draw (476.5,58.92) node [scale=0.7]  {$SO( ln+2)$};
\draw (556,60) node [scale=1.2]  {$T^{u}_{l+1}$};
\draw (626,58.92) node [scale=0.7]  {$Sp( 2N)$};
\draw (27,52.92) node  [align=left] {a)};
\draw (66,151) node [scale=1.2]  {$T^{u}_{1}$};
\draw (143.5,148.92) node [scale=0.7]  {$Sp( n-1)$};
\draw (221,151) node [scale=1.2]  {$T^{u}_{2}$};
\draw (298.5,148.92) node [scale=0.7]  {$SO( 2n+1)$};
\draw (401,150) node [scale=1.2]  {$T^{u}_{l}$};
\draw (478.5,147.92) node [scale=0.7]  {$SO( ln+1)$};
\draw (556,150) node [scale=1.2]  {$T^{u}_{l+1}$};
\draw (626,148.92) node [scale=0.7]  {$Sp( 2N)$};
\draw (27,142.92) node  [align=left] {b)};
\draw (27,78.92) node   {$l\ odd,~n\ odd$};
\draw (29,164.92) node   {$l\ even,~n\ odd$};

\end{tikzpicture}}

\caption{Quiver for twisted $sl_{2N+1}$ type theory defined using formula \ref{Ctwistedirr}. a) $l=\frac{2N}{n}-1$ b) $l=\frac{2N+1}{n}-1$, here $u=2n+2k+1$. For $l$ even and $u=1$, $T_1^u$ is a half hyper for the first gauge group.}
\label{C2type}
\end{figure}
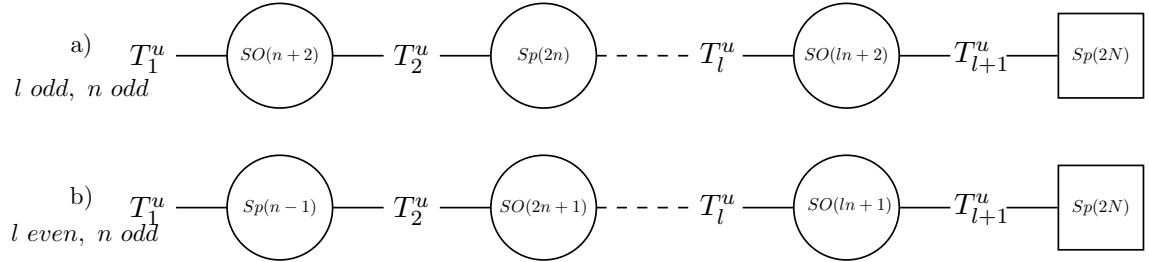

\textbf{Twisted  $\fsl_{2N+1}$ type theory}:
For the twisted $\fsl_{2N+1}$ theory, the Higgs field takes following form
\begin{equation}
\Phi={T^t\over z^{2+{k+1/2\over n}}}+\ldots
\label{Ctwistedirr}
\end{equation}
Here $n$ can be chosen as follows: either a) $n$ is an odd divisor of $2N$, or b) $n$ is a divisor of $2N+1$. The corresponding quiver is shown in figure \ref{C2type}. 
We can not find the Higgs branch for $u=1$ theory using previous methods as the quiver can not be realized as the quiver tail of a $B$ type partition, as the quiver constructed 
from $B$ type partition would have $D$ type and $C$ type quiver nodes, but our quiver shown in figure. \ref{C2type}  has $B$ type and $C$ type quiver nodes.


\begin{figure}

\tikzset{every picture/.style={line width=0.75pt}} 
\resizebox{\textwidth}{!}{
\begin{tikzpicture}[x=0.75pt,y=0.75pt,yscale=-1,xscale=1]

\draw    (72.5,78.92) -- (102.5,78.92) ;

\draw   (102.5,78.92) .. controls (102.5,61.8) and (116.38,47.92) .. (133.5,47.92) .. controls (150.62,47.92) and (164.5,61.8) .. (164.5,78.92) .. controls (164.5,96.04) and (150.62,109.92) .. (133.5,109.92) .. controls (116.38,109.92) and (102.5,96.04) .. (102.5,78.92) -- cycle ;
\draw    (164.5,78.92) -- (194.5,78.92) ;

\draw    (227.5,78.92) -- (257.5,78.92) ;

\draw   (257.5,78.92) .. controls (257.5,61.8) and (271.38,47.92) .. (288.5,47.92) .. controls (305.62,47.92) and (319.5,61.8) .. (319.5,78.92) .. controls (319.5,96.04) and (305.62,109.92) .. (288.5,109.92) .. controls (271.38,109.92) and (257.5,96.04) .. (257.5,78.92) -- cycle ;
\draw  [dash pattern={on 4.5pt off 4.5pt}]  (319.5,78.92) -- (373.5,78.92) ;

\draw    (405.5,78.92) -- (435.5,78.92) ;

\draw   (435.5,78.92) .. controls (435.5,61.8) and (449.38,47.92) .. (466.5,47.92) .. controls (483.62,47.92) and (497.5,61.8) .. (497.5,78.92) .. controls (497.5,96.04) and (483.62,109.92) .. (466.5,109.92) .. controls (449.38,109.92) and (435.5,96.04) .. (435.5,78.92) -- cycle ;
\draw    (497.5,78.92) -- (527.5,78.92) ;

\draw    (560.5,78.92) -- (590.5,78.92) ;

\draw   (591,53.92) -- (641,53.92) -- (641,103.92) -- (591,103.92) -- cycle ;
\draw    (72.5,168.92) -- (102.5,168.92) ;

\draw   (102.5,168.92) .. controls (102.5,151.8) and (116.38,137.92) .. (133.5,137.92) .. controls (150.62,137.92) and (164.5,151.8) .. (164.5,168.92) .. controls (164.5,186.04) and (150.62,199.92) .. (133.5,199.92) .. controls (116.38,199.92) and (102.5,186.04) .. (102.5,168.92) -- cycle ;
\draw    (164.5,168.92) -- (194.5,168.92) ;

\draw    (227.5,168.92) -- (257.5,168.92) ;

\draw   (257.5,168.92) .. controls (257.5,151.8) and (271.38,137.92) .. (288.5,137.92) .. controls (305.62,137.92) and (319.5,151.8) .. (319.5,168.92) .. controls (319.5,186.04) and (305.62,199.92) .. (288.5,199.92) .. controls (271.38,199.92) and (257.5,186.04) .. (257.5,168.92) -- cycle ;
\draw  [dash pattern={on 4.5pt off 4.5pt}]  (319.5,168.92) -- (373.5,168.92) ;

\draw    (405.5,168.92) -- (435.5,168.92) ;

\draw   (435.5,168.92) .. controls (435.5,151.8) and (449.38,137.92) .. (466.5,137.92) .. controls (483.62,137.92) and (497.5,151.8) .. (497.5,168.92) .. controls (497.5,186.04) and (483.62,199.92) .. (466.5,199.92) .. controls (449.38,199.92) and (435.5,186.04) .. (435.5,168.92) -- cycle ;
\draw    (497.5,168.92) -- (527.5,168.92) ;

\draw    (560.5,168.92) -- (590.5,168.92) ;

\draw   (591,143.92) -- (641,143.92) -- (641,193.92) -- (591,193.92) -- cycle ;

\draw (56,81) node [scale=1.2]  {$T^{u}_{1}$};
\draw (133.5,78.92) node [scale=0.7]  {$Sp( n-2)$};
\draw (211,81) node [scale=1.2]  {$T^{u}_{2}$};
\draw (288.5,75.92) node [scale=0.7]  {$SO( 2n)$};
\draw (391,80) node [scale=1.2]  {$T^{u}_{l}$};
\draw (468.5,77.92) node [scale=0.7]  {$SO( ln)$};
\draw (546,80) node [scale=1.2]  {$T^{u}_{l+1}$};
\draw (616,78.92) node [scale=0.6]  {$Sp( 2N-2)$};
\draw (56,171) node [scale=1.2]  {$T^{u}_{1}$};
\draw (133.5,168.92) node [scale=0.7]  {$SO( n+2)$};
\draw (211,171) node [scale=1.2]  {$T^{u}_{2}$};
\draw (288.5,168.92) node [scale=0.7]  {$Sp( 2n)$};
\draw (390,169) node [scale=1.2]  {$T^{u}_{l}$};
\draw (466.5,168.92) node [scale=0.7]  {$SO( ln+2)$};
\draw (546,170) node [scale=1.2]  {$T^{u}_{l+1}$};
\draw (616,168.92) node [scale=0.6]  {$Sp( 2N-2)$};
\draw (25,71.92) node  [align=left] {a)};
\draw (25,161.92) node  [align=left] {b)};
\draw (26,97.92) node   {$l \ even,~n \ even$};
\draw (27,183.92) node   {$l\ odd,~~n\  even$};

\end{tikzpicture}
}

\caption{Quiver for twisted $D_N$ type theory defined using formula \ref{twistedDirr}. Here we have a) $l=\frac{2N}{n}-1$, b) $l=\frac{2N-2}{n}-1$, and $u=n+k$.}
\label{twistedD}
\end{figure}
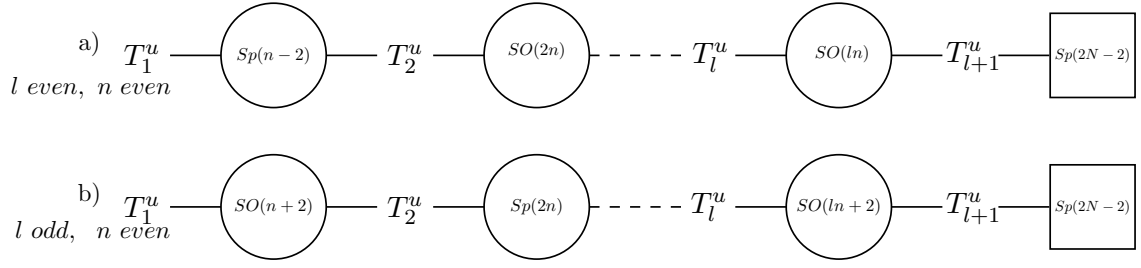

\begin{figure}

\tikzset{every picture/.style={line width=0.75pt}} 
\resizebox{\textwidth}{!}{
\begin{tikzpicture}[x=0.75pt,y=0.75pt,yscale=-1,xscale=1]

\draw    (82.5,58.92) -- (112.5,58.92) ;

\draw   (112.5,58.92) .. controls (112.5,41.8) and (126.38,27.92) .. (143.5,27.92) .. controls (160.62,27.92) and (174.5,41.8) .. (174.5,58.92) .. controls (174.5,76.04) and (160.62,89.92) .. (143.5,89.92) .. controls (126.38,89.92) and (112.5,76.04) .. (112.5,58.92) -- cycle ;
\draw    (174.5,58.92) -- (204.5,58.92) ;

\draw    (237.5,58.92) -- (267.5,58.92) ;

\draw   (267.5,58.92) .. controls (267.5,41.8) and (281.38,27.92) .. (298.5,27.92) .. controls (315.62,27.92) and (329.5,41.8) .. (329.5,58.92) .. controls (329.5,76.04) and (315.62,89.92) .. (298.5,89.92) .. controls (281.38,89.92) and (267.5,76.04) .. (267.5,58.92) -- cycle ;
\draw  [dash pattern={on 4.5pt off 4.5pt}]  (329.5,58.92) -- (383.5,58.92) ;

\draw    (415.5,58.92) -- (445.5,58.92) ;

\draw   (445.5,58.92) .. controls (445.5,41.8) and (459.38,27.92) .. (476.5,27.92) .. controls (493.62,27.92) and (507.5,41.8) .. (507.5,58.92) .. controls (507.5,76.04) and (493.62,89.92) .. (476.5,89.92) .. controls (459.38,89.92) and (445.5,76.04) .. (445.5,58.92) -- cycle ;
\draw    (507.5,58.92) -- (537.5,58.92) ;

\draw    (570.5,58.92) -- (600.5,58.92) ;

\draw   (601,33.92) -- (651,33.92) -- (651,83.92) -- (601,83.92) -- cycle ;
\draw    (82.5,148.92) -- (112.5,148.92) ;

\draw   (112.5,148.92) .. controls (112.5,131.8) and (126.38,117.92) .. (143.5,117.92) .. controls (160.62,117.92) and (174.5,131.8) .. (174.5,148.92) .. controls (174.5,166.04) and (160.62,179.92) .. (143.5,179.92) .. controls (126.38,179.92) and (112.5,166.04) .. (112.5,148.92) -- cycle ;
\draw    (174.5,148.92) -- (204.5,148.92) ;

\draw    (237.5,148.92) -- (267.5,148.92) ;

\draw   (267.5,148.92) .. controls (267.5,131.8) and (281.38,117.92) .. (298.5,117.92) .. controls (315.62,117.92) and (329.5,131.8) .. (329.5,148.92) .. controls (329.5,166.04) and (315.62,179.92) .. (298.5,179.92) .. controls (281.38,179.92) and (267.5,166.04) .. (267.5,148.92) -- cycle ;
\draw  [dash pattern={on 4.5pt off 4.5pt}]  (329.5,148.92) -- (383.5,148.92) ;

\draw    (415.5,148.92) -- (445.5,148.92) ;

\draw   (445.5,148.92) .. controls (445.5,131.8) and (459.38,117.92) .. (476.5,117.92) .. controls (493.62,117.92) and (507.5,131.8) .. (507.5,148.92) .. controls (507.5,166.04) and (493.62,179.92) .. (476.5,179.92) .. controls (459.38,179.92) and (445.5,166.04) .. (445.5,148.92) -- cycle ;
\draw    (507.5,148.92) -- (537.5,148.92) ;

\draw    (570.5,148.92) -- (600.5,148.92) ;

\draw   (601,123.92) -- (651,123.92) -- (651,173.92) -- (601,173.92) -- cycle ;

\draw (66,61) node [scale=1.2]  {$T^{u}_{1}$};
\draw (143.5,58.92) node [scale=0.7]  {$Sp( n-1)$};
\draw (221,61) node [scale=1.2]  {$T^{u}_{2}$};
\draw (298.5,58.92) node [scale=0.7]  {$SO( 2n+1)$};
\draw (401,60) node [scale=1.2]  {$T^{u}_{l}$};
\draw (476.5,58.92) node [scale=0.7]  {$Sp( ln-1)$};
\draw (556,60) node [scale=1.2]  {$T^{u}_{l+1}$};
\draw (626,58.92) node [scale=0.6]  {$SO( 2N+1)$};
\draw (27,52.92) node  [align=left] {a)};
\draw (66,151) node [scale=1.2]  {$T^{u}_{1}$};
\draw (143.5,148.92) node [scale=0.7]  {$SO( n+2)$};
\draw (221,151) node [scale=1.2]  {$T^{u}_{2}$};
\draw (298.5,148.92) node [scale=0.7]  {$Sp( 2n)$};
\draw (401,150) node [scale=1.2]  {$T^{u}_{l}$};
\draw (478.5,147.92) node [scale=0.7]  {$Sp( ln)$};
\draw (556,150) node [scale=1.2]  {$T^{u}_{l+1}$};
\draw (626,148.92) node [scale=0.6]  {$SO( 2N+1)$};
\draw (27,142.92) node  [align=left] {b)};
\draw (27,78.92) node   {$l\ odd,~n\ odd$};
\draw (29,164.92) node   {$l\ even,~n\ odd$};

\end{tikzpicture}}

\caption{Quiver for twisted $\fsl_{2N}$ type theory defined using formula \ref{Btwistedirr}. a) $l=\frac{2N}{n}-1$ b) $l=\frac{2N-1}{n}-1$., here $u=2n+2k+1$. When $u=1$ and $l$ odd, $T_1^u$ is not trivial, but a half hyper.}
\label{Btype}
\end{figure}
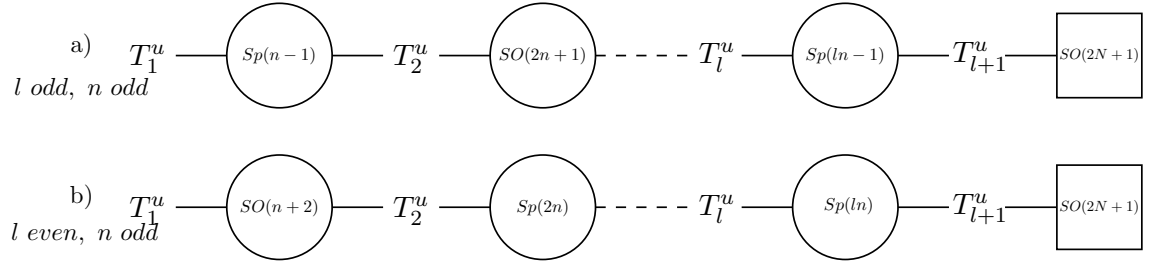

\newpage
\section{Non-admissible lisse W-algebra: Exceptional Lie algebra}
\label{sec:nonadm:exceptional}
\begin{table}[h]
\begin{center}
		\begin{tabular}{|c|l|l|}
			\hline
			$\mathfrak{g}$  & $n$ & VOA  \\ \hline
			$\mathfrak{e}_6$ & $12, 9, 6,3$ & $W^{-12+{n\over u}}(\mathfrak{e}_6,f_{prin})$\\ \hline
			$\mathfrak{e}_7$ & $18, 14, 6,2$ &$W^{-18+{n\over u}}(\mathfrak{e}_7,f_{prin})$ \\ \hline
                        	$\mathfrak{e}_8$ & $30,24,20,15,12,10,8,6,5,4,3,2$ &$W^{-30+{n\over u}}(\mathfrak{e}_8,f_{prin})$\\ \hline
			$\mathfrak{g}_2$ & $4,2,1$ &$W^{-4+{n\over u}}(\mathfrak{g}_2,f_{prin})$ \\ \hline
			$\mathfrak{f}_4$ & $9,6,4,3,2,1$ &$W^{-9+{n\over u}}(\mathfrak{f}_4,f_{prin})$ \\ \hline

		\end{tabular}
	
		\caption{Lisse W-algebras defined using exceptional Lie algebras. Here $u$ is coprime with $n$.}
		\label{table:lisseexceptional}
		
	\end{center}
\end{table}

In this section, we are going to classify non-admissible W-algebra whose corresponding four dimensional $\mathcal{N}=2$ SCFT is 
engineered from six dimensional $(2,0)$ theory of exceptional type.  The classification of  principal lisse W-algebra is equivalent to 
the classification of $E$ type irregular singularity which do not admit any mass deformation (the regular singularity is given by the principal nilpotent orbit), see table \ref{table:constraintADEirregular} and \ref{table:constraintwist} for the classification. 

The classification of lisse W-algebra from 4d theory engineered using non-trivial regular singularities is more difficult, as little is known in the literature about the associated variety of non-admissible W-algebra with exceptional Lie algebras. 
However, motivated by considerations of four dimensional physics, we develop a simple method to determine the associated variety of non-admissible AKM, and complete the full classification of those 4d theories whose 2d VOAs are lisse W-algebras.

\subsection{Principal lisse W-algebra}
First we consider theories engineered using the irregular singularity only. The lisse condition is equivalent to the 
absence of mass parameter of the  theory, and the constraint on irregular singularity is listed in \ref{table:constraintADEirregular} and \ref{table:constraintwist}.  The corresponding lisse W algebra is listed in table. \ref{table:lisseexceptional}.
Some of the physical information of these 4d theories are:
\begin{enumerate}
\item The W-algebra takes the following form
\begin{equation}
W^{-h^\vee+{n\over u}}(\mathfrak{g},f_{prin})
\end{equation}
with $h^\vee$ being the dual Coxeter number. The allowed value of $n$ is listed in table \ref{table:lisseexceptional}.  
\item The 4d theory has no Higgs branch, and the Coulomb branch spectrum can be easily found from the mini-versal deformation of the corresponding 3-fold singularity. 

\item We do not know the character of these W-algebras yet, but the growth of the character is given as ${\cal G}=-48(a_{4d}-c_{4d})$, which can 
be calculated using Coulomb branch data, and it takes the following form:
\begin{equation}
{\cal G}=\mathrm{rank}(\mathfrak{g})-{d_{\mathfrak{g}}(n)\over u},
\end{equation} 
Here $\mathrm{rank}(\mathfrak{g})$ is the rank of the lie algebra $\mathfrak{g}$, and $d_{\mathfrak{g}}(n)$ is a constant depending on $\mathfrak{g}$ and $n$ only which are summarized in table \ref{Growth}. 
\end{enumerate}

\begin{table}[h]
\begin{center}
\resizebox{\textwidth}{!}{
\begin{tabular}{|l|l|}
\hline
  $\mathfrak{g}$ & $d_{\mathfrak{g}}(n)$\\ \hline
$\mathfrak{e}_6$ &$ d(12)=78,~d(9)=56,~d(6)=36,~d(3)=24$ \\ \hline
$\mathfrak{e}_7$&  $d(18)=133,~d(14)=99,~d(6)=39,~d(2)=21$ \\ \hline
$\mathfrak{e}_8$&$d(30)=248,~d(24)=190,~d(20)=156,~d(15)=112,~d(12)=92,~d(10)=72$ \\ 
~& $~d(8)=66,~d(6)=40,~d(5)=48,~d(4)=36,~d(3)=32,~d(2)=24$ \\ \hline
$\mathfrak{g}_2$ &  $d(4)=14,~d(2)=12 (u=3k),~d(2)=4 (u\neq 3k),~d(1)=8$ \\ \hline
$\mathfrak{f}_4$ &  $d(9)=52,~d(6)=30,~d(3)=24 (u=3k-1),~d(3)=12(u=3k+1),~d(2)=18$ \\ \hline
\end{tabular}
}
\end{center}
\caption{Data used to compute the growth function ${\cal G}$ of VOA $W^{-h^\vee+{n\over u}}(\mathfrak{g},f_{prin})$.  ${\cal G}=\mathrm{rank}(\mathfrak{g})-{d_{\mathfrak{g}}(n)\over u}$.}
  \label{Growth}
\end{table}

Since there is very little study on non-admissible lisse W-algebras, it is definitely good to know some isomorphism between non-admissible and known lisse W-algebras. 
We can use the isomorphism between 3-fold singularities to infer isomorphisms between W-algebra.  The idea is that a single 3d singularity which defines our 4d theory can 
be written in different ways, which could have different 6d $(2,0)$ realization. One can read different W-algebra from these different 6d realizations, and therefore we 
can find non-trivial isomorphism between W-algebras. Some examples are shown in table \ref{table:exceptionaliso}.

\textbf{Example}: Consider a three-fold singularity $x_1^2+x_2^3+x_3^4+z^2=0$. This singularity can be written in three different ways:
\begin{equation}
I:~(x_1^2+z^2+x_2^3)+x_3^4=0,~~II:~(x_1^2+z^2+x_3^4)+x_2^3=0,~~III:~(x_1^2+x_2^3+x_3^4)+z^2=0
\end{equation}
The first one can be realized by the 6d $A_2$ $(2,0)$ theory,  the second by the 6d $A_3$ $(2,0)$ theory, and the third one by 6d $E_6$ $(2,0)$. We then get the following isomorphisms among
 W-algebras
 \begin{equation}
 W^{-3+{3\over 7}}(\mathfrak{sl}_3,f_{prin})= W^{-4+{4\over 7}}(\mathfrak{sl}_4,f_{prin})=W^{-12+{6\over 7}}(\mathfrak{e}_6,f_{prin})
 \end{equation}

\begin{table}[h]
\begin{center}
\resizebox{\textwidth}{!}{
		\begin{tabular}{|c|l|l|}
			\hline
			Singularity & Isomorphism & W-algebras \\ \hline
			$x_1^2+x_2^3+x_3^4+z^2=0$& $(A_2, A_3)=(E_6,A_1)$&  $W^{-3+{3\over 7}}(\mathfrak{sl}_3,f_{prin})=W^{-12+{8\over 7}}(\mathfrak{e}_6,f_{prin})$
  \\ \hline
$x_1^2+x_2^3+x_3^4+z^5=0$ & $(E_6,A_4)=(E_8,A_3)$& $W^{-12+{12\over 17}}(\mathfrak{e}_6,f_{prin})=W^{-30+{15\over 17}}(\mathfrak{e}_8,f_{prin})$  \\  \hline
$x_1^2+x_2^3+x_3^4+x_3z^2=0$& $(D_5, A_2)= (E_6, b=9,k=2)$& $W^{-8+{8\over 11}}(\mathfrak{so}_{10},f_{prin})=W^{-12+{9\over 11}}(\mathfrak{e}_6,f_{prin})$\\  \hline
$x_1^2+x_2^3+x_2x_3^3+z^5=0$& $(E_7, A_4)= (E_8, b=20,k=3)$& $W^{-23+{18\over 23}}(\mathfrak{e}_7,f_{prin})=W^{-30+{20\over 23}}(\mathfrak{e}_8,f_{prin})$\\  \hline
$x_1^2+x_2^3+x_3^5+z^2=0$ & $(A_2, A_4)=(E_8,A_1)$& $W^{-3+{3\over 8}}(\mathfrak{sl}_3,f_{prin})=W^{-30+{15\over 16}}(\mathfrak{e}_8,f_{prin})$ \\ \hline
$x_1^2+x_2^3+x_3^5+z^3=0$& $(D_4, A_4)=(E_8,A_2)$& $W^{-6+{6\over 11}}(\mathfrak{so}_8,f_{prin})=W^{-30+{10\over 11}}(\mathfrak{e}_8,f_{prin})$ \\  \hline
$x_1^2+x_2^3+x_3^5+x_3z^2=0$& $(D_6, A_2)=(E_8,b=24,k=2)$&$W^{-10+{10\over 13}}(\mathfrak{so}_{12},f_{prin})=W^{-30+{12\over 13}}(\mathfrak{e}_6,f_{prin})$ \\  \hline

		\end{tabular}
	}
		\caption{Here we first wrote the equivalence of the 3-fold singularity in the notation of $(G, G^{'})$, where $G$ denotes the corresponding ADE singularity of type $G$, then we listed the equivalence between W-algebras.}
		\label{table:exceptionaliso}
		
	\end{center}
\end{table}

\subsection{Adding regular singularity}
\label{associated}
Now consider 4d theories engineered using an irregular singularity used in the last subsection and a general regular singularity labeled by
a nilpotent orbit $f$ of $\mathfrak{g}$. We would like to find $f$ such that the corresponding W-algebra $W^{-h^\vee+ {n\over u}}(\mathfrak{g},f)$ is lisse. 
To find such $f$, we follow the strategy which is similar to what was done in the admissible case:
\begin{itemize}
\item First consider the theory constructed from using a trivial nilpotent orbit $f$, so that the 4d theory has a flavor symmetry group $G$. The corresponding VOA is the non-admissible AKM 
$V^{-h^\vee+ {n\over u}}(\mathfrak{g})$, and we \textbf{assume} that its associated variety $\overline{{\cal O}}_{\fg, n,u}$ is irreducible (see the discussion in \cite{Arakawa:2016aa}) and is the closure of a nilpotent orbit
  ${\cal O}_{\fg, n,u}$. 
\item Now for the W-algebra $W^{-h^\vee+ {n\over u}}(\mathfrak{g},f)$, its associated variety is 
\begin{equation}
S_f\cap   \overline{\cal O}_{\fg, n,u}
\end{equation}
so if we choose $f\in {\cal O}_{\fg, n,u}$, the above associated variety would be trivial, and the corresponding W-algebra is
lisse.
\end{itemize}
So the crucial part of finding $f$ is to determine the associated variety of non-admissible AKM, which has not been studied much yet. It might be possible to directly compute it using the 
method developed in \cite{MR3456698}, here we use a method which is motivated from four dimensional physics.  

To find the associated variety of a non-admissible AKM from using the data of four dimensional $\mathcal{N}=2$ SCFT, we need to use following facts:
\begin{itemize}
\item  First we need to use \textbf{anomaly matching}. The difference of central charges $a_{4d}-c_{4d}$ can be computed from the Coulomb branch data \cite{Shapere:2008zf, Xie:2013jc}, and is related 
to the growth function of W-algebra vacuum character as follows
\begin{equation}
{\cal G}=-48(a_{4d}-c_{4d}).
\label{growth}
\end{equation}
$a_{4d}$ and $c_{4d}$ can be explicitly calculated using the method developed in \cite{Wang:2015mra,Wang:2018gvb}, and it takes the following simple form
\begin{equation}
{\cal G}_{UV}(V^{-h^\vee+{n\over u}}(\mathfrak{g}))=\dim(\mathfrak{g})-{d_{\mathfrak{g}}(n)\over u}
\end{equation}
Here the data $d_{\mathfrak{g}}(n)$ is given in table. \ref{Growth}. 
 
 Now when we turn on expectation values of Higgs branch operators $\hat{{\cal B}}_R$, 
the IR theory could consist of $n_h$ free hypermultiplets, $n_v$ free vector multiplets and an interacting theory $T_{IR}$ which does not have a Higgs branch. To match the anomaly \cite{Shimizu:2017kzs}, we have the following
formula:
\begin{equation}
{\cal G}_{UV}=2n_h-2n_v+{\cal G}_{IR}.
\end{equation}
Here ${\cal G}_{IR}$ is the growth function of the IR theory ${\cal T}_{IR}$.
Since the Higgs branch is assumed to be the closure of a nilpotent orbit ${\cal O}_{\mathfrak{g},n,u}$,  $2n_h$ has to equal to the dimension of ${\cal O}_{\mathfrak{g},n,u}$, so we have 
\begin{equation}
\dim({\cal O}_{\mathfrak{g},n,u})={\cal G}_{UV}-{\cal G}_{IR}+2n_v.
\label{matching}
\end{equation}
One simple consequence of formula (\ref{matching}) is that if ${\cal G}_{UV}$ is fractional, then ${\cal T}_{IR}$ can not be trivial, as the contribution of 
free hypermultiplet and free vector multiplet has to be an integer. The theory ${\cal T}_{IR}$ is the four dimensional theory whose Higgs branch is trivial and therefore
whose 2d associated W-algebra is lisse. 

To determine the dimension of associated variety, we need to have some information of $n_v$ and ${\cal G}_{IR}$ in the IR.  Furthermore, even if we determine the dimension of 
${\cal O}_{\fg, n,u}$, it is still possible that there are more than one nilpotent orbit with same dimension. We need to use more four dimensional input.  

\item The ${\cal T}_{IR}$ in the Higgs branch can actually be identified as follows:  it is just the theory defined by adding a regular singularity $f_{IR}$ because of the fact that going to the Higgs branch of a class S theory 
is equivalent to closing the puncture. We
determine $f_{IR}$ using the Coulomb branch data:
\begin{enumerate}
\item If the UV growth ${\cal G}_{UV}$ is fractional, then the IR interacting theory ${\cal T}_{IR}$ has to be non-trivial and has Coulomb branch operator(s). We choose maximal $f_{IR}$ (using the order of nilpotent orbit) such that 
there are non-trivial Coulomb branch operators. To ensure that there are no free hypermultiplets for the theory defined using $f_{IR}$, we need to ensure that the leading order differential of the SW curve of corresponding 6d configuration has non-negative Coulomb branch operators.
\item  If the UV growth ${\cal G}_{UV}$ is integral, we can still find the maximal $f_{IR}$ such that there are non-trivial Coulomb branch operators. However, there is a further possibility that the interacting theory is trivial, and 
we actually need to consider larger $f_{IR}$, and there is no simple way to determine $f$ due to the possibility of existence of free vector multiplets. We will then need to use Coulomb branch data to further constrain $f_{IR}$. 
\end{enumerate}
In practice, we can first try $f_{IR}$ whose dimension is just below the growth function, and gradually increase $f_{IR}$. Our assumption is that the orbit of $f_{IR}$ which gives $T_{IR}$ is ${\cal O}_{\mathfrak{g},n,u}$
\begin{equation}
\CO_{f_{IR}}={\cal O}_{\mathfrak{g},n,u}.
\end{equation}

\item We cannot completely determine $f_{IR}$ and $n_v$ using a single equation \ref{matching}, so we need another equation.
There is one further constraint from the Coulomb branch: The dimension of Hitchin moduli space $d_{f_{IR}}$ should be equal to the number of free vector multiplets and the rank of ${\cal T}_{IR}$
\begin{equation}
d_{f_{IR}}=r_{IR}+n_v.
\end{equation}
On the other hand, the dimension of the Hitchin moduli space defined using the regular singularity $f$ is 
\begin{equation}
d_{f_{IR}}=r_{UV}+{1\over 2}\dim({\cal O}_{f_{IR}^d})-{1\over2}(\dim(\mathfrak{g})-\mathrm{rank}(\mathfrak{g})),
\end{equation}
where $r_{UV}$ is the Coulomb branch dimension of $UV$ theory, and $f_{IR}^d$ is the dual Hitchin label for the puncture. Hence we have the constraint
\begin{equation}
r_{IR}+n_v=r_{UV}+{1\over 2}\dim({\cal O}_{f_{IR}^d})-{1\over2}(\dim(\mathfrak{g})-\mathrm{rank}(\mathfrak{g})).
\label{coulombidentity}
\end{equation}
\end{itemize}
In practice, using equation \ref{matching} and \ref{coulombidentity}, one can determine the associated variety of AKM with exceptional Lie algebra as in table \ref{table:g2f4assoicate}, \ref{table:e6e7assoicate}, and \ref{table:e8assoicate}. Once we determine the associated variety ${\cal O}_{\mathfrak{g},n,u}$ of AKM $V^{-h^\vee+ {n\over u}}(\mathfrak{g})$,  we can find $f\in {\cal O}_{\mathfrak{g},n,u}$ such that $W^{-h^\vee+ {n\over u}}(\mathfrak{g},f)$ is lisse. Full results are listed in table \ref{table:e6e7lisse} and \ref{table:e8lisse}.

\textbf{Example 1}: Consider a theory engineered by an irregular singularity of the type $\mathfrak{g}=\mathfrak{e}_6, n=9,~u=8$,  and a regular singularity labeled by trivial nilpotent orbit. The growth function is ${\cal G}_{UV}=78-56/8=71$, 
we can try $f_{IR}=E_6(a_1)$ whose dimension is $70$ which is just below the growth function, and the theory has Coulomb branch spectrum $\frac{15}{8}, \frac{9}{8}, \frac{5}{4}, \frac{3}{2}$, which can actually be engineered by 
the singularity $F_{IR}:~x^2+y^2+z^3+w^5=0$.  Theory with $f_{IR}$ has no Higgs branch, so we conclude that the associated variety of the $V^{-12+9/8}(\mathfrak{e}_6)$ is just $E_6(a_1)$.  The growth function of IR interacting theory
is ${\cal G}_{IR}=1$, so using formula, \ref{matching}, we conclude that $n_v=0$, and there is no free vector multiplets in the generic point of Higgs branch. One can also check that formula \ref{coulombidentity} is also satisfied: we have $r_{UV}=29$, and $r_{IR}=4$,
and the dual of $E_6(a_1)$ is $A_1$ orbit, whose dimension is 22.

\textbf{Example 2}: Consider a theory engineered by an irregular singularity of the type $\mathfrak{g}=\mathfrak{e}_6, ~n=3,~u=2$, and a regular singularity labeled by trivial nilpotent orbit. 
The growth function is ${\cal G}=78-24/2=66$. Now change the regular singularity to $f_1=E_6(a_3)$ whose dimension is equal to $66$. The Coulomb branch spectrum of the IR theory is  $\{ 3, 2, (\frac{3}{2})^4\}$.
This theory has a gauge theory description which is just $SU(3)$ coupled with four copies of $D_2(SU(3))$ theory. This theory can also be engineered by a complete intersection singularity defined with two polynomials. The complete
intersection singularity is not terminal, and has a  crepant resolution, so it has a Higgs branch. In fact, the Higgs branch of the theory $F_{IR}$ has $n_h=1,n_v=1$, and no further interacting theory. This implies that 
\begin{equation}
\dim({\cal O}_{\mathfrak{e}_6,3,2})={\cal G}_{UV}-{\cal G}_{IR}+2n_v=68,
\end{equation}
so the associated variety $\overline{\cal O}_{\mathfrak{e}_6,3,2}$ is just closure of nilpotent orbit with the label $D_5$, as this is the unique $E_6$ nilpotent orbit whose dimension is 68. The Higgs branch of the original theory 
therefore has $34$ free hypermultipelts, and one free vector multiplet.  

\textbf{Example 3}:  Now consider a more complicated example.  Consider a theory engineered by an irregular singularity of the type $\mathfrak{g}=\mathfrak{e}_8, n=5,~q=3$,  we also have the regular singularity with label $f$ trivial. 
and the growth of this theory is 
${\cal G}=248-{48\over 3}=232$, which is an integer and we are not sure whether there would be an interacting piece in the Higgs branch. To find the effective theory 
on the Higgs  branch, we first choose $f_1=E_8(a_4)$ whose dimension is $232$, and the theory has Coulomb branch spectrum $\{5, 4, (\frac{10}{3})^3, 3, (\frac{7}{3})^3, 2, (\frac{5}{3})^4, (\frac{4}{3})^3\}$. It is interesting that this theory has the same Coulomb branch spectrum as the theory engineered using three-fold singularity $x^3+y^3+z^3+w^5=0$, which actually has a Higgs branch whose IR theory consists of 
$n_h=1, n_v=1$, together with an interacting theory engineered by the singularity $F_{IR}:~x^3+y^3+z^3+w^2=0$. The theory engineered by $F_{IR}$ has no Higgs branch, so there is no further Higgs branch deformation.   
Using the formula \ref{matching},
we see that (We use the fact ${\cal G}_{IR}=0$): 
\begin{equation}
\dim({\cal O}_{\mathfrak{e}_8,5,3})={\cal G}_{UV}-{\cal G}_{IR}+2n_v=234,
\end{equation}
and there is a unique nilpotent orbit $E_8(a_3)$ whose dimension is equal to $234$, so the associated variety $\overline{\cal O}_{\mathfrak{e}_8,5,3}$ is the closure of $E_8(a_3)$ orbit. The IR theory on the Higgs branch consists of 
free hypermultipelts, free vector multiplet, and an interacting theory.

\textbf{Example 4}:  For $\mathfrak{e}_6, n=3,~u=1$, we have ${\cal G}_{UV}=54$, so we
 first take $f=2A_2+A_1$ whose dimension is $54$, the Coulomb branch spectrum of the theory defined with $f$ is $\{2,3\}$. To match the $a-c$ anomaly, the theory should be $\mathcal{N}=4$ $SU(3)$ SYM. 
We can further Higgs $\mathcal{4}$ SYM, and in the IR, there are $n_v=2$ and $n_h=2$, and no interacting theory.  So the associated variety of original theory should have dimension $58$, and there is a unique 
$E_6$ nilpotent orbit with this dimension $D_4(a_1)$, and we conclude that this is the associated variety of the original theory.

\textbf{Example 5}:  Now we give a more detailed explanation for the case $\mathfrak{g}=E_8,~n=3,~u=1$, and ${\cal G}_{UV}=216$. We take $f=E_6(a_1)$, 
and the Coulomb branch spectrum is $\{2^4,3^7\}$. This theory is actually 
isomorphic to $A_2$ $(2,0)$ theory on a genus two Riemann surface with one regular puncture. It is known that one can further higgs this theory, and
the Higgs branch  is not a pure Higgs branch, and we 
have $n_v=4$ with no interacting part left. Using formula \ref{matching}, we can see that the dimension of the associated variety of original theory should be 224, and there 
are two choices for it: $E_8(a_6)$ and $E_7(a_2)$. Using formula \ref{coulombidentity}, we can conclude the associated variety should be $E_8(a_6)$.   

\begin{table}
\begin{center}
\begin{tabular}{cc}

\begin{tabular}{|c|c|c|}
\hline
  $G_2$ & $u$ & $\overline{{\cal O}}_u$\\ \hline
  $n=4$&$>6$ & $G_2$ \\ \hline
      ~&$5$ & $G_2(a_1)$ \\ \hline
   $n=2$ &$>4$ & $G_2$ \\ \hline
    ~&$3$ & $G_2(a_1)$ \\ \hline
  ~&$1^*$ & $G_2(a_1)$ \\ \hline
  $n=1$&$>2$ & $G_2$ \\ \hline
  ~&$2^\#$ $(n_v=1)$ & $G_2$ \\ \hline
  ~&$1^\#$ ($n_v=2$)&  $G_2(a_1)$\\ \hline
\end{tabular}

&

\begin{tabular}{|c|c|c|}
\hline
  $F_4$ & $u$ & $\overline{{\cal O}}_u$\\ \hline
  $n=9$&$>12$ & $F_4$ \\ \hline
      ~&$11$ & $F_4(a_1)$ \\ \hline
    ~&$7$ & $F_4(a_2)$ \\ \hline
    ~&$5$ & $F_4(a_3)$ \\ \hline
  $n=6$&$>6$ & $F_4$ \\ \hline
  ~&$5^*$ & $F_4(a_1)$ \\ \hline
  ~&$1^*$ & $\tilde{A_1}$ \\ \hline
   $n=3$ &$>3$ & $F_4$ \\ \hline
    ~&$2^{\#}(n_v=1)$ & $B_3$ \\ \hline
  ~&$1^*$ & $F_4(a_3)$ \\ \hline
  $n=2$&$>2$ & $F_4$ \\ \hline
  ~&$1^{\#}$ $(n_v=3)$ & $F_4(a_3)$\\ \hline
\end{tabular}

\end{tabular}
\end{center}
\caption{Associated variety for $G_2$ and $F_4$ type theory. 
 The rank of Coulomb branch of $G_2$ type theory: $r={4\over n} u-1$. The rank of Coulomb branch of $F_4$ type theory: $r={18 u\over n}-2$.
 The one with $\#$ implies that the IR theory has free vector multiplets in the Higgs branch, and $n_v$ is the number of free vector multiplets. The one with $*$ means that the theory has pure Higgs branch, i.e. the IR theory on the Higgs branch consists of free hypermultiplets only. }
  \label{table:g2f4assoicate}
\end{table}

\begin{table}[H]
\begin{center}
\begin{tabular}{cc}

\begin{tabular}{|c|c|c|}
\hline
  $E_6$ & $u$ & $\overline{{\cal O}}_u$\\ \hline
  $n=12$&$>12$ & $E_6$ \\ \hline
      ~&$11$ & $E_6(a_1)$ \\ \hline
    ~&$7$ & $E_6(a_3)$ \\ \hline
  ~&$5$ & $A_4+A_1$ \\ \hline
  $n=9$&$>9$ & $E_6$ \\ \hline
  ~&$7^*,8$ & $E_6(a_1)$ \\ \hline
  ~&$5$ & $E_6(a_3)$ \\ \hline
  ~&$4^*$ & $D_5(a_1)$ \\ \hline
  ~&$2^*$ &$A_2+2A_1$  \\ \hline
  ~&$1^*$ & $A_1$ \\ \hline
   $n=6$ &$>6$ & $E_6$ \\ \hline
    ~&$5$ & $E_6(a_1)$ \\ \hline
  ~&$1^*$ & $A_2$ \\ \hline
  $n=3$&$>3$ & $E_6$ \\ \hline
  ~&$2^\#$ $(n_v=1)$ & $D_5$ \\ \hline
  ~&$1^\#$ ($n_v=2$)&  $D_4(a_1)$\\ \hline
\end{tabular}

&

\begin{tabular}{|c|c|c|}
\hline
  $E_7$ & $u$ & $\overline{{\cal O}}_u$\\ \hline
  $n=18$&$>18$ & $E_7$ \\ \hline
      ~&$17$ & $E_7(a_1)$ \\ \hline
    ~&$13$ & $E_7(a_2)$ \\ \hline
    ~&$11$ & $E_7(a_3)$ \\ \hline
    ~&$7^*$ & $A_6$ \\ \hline
  ~&$5$ & $A_4+A_2$ \\ \hline
  $n=14$&$>14$ & $E_7$ \\ \hline
  ~&$11^*,13$ & $E_7(a_1)$ \\ \hline
  ~&$9^*$ & $E_7(a_2)$ \\ \hline
  ~&$5$ &$E_7(a_5)$  \\ \hline
  ~&$3^*$&$A_3+A_2+A_1$\\ \hline
  ~&$1^*$ & $A_1$ \\ \hline
   $n=6$ &$>6$ & $E_7$ \\ \hline
    ~&$5$ & $E_7(a_1)$ \\ \hline
  ~&$1^*$ & $D_4(a_1)$ \\ \hline
  $n=2$&$>2$ & $E_7$ \\ \hline
  ~&$1^{\#}$ $(n_v=3)$ & $E_6(a_1)$\\ \hline
\end{tabular}

\end{tabular}
\end{center}
\caption{Associated variety for $E_6$ and $E_7$ type AKM $V^{-h^\vee+{n\over u}}(\mathfrak{g})$. The one with star implies that 
that theory has a pure Higgs branch!  The one with $\#$ implies that the IR theory has  free vector multiplets in the Higgs branch, and $n_v$ is the number of free vector multiplets. The one with $*$ means that the theory has pure Higgs branch, i.e. the IR theory on the Higgs branch consists of free hypermultiplets only.
The IR theory in the Higgs branch of other theories has 
an interacting piece, which is listed in table \ref{table:e6e7lisse}.
 The rank of Coulomb branch of $E_6$ type theory: $r_{UV}={36\over n} u-3$. The rank of Coulomb branch of $E_7$ type theory: $r_{UV}={1\over 2}({126 u\over n}-7)$.}
  \label{table:e6e7assoicate}
\end{table}

\begin{table}[h]
\begin{center}
\resizebox{\textwidth}{!}{
\begin{tabular}{|l|l|c|c|c|}
\hline
 $(f,n,u)$ & Coulomb branch &  Isomorphism & W-algebra \\ \hline
 $(G_2(a_1),2,3)$ & $(2,({4\over 3})^3)$ & $x^3+y^3+z^3+2^2=0$ & $W^{-6+{2\over 3}}(\mathfrak{so}_8,f_{prin})$ \\ \hline 
$(E_6(a_1), 9,8)$ & $\frac{15}{8}, \frac{9}{8}, \frac{5}{4}, \frac{3}{2}$ & $(A_2,A_4)$ & ~ \\ \hline
$(E_6(a_3), 9,5)$ & $(\frac{6}{5})^2$  & $(A_1, A_2)^2$ & $\mathrm{Vir}_{2,5}^{\otimes 2}$ \\ \hline 
$(E_6(a_1), 6,5)$ & $(\frac{6}{5})^2$ & $(A_1, A_2)^2$ & $\mathrm{Vir}_{2,5}^{\otimes 2}$ \\ \hline \hline
$(E_7(a_1), 14,13)$ & $\left\{\frac{22}{13},\frac{20}{13},\frac{18}{13},\frac{30}{13},\frac{16}{13},\frac{28}{13},\frac{14}{13},\frac{38}{13},\frac{24}{13}\right\}$  & $~$ & ~ \\ \hline
$(E_7(a_5), 14,5)$ & $(\frac{6}{5})^3$  & $(A_1, A_2)^3$ & $\mathrm{Vir}_{2,5}^{\otimes 3}$ \\ \hline
$(E_7(a_1), 6,5)$ & $(\frac{6}{5})^3$  & $(A_1, A_2)^3$ & $\mathrm{Vir}_{2,5}^{\otimes 3}$ \\ \hline 
 \end{tabular} 
 }               
\end{center}
\caption{The physical data for 4d theory whose 2d VOA is $W^{-h^\vee+{n\over u}}(\mathfrak{g},f)$. }
  \label{table:e6e7lisse}
\end{table}

\newpage

\begin{table}[H]

\begin{center}
\begin{tabular}{cc}

\begin{tabular}[t]{|c|c|c|}
\hline
  $n$ & $u$ & $\overline{{\cal O}}_u$\\ \hline
  $n=30$&$>30$ & $E_8$ \\ \hline
  $~$&$29$ & $E_8(a_1)$ \\ \hline
  $~$&$23$ & $E_8(a_2)$ \\ \hline
  $~$&$19$ & $E_8(a_3)$ \\ \hline
    $~$&$17$ & $E_8(a_4)$ \\ \hline
  $~$&$13$ & $E_8(a_5)$ \\ \hline
    $~$&$11$ & $E_8(a_6)$ \\ \hline
  $~$&$7$ & $A_6+A_1$ \\ \hline
  $n=24$&$>24$ & $E_8$ \\ \hline
  ~&$23$, $19^*$ & $E_8(a_1)$ \\ \hline
  ~&$17$ & $E_8(a_2)$ \\ \hline
  ~&$13$ & $E_8(a_4)$ \\ \hline
  ~&$11$ & $E_8(b_4)$ \\ \hline
  ~&$7$ & $E_8(b_6)$ \\ \hline
  ~&$5^*$ & $D_6(a_1)$ \\ \hline
  ~&$1^*$ & $A_1$ \\ \hline
   $n=20$ &$>20$ & $E_8$ \\ \hline
    ~&$19$, $17$ & $E_8(a_1)$ \\ \hline
    ~&$13^*$ & $E_8(a_2)$ \\ \hline
    ~&$11$ &  $E_8(a_4)$ \\ \hline
    ~&$9$ & $E_8(b_4)$ \\ \hline
    ~&$7$ &  $E_8(a_6)$ \\ \hline
    ~&$3^*$ &  $A_4+A_2+A_1$ \\ \hline
   ~&$1^*$ & $2A_1$ \\ \hline
  $n=15$&$>15$ & $E_8$ \\ \hline
  ~&$14$ & $E_8(a_1)$\\ \hline
  ~&$13$ & $E_8(a_1)$\\ \hline
  ~&$11$ & $E_8(a_2)$\\ \hline
  ~&$8$ & $E_8(a_4)$\\ \hline
  ~&$7^*$ & $E_8(a_4)$\\ \hline
  ~&$4^*$ &  $E_8(b_6)$\\ \hline
  ~&$2^*$ & $A_4+2A_1$\\ \hline
  ~&$1^*$ & $A_2+A_1$\\ \hline
\end{tabular}

&

\begin{tabular}[t]{|c|c|c|}
\hline
  $n$ & $u$ & $\overline{{\cal O}}_u$\\ \hline
  $n=12$&$>12$ & $E_8$ \\ \hline
  $~$&$11$ & $E_8(a_1)$ \\ \hline
  $~$&$7$ & $E_8(a_3)$ \\ \hline
  $~$&$5$ & $E_8(a_5)$ \\ \hline
  $~$&$1^*$ & $2A_2$ \\ \hline
  $n=10$&$>10$ & $E_8$ \\ \hline
  ~&$9$ & $E_8(a_1)$ \\ \hline
  ~&$7$ & $E_8(a_2)$ \\ \hline
  ~&$3^*$ & $E_8(a_6)$ \\ \hline
  ~&$1^*$ & $D_4(a_1)+A_1$ \\ \hline
   $b=8$ &$>8$ & $E_8$ \\ \hline
    ~&$7$ & $E_8(a_1)$ \\ \hline
    ~&$5$  & $E_8(a_3)$ \\ \hline
    ~&$3$ & $E_8(b_5)$ \\ \hline
    ~&$1^\#$$(n_v=1)$ & $D_4(a_1)+A_2$ \\ \hline
  $n=6$&$>6$ & $E_8$ \\ \hline
  ~&$5$ & $E_8(a_1)$\\ \hline
  ~&$1^*$ &  $E_8(a_7)$\\ \hline
  $b=5$&$>5$ & $E_8$ \\ \hline
  ~&$4^\#$($n_v=1$) & $E_8(a_1)$\\ \hline
  ~&$3^\#$ $(n_v=1)$ & $E_8(a_3)$\\ \hline
  ~&$2^\#$ $(n_v=4)$ & $E_8(a_4)$\\ \hline
  ~&$1^\#$ $(n_v=4)$ & $E_8(a_7)$\\ \hline
  $n=4$&$>4$ & $E_8$ \\ \hline
  ~&$3$ & $E_8(a_2)$\\ \hline
  ~&$1^\#$ ($(n_v=1)$ &  $E_6(a_1)$\\ \hline
  $n=3$&$>3$ & $E_8$ \\ \hline
  ~&$2^\#$$(n_v=2)$ & $E_8(a_2)$\\ \hline
  ~&$1^\#$ $(n_v=4)$ & $E_8(a_6)$  \\ \hline
  $n=2$&$>2$ & $E_8$ \\ \hline
  ~&$1^\#$($n_v=4)$ &  $E_8(a_4)$\\ \hline
\end{tabular}
\end{tabular}

\end{center}

\caption{Associated variety for $E_8$ type theory. The rank of the Coulomb branch is: $r={120\over n}u-4$. The one with star implies that 
the theory has a pure Higgs branch!  The one with $\#$ implies that the IR theory in the Higgs branch has free vector multiplets, and $n_v$ is the number. For $n=24, u=5$, there 
are actually two possible choices of nilpotent orbits ($D_6(a_1)$ and $A_6$) which satisfy our formula \ref{matching} and \ref{coulombidentity} (with $n_v=0$); To pick $D_6(a_1)$, we use the following  closure relation: $D_5+A_1$ 
orbit is in the closure of $D_6(a_1)$ but is not in closure of $A_6$, now we can compute the Coulomb branch operator of the theory defined using $D_5+A_1$ which is non-empty, 
and we can further Higgs this particular theory. This observation then implies that we have to choose $D_6(a_1)$ as the Higgs branch of theory defined using regular puncture with trivial 
nilpotent orbit!
}
  \label{table:e8assoicate}
\end{table}

\begin{table}[H]
\begin{center}
\resizebox{\textwidth}{!}{
\begin{tabular}{|l|l|l|l|l|}
\hline
 $(f,n,u)$ & Coulomb branch &  Isomorphism & W-algebra \\ \hline
$(E_8(a_1), 24, 23)$ & \parbox[c]{4cm}{ $\frac{114}{23}, \frac{96}{23}, \frac{90}{23}, \frac{78}{23}, \frac{76}{23}, \frac{72}{23}$ \\ $\frac{66}{23}, \frac{60}{23}, \frac{58}{23}, \frac{54}{23}, \frac{52}{23}, \frac{48}{23}, (\frac{42}{23})^2$ \\ $\frac{40}{23}, \frac{36}{23}, \frac{34}{23}, \frac{30}{23}, \frac{28}{23}, \frac{24}{23}$ }  & ~ & ~\\ \hline
$(E_8(a_2), 24, 17)$ & \parbox[c]{3.5cm}{ $\frac{30}{17}, \frac{28}{17}, \frac{26}{17}, \frac{24}{17}, \frac{22}{17}, \frac{20}{17}, \frac{18}{17}$ } & $x^2+y^2+z^2+w^{15}=0$  & $\mathrm{Vir}_{2,15}$\\ \hline
$(E_8(a_4), 24, 13)$ & \parbox[c]{3cm}{ $\frac{30}{13}, \frac{27}{13}, \frac{24}{13}, \frac{21}{13}, \frac{20}{13}$ \\ $\frac{18}{13}, \frac{17}{13}, \frac{15}{13}, \frac{14}{13}$ }  & $x^2+y^2+z^3+w^{10}=0$ & ~ \\ \hline
$(E_8(b_4), 24, 11)$ & \parbox[c]{3cm}{ $\frac{12}{11},\frac{14}{11},\frac{16}{11},\frac{18}{11}$ }  & $x^2+y^2+z^2+w^{9}=0$ & $\mathrm{Vir}_{2,9}$ \\ \hline
$(E_8(b_6), 24, 7)$ & \parbox[c]{3cm}{ $\frac{8}{7},\frac{9}{7},\frac{12}{7}$ }  & $x^2+y^2+z^3+w^{4}=0$ & $\mathrm{Vir}_{3,14}\oplus L(13,1)$ \\ \hline
$(E_8(b_6), 24, 7)$ & \parbox[c]{3cm}{ $\frac{8}{7},\frac{9}{7},\frac{12}{7}$ }  & $x^2+y^2+z^3+w^{4}=0$ & $\mathrm{Vir}_{3,14}\oplus L(13,1)$ \\ \hline

$(E_8(a_1), 20, 19)$ & \parbox[c]{4cm}{ $\frac{90}{19}, 4, \frac{70}{19}, \frac{62}{19}, \frac{60}{19}, \frac{56}{19}, \frac{50}{19}$ \\ $\frac{48}{19}, \frac{46}{19}, \frac{42}{19}, \frac{40}{19}, \frac{36}{19}, \frac{34}{19}$ \\ $\frac{32}{19}, \frac{30}{19}, \frac{28}{19}, \frac{26}{19}, \frac{22}{19}, \frac{20}{19}$ } &  $x^2+y^3+yz^3+ zw^5=0$  & ~\\ \hline
$(E_8(a_1), 20, 17)$ & \parbox[c]{3.6cm}{ $\frac{30}{17}, \frac{28}{17}, \frac{26}{17}, \frac{24}{17}, \frac{22}{17}, \frac{20}{17},\frac{18}{17}$ } &  $x^2+y^2+z^2+w^{15}=0$  & $\mathrm{Vir}_{2,15}$\\ \hline
$(E_8(a_4), 20, 11)$ &  \parbox[c]{3cm}{ $\frac{30}{11}, \frac{25}{11}, \frac{24}{11}, \frac{20}{11}, \frac{19}{11}$ \\ $\frac{18}{11}, \frac{15}{11}, \frac{14}{11}, \frac{13}{11}, \frac{12}{11}$ }  & $x^2+y^2+z^5+w^6=0$ & ~\\ \hline
$(E_8(b_4), 20, 9)$ &  \parbox[c]{3cm}{ $\frac{10}{9},\frac{4}{3},\frac{14}{9}$ }  & $x^2+y^2+z^2+w^7=0$ & $\mathrm{Vir}_{2,7}$ \\ \hline

$(E_8(a_6), 20, 7)$ & $(\frac{10}{7})^3, (\frac{8}{7})^3$  &$3\times (x^2+y^2+z^2+w^5=0)$& 3$\times \mathrm{Vir}_{2,7}$\\ \hline

$(E_8(a_1), 15, 14)$ & \parbox[c]{3.0cm}{ $\frac{15}{14},\frac{8}{7},\frac{9}{7},\frac{3}{2},\frac{11}{7},\frac{12}{7},\frac{25}{14},\\ \frac{27}{14},\frac{15}{7},\frac{31}{14},\frac{33}{14},\frac{18}{7},  \\ \frac{20}{7},3,\frac{45}{14},\frac{51}{14},\frac{30}{7}$ } &  $x^2+y^3+z^4+z w^{5}=0$  & ~\\ \hline
$(E_8(a_1), 15, 13)$ & \parbox[c]{3.0cm}{ $\frac{30}{13}, \frac{27}{13}, \frac{24}{13}, \frac{21}{13}, \frac{20}{13}$ \\ $\frac{18}{13},\frac{17}{13}, \frac{15}{13}, \frac{14}{13}$ } &  $x^2+y^2+z^3+w^{10}=0$  & ~\\ \hline
$(E_8(a_2), 15, 11)$ & \parbox[c]{3cm}{ $(\frac{9}{8},\frac{5}{4},\frac{3}{2},\frac{15}{8} )^2$ } &  $x^2+y^2+z^3+w^5=0$  & ~\\ \hline

$(E_8(a_1), 12, 11)$ & \parbox[c]{3.6cm}{ $\frac{42}{11}, \frac{36}{11}, (\frac{30}{11})^2, \frac{28}{11}, (\frac{24}{11})^2$ \\ $2, (\frac{18}{11})^3, (\frac{16}{11})^2, (\frac{12}{11})^2$ }  & ~ & ~\\ \hline
$(E_8(a_3), 12,7)$ & $\frac{12}{7}, \frac{9}{7}, \frac{8}{7}$  & $x^2+y^2+z^3+w^{4}=0$ & $\mathrm{Vir}_{3,14}\oplus L(13,1)$ \\ \hline
$(E_8(a_5), 12, 5)$ & $(\frac{6}{5})^4$  & $4\times(x^2+y^2+z^2+w^3=0)$ &$4\times \mathrm{Vir}_{2,5}$ \\ \hline

$(E_8(a_1), 10, 9)$ & \parbox[c]{1.5cm}{ $\frac{10}{9},\frac{10}{9},\frac{4}{3},\frac{4}{3},\frac{14}{9},\frac{16}{9},\frac{16}{9},2
$ } & &\\ 
~&  \parbox[c]{1.5cm}{ $\frac{20}{9},\frac{20}{9},\frac{22}{9},\frac{26}{9},\frac{10}{3}
$ } &~&~\\ \hline
$(E_8(a_2), 10, 7)$ & \parbox[c]{1.5cm}{ $(\frac{10}{7})^3, (\frac{8}{7})^3$ } & $3\times (x^2+y^2+z^2+w^7=0)$  & $3\times \mathrm{Vir}_{2,9}$\\ \hline

$(E_8(a_1), 8, 7)$ & \parbox[c]{4cm}{ $\frac{18}{7}, \frac{16}{7}, 2, (\frac{12}{7})^2, (\frac{10}{7})^3, (\frac{8}{7})^2$ }  &  $D_7/Z_2:x^2+y^6+yz^2+w^2=0$ & ~\\ \hline
$(E_8(a_3), 8,5)$ & $\frac{12}{5}, 2, (\frac{8}{5})^2, \frac{7}{5}, (\frac{6}{5})^2$  & $D_5/Z_2:x^2+y^4+yz^2+w^2=0$ & ~ \\ \hline
$(E_8(b_5), 8,3)$ & $2,({4\over3})^3$  & $x^3+y^3+z^3+w^2=0$ & ~ \\ \hline

$(E_8(a_1), 6, 5)$ & $({6\over 5})^5$  & $5\times(x^2+y^2+z^2+w^3)=0$ & $5\times  \mathrm{Vir}_{2,5}$\\ \hline

$(E_8(a_3), 5, 3)$ & $2,({4\over3})^3,$  & $x^3+y^3+z^3+w^2=0$ & ~ \\ \hline

$(E_8(a_2), 4, 3)$ & \parbox[c]{3.5cm}{ $\frac{10}{3}, (\frac{8}{3})^2, 2^4, (\frac{4}{3})^5$ }  & ~  & ~\\ \hline

 \end{tabular} 
 }               
\end{center}
\caption{Physical data for 4d theories whose VOA is $W^{-h^\vee+{n\over u}}(\mathfrak{e}_8,f)$. Here $f$ denotes regular singularity. }
  \label{table:e8lisse}
\end{table}

\subsection{A proposal for the character}
Now we discuss the character of lisse W-algebra with exceptional Lie algebra. We first consider the case when the regular puncture has full flavor symmetry group (the nilpotent orbit $f$ is trivial). The corresponding VOA is  
the affine Kac-Moody algebra $V^k(\mathfrak{g})$, with the level $k$ takes the following form
\begin{equation}
k=-h^\vee+{n\over u},
\end{equation}
Here $(n,u)=1$ and $n$ is chosen such that there is no mass parameter encoded in irregular singularity. In principle, the vacuum character can be computed using the Kazhdan-Lusztig formula, 
however, the computation is quite involved. Here we propose a much simpler formula inspired by physics  and some observations on known formula found in the classical group case.

The formula we are going to propose for $V^{-h^\vee+{n\over u}}(\mathfrak{g})$ is
\begin{equation}
\ch_{\mathfrak{g},n,u}(q)=\pe{{(q-q^u)\chi^{adj}_{\mathfrak{g}}\over(1-q)(1-q^u) }} \chi_{\mathfrak{g},n,u=1}(q^u),
\end{equation}
where $\chi_{\mathfrak{g},n,u=1}(q)$ is the character for the theory defined at $u=1$.  This formula is confirmed for the classical group case by using the weakly coupled gauge theory description, and 
we propose that this is true for arbitrary case. This also works for the case $n=h^\vee$ as the $u=1$ theory is trivial so the index is just the $\mathrm{PE}$ term which recovers the general formula found in \cite{Xie:2019zlb}. 

 If this general proposal is correct, the computation of index of AKM is reduced to find the index for the theory for $u=1$, which is known for many cases. Firstly, let us set up some notations. For a semi-simple Lie algebra $\fg$ with rank $r$, define
\begin{equation}
\eta^{\fg}(\mathbf{a})=\prod_{w\in \mathrm{adj} \fg} (\mathbf{a}^w q;q),
\end{equation}
with $\mathbf{a}^w=\prod_{i=1}^r a_i^{w_i}$. The character, or Schur polynomial, of the representation $\lambda$ of $\fg$ is denoted as $\chi_{\lambda}^{\fg}(\mathbf{a})$ while the $q$-dimension of the representation $\lambda$ is just $\chi_{\lambda}^{\fg}(q^{\rho})$ with $\rho$ being the Weyl vector (half sum of positive roots) of $\fg$.

\subsubsection*{$E_6$: $n=9,~u=1$, $E_6$ MN~theory}
This is realized by the 6d $A_2$ $(2,0)$ theory on a sphere with three full punctures, and the Schur index is
\begin{equation}
\CI^{E_6,n=9}=\frac{(q^2;q)(q^3;q)}{\eta^{\fsu_3}(\mathbf{a})\eta^{\fsu_3}(\mathbf{b})\eta^{\fsu_3}(\mathbf{c})}
\sum_{\lambda}\frac{\chi^{\fsu_3}_{\lambda}(\mathbf{a})\chi^{\fsu_3}_{\lambda}(\mathbf{b})\chi^{\fsu_3}_{\lambda}(\mathbf{c})}{\chi^{\fsu_3}_{\lambda}(q^\rho)}.
\end{equation}
The sum is over all highest weight representations of $\fsu_3$. Although not manifest, the above formula has an $E_6$ symmetry in the end.

\subsubsection*{$E_7$: $n=14,~u=1$, $E_7$ MN~theory}

It is realized by the 6d $A_3$ $(2,0)$ theory on a sphere with Young tableaux $[1^4]$, $[1^4]$ and $[2^2]$ (two full punctures and one partially closed puncture). The Schur index is
\begin{equation}
\CI^{E_7,n=14}=\frac{(q^3;q)(q^4;q)}{\eta^{\fsu_4}(\mathbf{a})\eta^{\fsu_4}(\mathbf{b})\eta^{[2^2]}(c)}
\sum_{\lambda}\frac{\chi^{\fsu_4}_{\lambda}(\mathbf{a})\chi^{\fsu_4}_{\lambda}(\mathbf{b})\chi^{\fsu_4}_{\lambda}(\tilde{\mathbf{c}})}{\chi^{\fsu_4}_{\lambda}(q^\rho)},
\end{equation}
with
\begin{equation}
\eta^{[2^2]}(c)=(q;q)(qc^{\pm};q)(q^2;q)(q^2c^{\pm},q)=\prod_{w\in\mathrm{adj}~\fsu(2)}(qc^w;q)(q^2c^w;q)
\end{equation}
and
\begin{equation}
\tilde{\mathbf{c}}=(q^{\half}c,q,q^{\half}c).
\end{equation}
We use the shorthand notation $(qa^\pm;q)=(qa;q)(qa^{-1};q)$. Again the sum is over all highest weight representations of $SU(4)$ and the result has an $E_7$ symmetry.

\subsubsection*{$E_8$: $n=24,~u=1$, $E_8$ MN~theory}

It is realized by the 6d $A_5$ $(2,0)$ theory on a sphere with Young tableaux $[1^6]$, $[3^2]$ and $[2^3]$ (one full puncture and two partially closed punctures). The Schur index is
\begin{equation}
\CI^{E_8,n=24}=\frac{(q^4;q)(q^5;q)(q^6;q)}{(q^3;q)\eta^{\fsu_6}(\mathbf{a})\eta^{[3^2]}(\mathbf{b})\eta^{[2^3]}(c)}
\sum_{\lambda}\frac{\chi^{\fsu_6}_{\lambda}(\mathbf{a})\chi^{\fsu_6}_{\lambda}(\tilde{\mathbf{b}})\chi^{\fsu_6}_{\lambda}(\tilde{\mathbf{c}})}{\chi^{\fsu_6}_{\lambda}(q^\rho)}
\end{equation}
where
\begin{equation}
\begin{split}
\eta^{[3^2]}(\mathbf{b})=&\prod_{w\in\mathrm{adj}~\fsu_3}(q\mathbf{b}^w;q)(q^2\mathbf{b}^w;q),\\
\eta^{[2^3]}(c)=&\prod_{w\in\mathrm{adj}~\fsu_2}(qc^w;q)(q^2c^w;q)(q^3c^w;q),
\end{split}
\end{equation}
and
\begin{equation}
\begin{split}
\tilde{\mathbf{b}}=&(q^{\half}b_1,qb_2,q^{\frac{3}{2}},qb^{-1}_2,q^{\half}b^{-1}_1),\\
\tilde{\mathbf{c}}=&(q^{\half} c, q, q^{\half}).
\end{split}
\end{equation}
Now the sum is over all highest weight representations of $SU(6)$. Again, there is an enhancement to $E_8$ symmetry.

\section{Conclusion}
\label{sec:conclusion}

We continue our studies of the correspondence between 2d W-algebras and 4d $\mathcal{N}=2$ SCFTs. 
Lisse  W-algebras are defined as those W-algebras whose corresponding Zhu's $C_2$ algebra are finite dimensional. As the associated variety of Zhu's $C_2$ algebra is identified with the 
Higgs branch in the 4d/2d correspondence, the lisse condition is equivalent to the absence of Higgs branch of the 4d theory. We classify 4d $\mathcal{N}=2$ SCFTs which 
do not admit a Higgs branch, and the corresponding 2d W-algebra of these theories should be lisse. In particular, we predict the existence of a large class of new 
non-admissible lisse W-algebra which has not been found before (see few examples considered in \cite{Arakawa:2016ad}). These 4d theories can appear in the IR theory of the Higgs branch of a general 4d 
$\mathcal{N}=2$ SCFT, and therefore they are crucial to understand the behavior of Higgs branch of general 4d $\mathcal{N}=2$ SCFTs. 

In our study of 4d $\mathcal{N}=2$ theory with no Higgs branch, we encounter the following generic situation.
Consider 4d $\mathcal{N}=2$ SCFT constructed from  6d $(2,0)$ theory on a sphere with one irregular singularity and one regular singularity labeled by trivial nilpotent orbit. 
We call this theory $UV$ theory, and the corresponding VOA is just the AKM algebra. 
 It is easy to study the IR theory of its Higgs branch: one simply change the type of regular puncture to the one labeled by the nilpotent element $f$, whose orbit leads to the associated variety of the
 AKM. The IR theory generically consists of free hypermultiplets, free vectormultiplets, and an interacting theory which has no Higgs branch. 
 This suggests that the UV AKM might be related to the VOA of free hypermultiplets and vectormultiplets, plus a lisse W-algebra. Indeed such realizations are discussed in \cite{Semikhatov:1993pr,2004math1023A,Adamovic:2014lra,2017arXiv171111342A,Beem:2019tfp}. 
 Our study reveals that such constructions are much more general and we can identify all the IR pieces in the Higgs branch using $(2,0)$ construction. Moreover, our results
 suggest that it is also possible to have free vectormultiplets in the IR. It would be interesting to further study the relations between these W-algebras. 
 
 We found new non-admissible lisse W-algebras, which deserve further studies from physical and mathematical point of view. We give a formula for vacuum characters for these W-algebras in the classical Lie algebra case, and it would 
 be interesting to verify them using 2d methods. The module structure is also very interesting to study, and some results will appear in \cite{Xie:2019}.

The lisse W-algebra in our context has interesting connection to three dimensional Q-factorial terminal singularity. In our paper, we mainly study the Gorenstein 
Q-terminal singularity, which gives the results in table \ref{table:constraintADEirregular}. The non-Gorenstein ones needs further study, and might be related to the lisse W-algebra  with non-simply laced 
Lie algebra, and it is interesting to further study them as well. 

\section*{Acknowledgements}
The authors would like to thank Tomoyuki Arakawa, Jethro van Ekeren, Reimundo Heluani, and Anne Moreau for helpful discussions. DX and WY are supported by Yau mathematical science center at Tsinghua University. WY is also supported by the Young overseas high-level talents introduction plan. WY would also like to thank the 2019 Pollica summer workshop, which was supported in part by the Simons Foundation (Simons Collaboration on the Non-perturbative Bootstrap) and in part by the INFN, and also appreciated the hospitality of the department of physics, University of Toronto during the final stage of this work. The authors are grateful for this support.

\appendix
\section{Review on the Higgs branch of a quiver with classical gauge groups}
This is a review of associating a quiver to a nilpotent orbit of a classical Lie algebra which were discussed in \cite{Gaiotto:2008ak}. This construction was used to find 
the Higgs branch of quiver gauge theories discussed in section \ref{sec:nonadm:classical}. The basic idea is the following: the quiver has a flavor symmetry whose Lie algebra $\fg$ is a classical semi-simple one, and its Higgs branch is assumed to be the closure of nilpotent orbit $e^d$ in $\fg$. To find $e^d$, we first find a nilpotent orbit $e$ in the Langlands dual $\fg^\vee$ Lie algebra, whose associated 
quiver is just the quiver we started. The nilpotent $e^d$ can be found from $e$ by using the Spaltenstein map.

\begin{figure}
\begin{center}
\tikzset{every picture/.style={line width=0.75pt}} 

\begin{tikzpicture}[x=0.75pt,y=0.75pt,yscale=-1,xscale=1]

\draw    (104.08,79.44) -- (104.08,135.26) ;

\draw    (244.41,79.44) -- (244.41,135.26) ;

\draw    (174.25,79.44) -- (174.25,135.26) ;

\draw    (349.66,79.44) -- (349.66,135.26) ;

\draw  [dash pattern={on 4.5pt off 4.5pt}]  (255.64,107.35) -- (344.05,107.35) ;

\draw    (104.08,98.05) -- (174.25,98.05) ;

\draw    (104.08,116.65) -- (174.25,116.65) ;

\draw  [dash pattern={on 0.84pt off 2.51pt}]  (139.16,98.05) -- (139.16,116.65) ;

\draw    (174.25,88.74) -- (244.41,88.74) ;

\draw    (174.25,125.95) -- (244.41,125.95) ;

\draw  [dash pattern={on 0.84pt off 2.51pt}]  (209.33,88.74) -- (209.33,125.95) ;

\draw    (349.66,88.74) -- (419.83,88.74) ;

\draw    (349.66,125.95) -- (419.83,125.95) ;

\draw  [dash pattern={on 0.84pt off 2.51pt}]  (384.75,88.74) -- (384.75,125.95) ;

\draw    (419.83,88.74) -- (490,88.74) ;

\draw    (419.83,125.95) -- (490,125.95) ;

\draw  [dash pattern={on 0.84pt off 2.51pt}]  (454.92,88.74) -- (454.92,125.95) ;

\draw    (419.83,79.44) -- (419.83,135.26) ;

\draw    (290,160) -- (290,218) ;
\draw [shift={(290,220)}, rotate = 270] [color={rgb, 255:red, 0; green, 0; blue, 0 }  ][line width=0.75]    (10.93,-3.29) .. controls (6.95,-1.4) and (3.31,-0.3) .. (0,0) .. controls (3.31,0.3) and (6.95,1.4) .. (10.93,3.29)   ;

\draw   (330.51,254.21) .. controls (339.07,245.48) and (353.05,245.32) .. (361.72,253.83) .. controls (370.39,262.35) and (370.48,276.32) .. (361.91,285.05) .. controls (353.35,293.77) and (339.37,293.93) .. (330.7,285.42) .. controls (322.03,276.9) and (321.94,262.93) .. (330.51,254.21) -- cycle ; \draw   (330.51,254.21) -- (361.91,285.05) ; \draw   (361.72,253.83) -- (330.7,285.42) ;
\draw   (86.09,254.95) .. controls (94.65,246.23) and (108.59,246.02) .. (117.21,254.49) .. controls (125.83,262.96) and (125.87,276.89) .. (117.31,285.61) .. controls (108.74,294.34) and (94.81,294.54) .. (86.19,286.08) .. controls (77.57,277.61) and (77.52,263.68) .. (86.09,254.95) -- cycle ; \draw   (86.09,254.95) -- (117.31,285.61) ; \draw   (117.21,254.49) -- (86.19,286.08) ;
\draw   (168.51,254.21) .. controls (177.07,245.48) and (191.05,245.32) .. (199.72,253.83) .. controls (208.39,262.35) and (208.48,276.32) .. (199.91,285.05) .. controls (191.35,293.77) and (177.37,293.93) .. (168.7,285.42) .. controls (160.03,276.9) and (159.94,262.93) .. (168.51,254.21) -- cycle ; \draw   (168.51,254.21) -- (199.91,285.05) ; \draw   (199.72,253.83) -- (168.7,285.42) ;
\draw   (248.51,254.21) .. controls (257.07,245.48) and (271.05,245.32) .. (279.72,253.83) .. controls (288.39,262.35) and (288.48,276.32) .. (279.91,285.05) .. controls (271.35,293.77) and (257.37,293.93) .. (248.7,285.42) .. controls (240.03,276.9) and (239.94,262.93) .. (248.51,254.21) -- cycle ; \draw   (248.51,254.21) -- (279.91,285.05) ; \draw   (279.72,253.83) -- (248.7,285.42) ;
\draw   (407.11,253.37) .. controls (415.67,244.65) and (429.64,244.48) .. (438.32,252.99) .. controls (446.99,261.51) and (447.07,275.48) .. (438.51,284.21) .. controls (429.94,292.93) and (415.97,293.1) .. (407.3,284.58) .. controls (398.63,276.07) and (398.54,262.09) .. (407.11,253.37) -- cycle ; \draw   (407.11,253.37) -- (438.51,284.21) ; \draw   (438.32,252.99) -- (407.3,284.58) ;
\draw    (122.83,261.4) -- (163,261.4) ;

\draw    (122.83,280) -- (163,280) ;

\draw  [dash pattern={on 0.84pt off 2.51pt}]  (142.92,261.4) -- (142.92,280) ;

\draw    (203.83,255) -- (244,255) ;

\draw    (203.83,285) -- (244,285) ;

\draw  [dash pattern={on 0.84pt off 2.51pt}]  (223.92,255) -- (223.92,285) ;

\draw    (363.83,254) -- (404,254) ;

\draw    (363.83,284) -- (404,284) ;

\draw  [dash pattern={on 0.84pt off 2.51pt}]  (383.92,254) -- (383.92,284) ;

\draw    (439.83,254) -- (480,254) ;

\draw    (439.83,284) -- (480,284) ;

\draw  [dash pattern={on 0.84pt off 2.51pt}]  (459.92,254) -- (459.92,284) ;

\draw  [dash pattern={on 4.5pt off 4.5pt}]  (289,270) -- (323,270) ;

\draw    (290,310) -- (290,358) ;
\draw [shift={(290,360)}, rotate = 270] [color={rgb, 255:red, 0; green, 0; blue, 0 }  ][line width=0.75]    (10.93,-3.29) .. controls (6.95,-1.4) and (3.31,-0.3) .. (0,0) .. controls (3.31,0.3) and (6.95,1.4) .. (10.93,3.29)   ;

\draw   (120,395) .. controls (120,381.19) and (131.19,370) .. (145,370) .. controls (158.81,370) and (170,381.19) .. (170,395) .. controls (170,408.81) and (158.81,420) .. (145,420) .. controls (131.19,420) and (120,408.81) .. (120,395) -- cycle ;
\draw   (200,395) .. controls (200,381.19) and (211.19,370) .. (225,370) .. controls (238.81,370) and (250,381.19) .. (250,395) .. controls (250,408.81) and (238.81,420) .. (225,420) .. controls (211.19,420) and (200,408.81) .. (200,395) -- cycle ;
\draw   (360,395) .. controls (360,381.19) and (371.19,370) .. (385,370) .. controls (398.81,370) and (410,381.19) .. (410,395) .. controls (410,408.81) and (398.81,420) .. (385,420) .. controls (371.19,420) and (360,408.81) .. (360,395) -- cycle ;
\draw   (440,370) -- (490,370) -- (490,420) -- (440,420) -- cycle ;
\draw    (170,395) -- (200,395) ;

\draw    (250,395) -- (280,395) ;

\draw    (330,395) -- (360,395) ;

\draw    (410,395) -- (440,395) ;

\draw  [dash pattern={on 4.5pt off 4.5pt}]  (280,395) -- (330,395) ;

\draw (454.5,148) node [scale=0.7]  {$N\ D3$};
\draw (383.5,148) node [scale=0.7]  {$r_{1} \ D3$};
\draw (207.5,148) node [scale=0.7]  {$r_{s-2} \ D3$};
\draw (136,148) node [scale=0.7]  {$r_{s-1} \ D3$};
\draw (302,21) node   {$A_{N-1} \ \mathrm{type} :[ h_{1} ,h_{2} ,\cdots ,h_{s}] ,\ \mathrm{with} \ r_{i} =N-\sum ^{i}_{j=1} h_{j}$};
\draw (340.5,190) node   {$S-dual$};
\draw (108.71,59) node [scale=0.8]  {$D5$};
\draw (175.86,58.5) node [scale=0.8]  {$D5$};
\draw (243.78,59.5) node [scale=0.8]  {$D5$};
\draw (350.3,59.5) node [scale=0.8]  {$D5$};
\draw (422.85,59.5) node [scale=0.8]  {$D5$};
\draw (141,302) node [scale=0.7]  {$r_{s-1} \ D3$};
\draw (220,302) node [scale=0.7]  {$r_{s-2} \ D3$};
\draw (384.5,303) node [scale=0.7]  {$r_{1} \ D3$};
\draw (460.5,302) node [scale=0.7]  {$N \ D3$};
\draw (100.5,221.5) node [scale=0.8]  {$NS5$};
\draw (184.5,221.5) node [scale=0.8]  {$NS5$};
\draw (262.5,221.5) node [scale=0.8]  {$NS5$};
\draw (344.5,221.5) node [scale=0.8]  {$NS5$};
\draw (424.5,221.5) node [scale=0.8]  {$NS5$};
\draw (339.5,340) node   {$quiver$};
\draw (465,395) node   {$N$};
\draw (145,395) node [scale=0.7]  {$U( r_{s-1})$};
\draw (225,395) node [scale=0.7]  {$U( r_{s-2})$};
\draw (385,395) node [scale=0.7]  {$U( r_{1})$};

\end{tikzpicture}
\caption{Upper row: We construct a $D5-D3$ brane system for a $A$ type partition. Middle row: We perform S duality on the brane system in upper row, which turns a D5 brane to a NS5 brane, and a D3 brane to a D3 brane. Bottom row: We read a quiver gauge theory from the middle row. The rule is following, we associate a $U(r)$ gauge group for $r$ D3 branes suspending between NS5 brane, and each NS5 brane contributes a hypermultiplet in bi-fundamental representation. }
\label{Atype}
\end{center}
\end{figure}
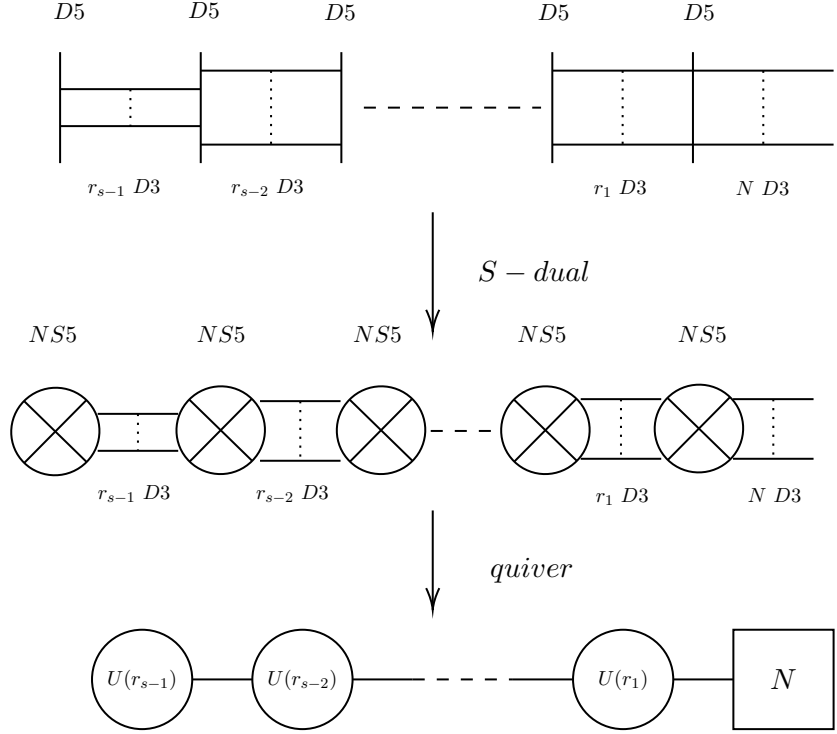

First consider $\fg=A_{N-1}$, and its Langlands dual is just itself. A nilpotent orbit in $A_{N-1}$ Lie algebra is specified by a partition $[h_1,\ldots, h_s]$, and to find its associated quiver, 
we first construct a $D3-D5$ brane system as in figure \ref{Atype}.  The number of D3 branes suspended between D5 branes are $r_1=N-h_1$, and $r_2=N-h_1-h_2$, and etc. The quiver gauge theory is found by doing S duality on above brane configuration: $D3$ branes stay as $D3$ branes, while $D5$ branes become $NS5$ branes.  The quiver gauge theory can be easily read from the $NS5-D3$ brane system. The process was summarized in figure \ref{Atype}.

For other classical Lie algebras, a nilpotent orbit is also labeled by a partition $[h_1,\ldots, h_s]$, but with following constraints: for $(B_N,D_N)$, even parts appear even time; for $C_N$, odd parts appear even times. To engineer gauge theories with orthogonal and symplectic gauge groups, we need to use O3 planes. There are a total of four kinds of $O3$ planes, and we call them $B$ $D$, and $C$, $C^{'}$ type. The name of such O3 planes are clear: if we put D3 branes on certain type of O3 planes, we would get corresponding gauge theory types (the number of D3 branes on corresponding O3 plane is then constrained by the type, i.e. there are $2r$ for D type, $2r+1$ for $B$ type, and $2r$ for C type).  S duality exchange those O3 planes as follows:
\begin{equation}
D\leftrightarrow D,~~B\leftrightarrow C~,~~C\leftrightarrow B,~~~C^{'}\leftrightarrow C^{'}
\end{equation}

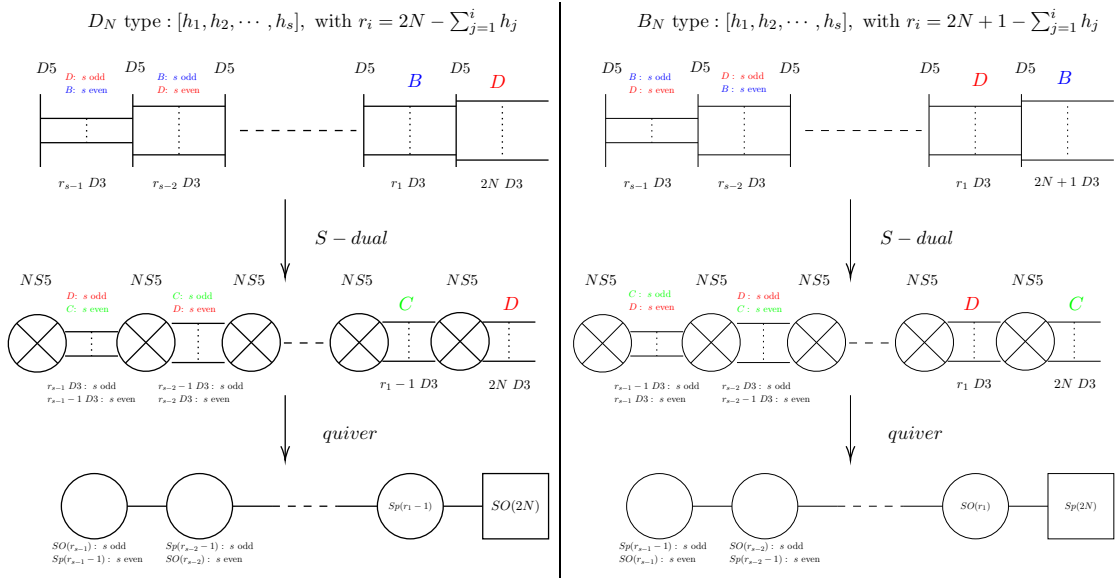
\begin{figure}
\begin{center}
\resizebox{\textwidth}{!}{
\begin{tabular}{c|c}
\tikzset{every picture/.style={line width=0.75pt}} 

\begin{tikzpicture}[x=0.75pt,y=0.75pt,yscale=-1,xscale=1]

\draw    (104.08,79.44) -- (104.08,135.26) ;

\draw    (244.41,79.44) -- (244.41,135.26) ;

\draw    (174.25,79.44) -- (174.25,135.26) ;

\draw    (349.66,79.44) -- (349.66,135.26) ;

\draw  [dash pattern={on 4.5pt off 4.5pt}]  (255.64,107.35) -- (344.05,107.35) ;

\draw    (104.08,98.05) -- (174.25,98.05) ;

\draw    (104.08,116.65) -- (174.25,116.65) ;

\draw  [dash pattern={on 0.84pt off 2.51pt}]  (139.16,98.05) -- (139.16,116.65) ;

\draw    (174.25,88.74) -- (244.41,88.74) ;

\draw    (174.25,125.95) -- (244.41,125.95) ;

\draw  [dash pattern={on 0.84pt off 2.51pt}]  (209.33,88.74) -- (209.33,125.95) ;

\draw    (349.66,88.74) -- (419.83,88.74) ;

\draw    (349.66,125.95) -- (419.83,125.95) ;

\draw  [dash pattern={on 0.84pt off 2.51pt}]  (384.75,88.74) -- (384.75,125.95) ;

\draw    (419.83,84.74) -- (490,84.74) ;

\draw    (419.83,129.95) -- (490,129.95) ;

\draw  [dash pattern={on 0.84pt off 2.51pt}]  (454.92,88.74) -- (454.92,125.95) ;

\draw    (419.83,79.44) -- (419.83,135.26) ;

\draw    (290,160) -- (290,218) ;
\draw [shift={(290,220)}, rotate = 270] [color={rgb, 255:red, 0; green, 0; blue, 0 }  ][line width=0.75]    (10.93,-3.29) .. controls (6.95,-1.4) and (3.31,-0.3) .. (0,0) .. controls (3.31,0.3) and (6.95,1.4) .. (10.93,3.29)   ;

\draw   (330.51,254.21) .. controls (339.07,245.48) and (353.05,245.32) .. (361.72,253.83) .. controls (370.39,262.35) and (370.48,276.32) .. (361.91,285.05) .. controls (353.35,293.77) and (339.37,293.93) .. (330.7,285.42) .. controls (322.03,276.9) and (321.94,262.93) .. (330.51,254.21) -- cycle ; \draw   (330.51,254.21) -- (361.91,285.05) ; \draw   (361.72,253.83) -- (330.7,285.42) ;
\draw   (86.09,254.95) .. controls (94.65,246.23) and (108.59,246.02) .. (117.21,254.49) .. controls (125.83,262.96) and (125.87,276.89) .. (117.31,285.61) .. controls (108.74,294.34) and (94.81,294.54) .. (86.19,286.08) .. controls (77.57,277.61) and (77.52,263.68) .. (86.09,254.95) -- cycle ; \draw   (86.09,254.95) -- (117.31,285.61) ; \draw   (117.21,254.49) -- (86.19,286.08) ;
\draw   (168.51,254.21) .. controls (177.07,245.48) and (191.05,245.32) .. (199.72,253.83) .. controls (208.39,262.35) and (208.48,276.32) .. (199.91,285.05) .. controls (191.35,293.77) and (177.37,293.93) .. (168.7,285.42) .. controls (160.03,276.9) and (159.94,262.93) .. (168.51,254.21) -- cycle ; \draw   (168.51,254.21) -- (199.91,285.05) ; \draw   (199.72,253.83) -- (168.7,285.42) ;
\draw   (248.51,254.21) .. controls (257.07,245.48) and (271.05,245.32) .. (279.72,253.83) .. controls (288.39,262.35) and (288.48,276.32) .. (279.91,285.05) .. controls (271.35,293.77) and (257.37,293.93) .. (248.7,285.42) .. controls (240.03,276.9) and (239.94,262.93) .. (248.51,254.21) -- cycle ; \draw   (248.51,254.21) -- (279.91,285.05) ; \draw   (279.72,253.83) -- (248.7,285.42) ;
\draw   (407.11,253.37) .. controls (415.67,244.65) and (429.64,244.48) .. (438.32,252.99) .. controls (446.99,261.51) and (447.07,275.48) .. (438.51,284.21) .. controls (429.94,292.93) and (415.97,293.1) .. (407.3,284.58) .. controls (398.63,276.07) and (398.54,262.09) .. (407.11,253.37) -- cycle ; \draw   (407.11,253.37) -- (438.51,284.21) ; \draw   (438.32,252.99) -- (407.3,284.58) ;
\draw    (122.83,261.4) -- (163,261.4) ;

\draw    (122.83,280) -- (163,280) ;

\draw  [dash pattern={on 0.84pt off 2.51pt}]  (142.92,261.4) -- (142.92,280) ;

\draw    (203.83,255) -- (244,255) ;

\draw    (203.83,285) -- (244,285) ;

\draw  [dash pattern={on 0.84pt off 2.51pt}]  (223.92,255) -- (223.92,285) ;

\draw    (363.83,254) -- (404,254) ;

\draw    (363.83,284) -- (404,284) ;

\draw  [dash pattern={on 0.84pt off 2.51pt}]  (383.92,254) -- (383.92,284) ;

\draw    (439.83,254) -- (480,254) ;

\draw    (439.83,284) -- (480,284) ;

\draw  [dash pattern={on 0.84pt off 2.51pt}]  (459.92,254) -- (459.92,284) ;

\draw  [dash pattern={on 4.5pt off 4.5pt}]  (289,270) -- (323,270) ;

\draw    (290,310) -- (290,358) ;
\draw [shift={(290,360)}, rotate = 270] [color={rgb, 255:red, 0; green, 0; blue, 0 }  ][line width=0.75]    (10.93,-3.29) .. controls (6.95,-1.4) and (3.31,-0.3) .. (0,0) .. controls (3.31,0.3) and (6.95,1.4) .. (10.93,3.29)   ;

\draw   (120,395) .. controls (120,381.19) and (131.19,370) .. (145,370) .. controls (158.81,370) and (170,381.19) .. (170,395) .. controls (170,408.81) and (158.81,420) .. (145,420) .. controls (131.19,420) and (120,408.81) .. (120,395) -- cycle ;
\draw   (200,395) .. controls (200,381.19) and (211.19,370) .. (225,370) .. controls (238.81,370) and (250,381.19) .. (250,395) .. controls (250,408.81) and (238.81,420) .. (225,420) .. controls (211.19,420) and (200,408.81) .. (200,395) -- cycle ;
\draw   (360,395) .. controls (360,381.19) and (371.19,370) .. (385,370) .. controls (398.81,370) and (410,381.19) .. (410,395) .. controls (410,408.81) and (398.81,420) .. (385,420) .. controls (371.19,420) and (360,408.81) .. (360,395) -- cycle ;
\draw   (440,370) -- (490,370) -- (490,420) -- (440,420) -- cycle ;
\draw    (170,395) -- (200,395) ;

\draw    (250,395) -- (280,395) ;

\draw    (330,395) -- (360,395) ;

\draw    (410,395) -- (440,395) ;

\draw  [dash pattern={on 4.5pt off 4.5pt}]  (280,395) -- (330,395) ;

\draw (454.5,148) node [scale=0.7]  {$2N\ D3$};
\draw (383.5,148) node [scale=0.7]  {$r_{1} \ D3$};
\draw (207.5,148) node [scale=0.7]  {$r_{s-2} \ D3$};
\draw (136,148) node [scale=0.7]  {$r_{s-1} \ D3$};
\draw (303.5,24) node   {$D_{N} \ \mathrm{type} :[ h_{1} ,h_{2} ,\cdots ,h_{s}] ,\ \mathrm{with} \ r_{i} =2N-\sum ^{i}_{j=1} h_{j}$};
\draw (340.5,190) node   {$S-dual$};
\draw (108.71,59) node [scale=0.8]  {$D5$};
\draw (175.86,58.5) node [scale=0.8]  {$D5$};
\draw (243.78,59.5) node [scale=0.8]  {$D5$};
\draw (350.3,59.5) node [scale=0.8]  {$D5$};
\draw (422.85,59.5) node [scale=0.8]  {$D5$};
\draw (143,308) node [scale=0.5]  {$ \begin{array}{l}
r_{s-1} \ D3:\ s\ \mathrm{odd}\\
r_{s-1}-1 \ D3:\ s\ \mathrm{even}
\end{array}$};
\draw (226,308) node [scale=0.5]  {$ \begin{array}{l}
r_{s-2} -1\ D3:\ s\ \mathrm{odd}\\
r_{s-2} \ D3:\ s\ \mathrm{even}
\end{array}$};
\draw (384,302) node [scale=0.7]  {$r_{1} -1\ D3$};
\draw (460.5,302) node [scale=0.7]  {$2N\ D3$};
\draw (100.5,221.5) node [scale=0.8]  {$NS5$};
\draw (184.5,221.5) node [scale=0.8]  {$NS5$};
\draw (262.5,221.5) node [scale=0.8]  {$NS5$};
\draw (344.5,221.5) node [scale=0.8]  {$NS5$};
\draw (424.5,221.5) node [scale=0.8]  {$NS5$};
\draw (339.5,340) node   {$quiver$};
\draw (465,395) node [scale=0.7]  {$SO( 2N)$};
\draw (147.5,431) node [scale=0.5]  {$ \begin{array}{l}
SO( r_{s-1}) :\ s\ \mathrm{odd}\\
Sp( r_{s-1} -1) :\ s\ \mathrm{even}
\end{array}$};
\draw (385,395) node [scale=0.5]  {$Sp( r_{1} -1)$};
\draw (451.5,70) node [color={rgb, 255:red, 255; green, 0; blue, 0 }  ,opacity=1 ]  {$D$};
\draw (388.5,69) node [color={rgb, 255:red, 0; green, 0; blue, 255 }  ,opacity=1 ]  {$B$};
\draw (461.5,240) node   {$\textcolor[rgb]{1,0,0}{D}$};
\draw (382,240) node   {$\textcolor[rgb]{0,1,0}{C}$};
\draw (139,72) node [scale=0.5]  {$ \begin{array}{l}
\textcolor[rgb]{1,0,0}{D}\textcolor[rgb]{1,0,0}{:\ s\ \mathrm{odd}}\\
\textcolor[rgb]{0,0,1}{B}\textcolor[rgb]{0,0,1}{:\ s\ \mathrm{even}}
\end{array}$};
\draw (209,72) node [scale=0.5]  {$ \begin{array}{l}
\textcolor[rgb]{0,0,1}{B}\textcolor[rgb]{0,0,1}{:\ s\ \mathrm{odd}}\\
\textcolor[rgb]{1,0,0}{D}\textcolor[rgb]{1,0,0}{:\ s\ \mathrm{even}}
\end{array}$};
\draw (140,239) node [scale=0.5]  {$ \begin{array}{l}
\textcolor[rgb]{1,0,0}{D}\textcolor[rgb]{1,0,0}{:\ s\ \mathrm{odd}}\\
\textcolor[rgb]{0,1,0}{C}\textcolor[rgb]{0,1,0}{:\ s\ \mathrm{even}}
\end{array}$};
\draw (221,239) node [scale=0.5]  {$ \begin{array}{l}
\textcolor[rgb]{0,1,0}{C}\textcolor[rgb]{0,1,0}{:\ s\ \mathrm{odd}}\\
\textcolor[rgb]{1,0,0}{D}\textcolor[rgb]{1,0,0}{:\ s\ \mathrm{even}}
\end{array}$};
\draw (232.5,431) node [scale=0.5]  {$ \begin{array}{l}
Sp( r_{s-2} -1) :\ s\ \mathrm{odd}\\
SO( r_{s-2}) :\ s\ \mathrm{even}
\end{array}$};

\end{tikzpicture}
&
\begin{tikzpicture}[x=0.75pt,y=0.75pt,yscale=-1,xscale=1]

\draw    (104.08,79.44) -- (104.08,135.26) ;

\draw    (244.41,79.44) -- (244.41,135.26) ;

\draw    (174.25,79.44) -- (174.25,135.26) ;

\draw    (349.66,79.44) -- (349.66,135.26) ;

\draw  [dash pattern={on 4.5pt off 4.5pt}]  (255.64,107.35) -- (344.05,107.35) ;

\draw    (104.08,98.05) -- (174.25,98.05) ;

\draw    (104.08,116.65) -- (174.25,116.65) ;

\draw  [dash pattern={on 0.84pt off 2.51pt}]  (139.16,98.05) -- (139.16,116.65) ;

\draw    (174.25,88.74) -- (244.41,88.74) ;

\draw    (174.25,125.95) -- (244.41,125.95) ;

\draw  [dash pattern={on 0.84pt off 2.51pt}]  (209.33,88.74) -- (209.33,125.95) ;

\draw    (349.66,88.74) -- (419.83,88.74) ;

\draw    (349.66,125.95) -- (419.83,125.95) ;

\draw  [dash pattern={on 0.84pt off 2.51pt}]  (384.75,88.74) -- (384.75,125.95) ;

\draw    (419.83,84.74) -- (490,84.74) ;

\draw    (419.83,129.95) -- (490,129.95) ;

\draw  [dash pattern={on 0.84pt off 2.51pt}]  (454.92,88.74) -- (454.92,125.95) ;

\draw    (419.83,79.44) -- (419.83,135.26) ;

\draw    (290,160) -- (290,218) ;
\draw [shift={(290,220)}, rotate = 270] [color={rgb, 255:red, 0; green, 0; blue, 0 }  ][line width=0.75]    (10.93,-3.29) .. controls (6.95,-1.4) and (3.31,-0.3) .. (0,0) .. controls (3.31,0.3) and (6.95,1.4) .. (10.93,3.29)   ;

\draw   (330.51,254.21) .. controls (339.07,245.48) and (353.05,245.32) .. (361.72,253.83) .. controls (370.39,262.35) and (370.48,276.32) .. (361.91,285.05) .. controls (353.35,293.77) and (339.37,293.93) .. (330.7,285.42) .. controls (322.03,276.9) and (321.94,262.93) .. (330.51,254.21) -- cycle ; \draw   (330.51,254.21) -- (361.91,285.05) ; \draw   (361.72,253.83) -- (330.7,285.42) ;
\draw   (86.09,254.95) .. controls (94.65,246.23) and (108.59,246.02) .. (117.21,254.49) .. controls (125.83,262.96) and (125.87,276.89) .. (117.31,285.61) .. controls (108.74,294.34) and (94.81,294.54) .. (86.19,286.08) .. controls (77.57,277.61) and (77.52,263.68) .. (86.09,254.95) -- cycle ; \draw   (86.09,254.95) -- (117.31,285.61) ; \draw   (117.21,254.49) -- (86.19,286.08) ;
\draw   (168.51,254.21) .. controls (177.07,245.48) and (191.05,245.32) .. (199.72,253.83) .. controls (208.39,262.35) and (208.48,276.32) .. (199.91,285.05) .. controls (191.35,293.77) and (177.37,293.93) .. (168.7,285.42) .. controls (160.03,276.9) and (159.94,262.93) .. (168.51,254.21) -- cycle ; \draw   (168.51,254.21) -- (199.91,285.05) ; \draw   (199.72,253.83) -- (168.7,285.42) ;
\draw   (248.51,254.21) .. controls (257.07,245.48) and (271.05,245.32) .. (279.72,253.83) .. controls (288.39,262.35) and (288.48,276.32) .. (279.91,285.05) .. controls (271.35,293.77) and (257.37,293.93) .. (248.7,285.42) .. controls (240.03,276.9) and (239.94,262.93) .. (248.51,254.21) -- cycle ; \draw   (248.51,254.21) -- (279.91,285.05) ; \draw   (279.72,253.83) -- (248.7,285.42) ;
\draw   (407.11,253.37) .. controls (415.67,244.65) and (429.64,244.48) .. (438.32,252.99) .. controls (446.99,261.51) and (447.07,275.48) .. (438.51,284.21) .. controls (429.94,292.93) and (415.97,293.1) .. (407.3,284.58) .. controls (398.63,276.07) and (398.54,262.09) .. (407.11,253.37) -- cycle ; \draw   (407.11,253.37) -- (438.51,284.21) ; \draw   (438.32,252.99) -- (407.3,284.58) ;
\draw    (122.83,261.4) -- (163,261.4) ;

\draw    (122.83,280) -- (163,280) ;

\draw  [dash pattern={on 0.84pt off 2.51pt}]  (142.92,261.4) -- (142.92,280) ;

\draw    (203.83,255) -- (244,255) ;

\draw    (203.83,285) -- (244,285) ;

\draw  [dash pattern={on 0.84pt off 2.51pt}]  (223.92,255) -- (223.92,285) ;

\draw    (363.83,254) -- (404,254) ;

\draw    (363.83,284) -- (404,284) ;

\draw  [dash pattern={on 0.84pt off 2.51pt}]  (383.92,254) -- (383.92,284) ;

\draw    (439.83,254) -- (480,254) ;

\draw    (439.83,284) -- (480,284) ;

\draw  [dash pattern={on 0.84pt off 2.51pt}]  (459.92,254) -- (459.92,284) ;

\draw  [dash pattern={on 4.5pt off 4.5pt}]  (289,270) -- (323,270) ;

\draw    (290,310) -- (290,358) ;
\draw [shift={(290,360)}, rotate = 270] [color={rgb, 255:red, 0; green, 0; blue, 0 }  ][line width=0.75]    (10.93,-3.29) .. controls (6.95,-1.4) and (3.31,-0.3) .. (0,0) .. controls (3.31,0.3) and (6.95,1.4) .. (10.93,3.29)   ;

\draw   (120,395) .. controls (120,381.19) and (131.19,370) .. (145,370) .. controls (158.81,370) and (170,381.19) .. (170,395) .. controls (170,408.81) and (158.81,420) .. (145,420) .. controls (131.19,420) and (120,408.81) .. (120,395) -- cycle ;
\draw   (200,395) .. controls (200,381.19) and (211.19,370) .. (225,370) .. controls (238.81,370) and (250,381.19) .. (250,395) .. controls (250,408.81) and (238.81,420) .. (225,420) .. controls (211.19,420) and (200,408.81) .. (200,395) -- cycle ;
\draw   (360,395) .. controls (360,381.19) and (371.19,370) .. (385,370) .. controls (398.81,370) and (410,381.19) .. (410,395) .. controls (410,408.81) and (398.81,420) .. (385,420) .. controls (371.19,420) and (360,408.81) .. (360,395) -- cycle ;
\draw   (440,370) -- (490,370) -- (490,420) -- (440,420) -- cycle ;
\draw    (170,395) -- (200,395) ;

\draw    (250,395) -- (280,395) ;

\draw    (330,395) -- (360,395) ;

\draw    (410,395) -- (440,395) ;

\draw  [dash pattern={on 4.5pt off 4.5pt}]  (280,395) -- (330,395) ;

\draw (454.5,147) node [scale=0.7]  {$2N+1\ D3$};
\draw (383.5,148) node [scale=0.7]  {$r_{1} \ D3$};
\draw (207.5,148) node [scale=0.7]  {$r_{s-2} \ D3$};
\draw (136,148) node [scale=0.7]  {$r_{s-1} \ D3$};
\draw (303.5,24) node   {$B_{N} \ \mathrm{type} :[ h_{1} ,h_{2} ,\cdots ,h_{s}] ,\ \mathrm{with} \ r_{i} =2N+1-\sum ^{i}_{j=1} h_{j}$};
\draw (340.5,190) node   {$S-dual$};
\draw (108.71,59) node [scale=0.8]  {$D5$};
\draw (175.86,58.5) node [scale=0.8]  {$D5$};
\draw (243.78,59.5) node [scale=0.8]  {$D5$};
\draw (350.3,59.5) node [scale=0.8]  {$D5$};
\draw (422.85,59.5) node [scale=0.8]  {$D5$};
\draw (143,308) node [scale=0.5]  {$ \begin{array}{l}
r_{s-1} -1\ D3:\ s\ \mathrm{odd}\\
r_{s-1} \ D3:\ s\ \mathrm{even}
\end{array}$};
\draw (226,308) node [scale=0.5]  {$ \begin{array}{l}
r_{s-2} \ D3:\ s\ \mathrm{odd}\\
r_{s-2} -1\ D3:\ s\ \mathrm{even}
\end{array}$};
\draw (384,302) node [scale=0.7]  {$r_{1} \ D3$};
\draw (460.5,301) node [scale=0.7]  {$2N\ D3$};
\draw (100.5,221.5) node [scale=0.8]  {$NS5$};
\draw (184.5,221.5) node [scale=0.8]  {$NS5$};
\draw (262.5,221.5) node [scale=0.8]  {$NS5$};
\draw (344.5,221.5) node [scale=0.8]  {$NS5$};
\draw (424.5,221.5) node [scale=0.8]  {$NS5$};
\draw (339.5,340) node   {$quiver$};
\draw (465,395) node [scale=0.5]  {$Sp( 2N)$};
\draw (147.5,431) node [scale=0.5]  {$ \begin{array}{l}
Sp( r_{s-1} -1) :\ s\ \mathrm{odd}\\
SO( r_{s-1}) :\ s\ \mathrm{even}
\end{array}$};
\draw (385,395) node [scale=0.5]  {$SO( r_{1})$};
\draw (452.5,67) node [color={rgb, 255:red, 255; green, 0; blue, 0 }  ,opacity=1 ]  {$\textcolor[rgb]{0,0,1}{B}$};
\draw (388.5,68) node [color={rgb, 255:red, 0; green, 0; blue, 255 }  ,opacity=1 ]  {$\textcolor[rgb]{1,0,0}{D}$};
\draw (461.5,240) node   {$\textcolor[rgb]{0,1,0}{C}$};
\draw (382,240) node   {$\textcolor[rgb]{1,0,0}{D}$};
\draw (139,72) node [scale=0.5]  {$ \begin{array}{l}
\textcolor[rgb]{0,0,1}{B:\ s\ \mathrm{odd}}\\
\textcolor[rgb]{1,0,0}{D:\ s\ \mathrm{even}}
\end{array}$};
\draw (209,71) node [scale=0.5]  {$ \begin{array}{l}
\textcolor[rgb]{1,0,0}{D:\ s\ \mathrm{odd}}\\
\textcolor[rgb]{0,0,1}{B:\ s\ \mathrm{even}}
\end{array}$};
\draw (139,238) node [scale=0.5]  {$ \begin{array}{l}
\textcolor[rgb]{0,1,0}{C:\ s\ \mathrm{odd}}\\
\textcolor[rgb]{1,0,0}{D:\ s\ \mathrm{even}}
\end{array}$};
\draw (221,239) node [scale=0.5]  {$ \begin{array}{l}
\textcolor[rgb]{1,0,0}{D:\ s\ \mathrm{odd}}\\
\textcolor[rgb]{0,1,0}{C:\ s\ \mathrm{even}}
\end{array}$};
\draw (232.5,431) node [scale=0.5]  {$ \begin{array}{l}
SO( r_{s-2}) :\ s\ \mathrm{odd}\\
Sp( r_{s-2} -1) :\ s\ \mathrm{even}
\end{array}$};

\end{tikzpicture}
\end{tabular}
}
\caption{Upper row: We construct a $D5-D3$ brane system for a $D$ type partition and $B$ type partition respectively. There  should be even number of D3 branes for a D type O3 plane, and odd number of D3 branes for a B type O3 plane.  For a general D type or B type partition, the brane configuration would violate the rule for D type or B type O3 planes, and 
we add one more D3 brane so that the number is consistent with the corresponding O3 plane. 
Middle: We perform S duality on brane system in upper row, which turn a D5 brane to a NS5 brane, and D3 brane to a D3 brane. Bottom row: We read a quiver gauge theory from brane configuration of middle row.}
\label{Dtype}
\end{center}
\end{figure}

Now we construct a quiver from a $D$ type or $B$ type partition, where the even parts appear even times. We first start with a $D3-O3-D5$ brane system, and there are two new facts:
\begin{itemize}
\item The type of O3 planes are alternating between $D3$ branes. Here we have D type and B type alternating. 
\item There are only an odd number of D3 branes ending on a half D5 brane. 
\end{itemize}
For the $B$ type or $D$ type partition where parts are all odd, we can construct a $D3-O3-D5$ brane system using the same rule as $A$ type partition. The quiver gauge theory can be found from S dual brane configuration, see figure \ref{Dtype}.  To deal with a partition with even parts, we need to do following modification:

$\bullet$  We add one more D3 brane to O3 brane for the segment which violates the corresponding rule for O3 plane. For example: if there are $2r$ D3 brane on a B type O3 plane, we add one more D3 brane, or if there are $2r+1$ D3 brane on D type O3 plane, we add one more D3 brane. 

After this modification, it is straightforward to find the quiver gauge theory by doing S duality on $D3-O3-D5$ brane system. Notice that the corresponding quiver gauge theory involves only $D$ type and $C$ type gauge groups (the flavor group is $D$ type or $C$ type).

\begin{figure}
\centering
\tikzset{every picture/.style={line width=0.75pt}} 

\begin{tikzpicture}[x=0.75pt,y=0.75pt,yscale=-1,xscale=1]

\draw    (104.08,79.44) -- (104.08,135.26) ;

\draw    (244.41,79.44) -- (244.41,135.26) ;

\draw    (174.25,79.44) -- (174.25,135.26) ;

\draw    (349.66,79.44) -- (349.66,135.26) ;

\draw  [dash pattern={on 4.5pt off 4.5pt}]  (255.64,107.35) -- (344.05,107.35) ;

\draw    (104.08,98.05) -- (174.25,98.05) ;

\draw    (104.08,116.65) -- (174.25,116.65) ;

\draw  [dash pattern={on 0.84pt off 2.51pt}]  (139.16,98.05) -- (139.16,116.65) ;

\draw    (174.25,88.74) -- (244.41,88.74) ;

\draw    (174.25,125.95) -- (244.41,125.95) ;

\draw  [dash pattern={on 0.84pt off 2.51pt}]  (209.33,88.74) -- (209.33,125.95) ;

\draw    (349.66,88.74) -- (419.83,88.74) ;

\draw    (349.66,125.95) -- (419.83,125.95) ;

\draw  [dash pattern={on 0.84pt off 2.51pt}]  (384.75,88.74) -- (384.75,125.95) ;

\draw    (419.83,84.74) -- (490,84.74) ;

\draw    (419.83,129.95) -- (490,129.95) ;

\draw  [dash pattern={on 0.84pt off 2.51pt}]  (454.92,88.74) -- (454.92,125.95) ;

\draw    (419.83,79.44) -- (419.83,135.26) ;

\draw    (290,160) -- (290,218) ;
\draw [shift={(290,220)}, rotate = 270] [color={rgb, 255:red, 0; green, 0; blue, 0 }  ][line width=0.75]    (10.93,-3.29) .. controls (6.95,-1.4) and (3.31,-0.3) .. (0,0) .. controls (3.31,0.3) and (6.95,1.4) .. (10.93,3.29)   ;

\draw   (330.51,254.21) .. controls (339.07,245.48) and (353.05,245.32) .. (361.72,253.83) .. controls (370.39,262.35) and (370.48,276.32) .. (361.91,285.05) .. controls (353.35,293.77) and (339.37,293.93) .. (330.7,285.42) .. controls (322.03,276.9) and (321.94,262.93) .. (330.51,254.21) -- cycle ; \draw   (330.51,254.21) -- (361.91,285.05) ; \draw   (361.72,253.83) -- (330.7,285.42) ;
\draw   (86.09,254.95) .. controls (94.65,246.23) and (108.59,246.02) .. (117.21,254.49) .. controls (125.83,262.96) and (125.87,276.89) .. (117.31,285.61) .. controls (108.74,294.34) and (94.81,294.54) .. (86.19,286.08) .. controls (77.57,277.61) and (77.52,263.68) .. (86.09,254.95) -- cycle ; \draw   (86.09,254.95) -- (117.31,285.61) ; \draw   (117.21,254.49) -- (86.19,286.08) ;
\draw   (168.51,254.21) .. controls (177.07,245.48) and (191.05,245.32) .. (199.72,253.83) .. controls (208.39,262.35) and (208.48,276.32) .. (199.91,285.05) .. controls (191.35,293.77) and (177.37,293.93) .. (168.7,285.42) .. controls (160.03,276.9) and (159.94,262.93) .. (168.51,254.21) -- cycle ; \draw   (168.51,254.21) -- (199.91,285.05) ; \draw   (199.72,253.83) -- (168.7,285.42) ;
\draw   (248.51,254.21) .. controls (257.07,245.48) and (271.05,245.32) .. (279.72,253.83) .. controls (288.39,262.35) and (288.48,276.32) .. (279.91,285.05) .. controls (271.35,293.77) and (257.37,293.93) .. (248.7,285.42) .. controls (240.03,276.9) and (239.94,262.93) .. (248.51,254.21) -- cycle ; \draw   (248.51,254.21) -- (279.91,285.05) ; \draw   (279.72,253.83) -- (248.7,285.42) ;
\draw   (407.11,253.37) .. controls (415.67,244.65) and (429.64,244.48) .. (438.32,252.99) .. controls (446.99,261.51) and (447.07,275.48) .. (438.51,284.21) .. controls (429.94,292.93) and (415.97,293.1) .. (407.3,284.58) .. controls (398.63,276.07) and (398.54,262.09) .. (407.11,253.37) -- cycle ; \draw   (407.11,253.37) -- (438.51,284.21) ; \draw   (438.32,252.99) -- (407.3,284.58) ;
\draw    (122.83,261.4) -- (163,261.4) ;

\draw    (122.83,280) -- (163,280) ;

\draw  [dash pattern={on 0.84pt off 2.51pt}]  (142.92,261.4) -- (142.92,280) ;

\draw    (203.83,255) -- (244,255) ;

\draw    (203.83,285) -- (244,285) ;

\draw  [dash pattern={on 0.84pt off 2.51pt}]  (223.92,255) -- (223.92,285) ;

\draw    (363.83,254) -- (404,254) ;

\draw    (363.83,284) -- (404,284) ;

\draw  [dash pattern={on 0.84pt off 2.51pt}]  (383.92,254) -- (383.92,284) ;

\draw    (439.83,254) -- (480,254) ;

\draw    (439.83,284) -- (480,284) ;

\draw  [dash pattern={on 0.84pt off 2.51pt}]  (459.92,254) -- (459.92,284) ;

\draw  [dash pattern={on 4.5pt off 4.5pt}]  (289,270) -- (323,270) ;

\draw    (290,310) -- (290,358) ;
\draw [shift={(290,360)}, rotate = 270] [color={rgb, 255:red, 0; green, 0; blue, 0 }  ][line width=0.75]    (10.93,-3.29) .. controls (6.95,-1.4) and (3.31,-0.3) .. (0,0) .. controls (3.31,0.3) and (6.95,1.4) .. (10.93,3.29)   ;

\draw   (120,395) .. controls (120,381.19) and (131.19,370) .. (145,370) .. controls (158.81,370) and (170,381.19) .. (170,395) .. controls (170,408.81) and (158.81,420) .. (145,420) .. controls (131.19,420) and (120,408.81) .. (120,395) -- cycle ;
\draw   (200,395) .. controls (200,381.19) and (211.19,370) .. (225,370) .. controls (238.81,370) and (250,381.19) .. (250,395) .. controls (250,408.81) and (238.81,420) .. (225,420) .. controls (211.19,420) and (200,408.81) .. (200,395) -- cycle ;
\draw   (360,395) .. controls (360,381.19) and (371.19,370) .. (385,370) .. controls (398.81,370) and (410,381.19) .. (410,395) .. controls (410,408.81) and (398.81,420) .. (385,420) .. controls (371.19,420) and (360,408.81) .. (360,395) -- cycle ;
\draw   (440,370) -- (490,370) -- (490,420) -- (440,420) -- cycle ;
\draw    (170,395) -- (200,395) ;

\draw    (250,395) -- (280,395) ;

\draw    (330,395) -- (360,395) ;

\draw    (410,395) -- (440,395) ;

\draw  [dash pattern={on 4.5pt off 4.5pt}]  (280,395) -- (330,395) ;

\draw (454.5,147) node [scale=0.7]  {$2N\ D3$};
\draw (383.5,148) node [scale=0.7]  {$r_{1} \ D3$};
\draw (207.5,148) node [scale=0.7]  {$r_{s-2} \ D3$};
\draw (136,148) node [scale=0.7]  {$r_{s-1} \ D3$};
\draw (303.5,24) node   {$C_{N} \ \mathrm{type} :[ h_{1} ,h_{2} ,\cdots ,h_{s}] ,\ \mathrm{with} \ r_{i} =2N-\sum ^{i}_{j=1} h_{j}$};
\draw (340.5,190) node   {$S-dual$};
\draw (108.71,59) node [scale=0.8]  {$D5$};
\draw (175.86,58.5) node [scale=0.8]  {$D5$};
\draw (243.78,59.5) node [scale=0.8]  {$D5$};
\draw (350.3,59.5) node [scale=0.8]  {$D5$};
\draw (422.85,59.5) node [scale=0.8]  {$D5$};
\draw (143,308) node [scale=0.5]  {$ \begin{array}{l}
r_{s-1} +1\ D3:\ s\ \mathrm{odd}\\
r_{s-1} \ D3:\ s\ \mathrm{even}
\end{array}$};
\draw (226,308) node [scale=0.5]  {$ \begin{array}{l}
r_{s-2} \ D3:\ s\ \mathrm{odd}\\
r_{s-2} +1\ D3:\ s\ \mathrm{even}
\end{array}$};
\draw (384,302) node [scale=0.7]  {$r_{1} \ D3$};
\draw (460.5,301) node [scale=0.7]  {$2N+1\ D3$};
\draw (100.5,221.5) node [scale=0.8]  {$NS5$};
\draw (184.5,221.5) node [scale=0.8]  {$NS5$};
\draw (262.5,221.5) node [scale=0.8]  {$NS5$};
\draw (344.5,221.5) node [scale=0.8]  {$NS5$};
\draw (424.5,221.5) node [scale=0.8]  {$NS5$};
\draw (339.5,340) node   {$quiver$};
\draw (465,395) node [scale=0.5]  {$SO( 2N+1)$};
\draw (147.5,431) node [scale=0.5]  {$ \begin{array}{l}
SO( r_{s-1} +1) :\ s\ \mathrm{odd}\\
Sp( r_{s-1}) :\ s\ \mathrm{even}
\end{array}$};
\draw (385,395) node [scale=0.5]  {$Sp( r_{1})$};
\draw (450.5,70) node [color={rgb, 255:red, 255; green, 0; blue, 0 }  ,opacity=1 ]  {$C$};
\draw (388.5,69) node [color={rgb, 255:red, 0; green, 0; blue, 255 }  ,opacity=1 ]  {$C'$};
\draw (461.5,240) node   {$\textcolor[rgb]{0,1,0}{B}$};
\draw (382,240) node   {$\textcolor[rgb]{0,0,1}{C'}$};
\draw (139,72) node [scale=0.5]  {$ \begin{array}{l}
\textcolor[rgb]{1,0,0}{C:\ s\ \mathrm{odd}}\\
\textcolor[rgb]{0,0,1}{C':\ s\ \mathrm{even}}
\end{array}$};
\draw (209,72) node [scale=0.5]  {$ \begin{array}{l}
\textcolor[rgb]{0,0,1}{C':\ s\ \mathrm{odd}}\\
\textcolor[rgb]{1,0,0}{C:\ s\ \mathrm{even}}
\end{array}$};
\draw (139,239) node [scale=0.5]  {$ \begin{array}{l}
\textcolor[rgb]{0,1,0}{B:\ s\ \mathrm{odd}}\\
\textcolor[rgb]{0,0,1}{C':\ s\ \mathrm{even}}
\end{array}$};
\draw (221,239) node [scale=0.5]  {$ \begin{array}{l}
\textcolor[rgb]{0,0,1}{C'}\textcolor[rgb]{0,0,1}{:\ s\ \mathrm{odd}}\\
\textcolor[rgb]{0,1,0}{B:\ s\ \mathrm{even}}
\end{array}$};
\draw (232.5,431) node [scale=0.5]  {$ \begin{array}{l}
Sp( r_{s-2}) :\ s\ \mathrm{odd}\\
SO( r_{s-2} +1) :\ s\ \mathrm{even}
\end{array}$};

\end{tikzpicture}
\caption{Upper row: We construct a $D5-D3$ brane system for a $C$ type partition. There  should be even number of D3 branes for  $C$ type or $C^{'}$ type O3 planes.  For a general C type partition, the brane configuration would violate the rule for O3 planes, and 
we add one more D3 brane so that the number is consistent with the O plane type. 
Middle: We perform S duality on brane system in upper row, which turn a D5 brane to a NS5 brane, and D3 brane to a D3 brane. Bottom row: We read a quiver gauge theory from brane configuration of middle row.}
\label{Ctype}
\end{figure}
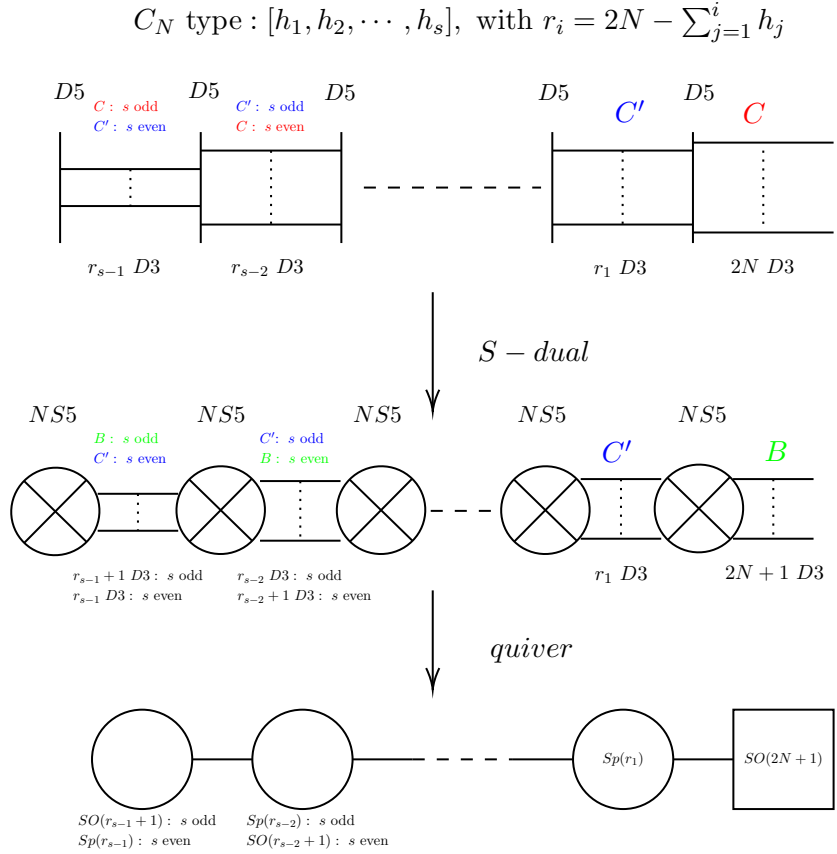

Now we consider how to construct a quiver from a $C$ type partition, where the even parts appear even times. We first start with a $D3-O3-D5$ brane system, and there are two new facts
\begin{itemize}
\item The type of O3 planes are alternating between $D3$ branes. Here we have $C$ type and $C^{'}$ type alternating. 
\item There are only an \textbf{even} number of D3 branes ending on a half D5 brane. 
\end{itemize}
For the $C$ type partition where the parts are all even, we can construct a $D3-O3-D5$ brane system using the same rule as $A$ type partition. The quiver gauge theory can be found from S dual brane configuration, see figure \ref{Ctype}.  To deal with a partition with even parts, we need to do following modification:

$\bullet$  We add one more D3 brane to O3 brane for the segment which violates the corresponding rule for O3 plane. For example: if there are $2r+1$ D3 brane on a C type or $C^{'}$ O3 plane, we add one more D3 brane. 

After this modification, it is straightforward to find the quiver gauge theory by doing S duality on $D3-O3-D5$ brane system. Notice that the corresponding quiver gauge theory involves only $B$ type and $C$ type gauge groups (flavor group is of $C$ type).  Notice that if we fixe the flavor group to be $C$ type, we do not get a quiver with alternating $D$ type and $C$ type gauge groups by using this method.

Given a nilpotent orbit $e$ in classical algebra $g$, and one can find its dual in $g^{\vee}$ using the Spaltenstein map. The rule is 
\begin{itemize}
\item For $A$ type: the partition of $e^d$ is given by the transpose of partition of $e$. 
\item For $D$ type: the partition of $e^d$ is given as follows: first find the transpose of partition $e$, and then do D collapse to get a D partition.
\item For $B$ type: the partition of $e^d$ is given as follows: first subtract one box from last row of $e$, and then do the transpose, and then do C collapse to get a C partition.
\item For $C$ type: the partition of $e^d$ is given as follows: first add a box to first row of $e$ and  find its transpose, and then do $B$ collapse to get a $B$ partition.
\end{itemize}

\newpage
\bibliographystyle{utphys}

\bibliography{ADhigher}

\providecommand{\href}[2]{#2}\begingroup\raggedright\begin{thebibliography}{100}

\bibitem{Beem:2013sza}
C.~Beem, M.~Lemos, P.~Liendo, W.~Peelaers, L.~Rastelli, and B.~C. van Rees,
  ``{Infinite Chiral Symmetry in Four Dimensions},''
  \href{http://dx.doi.org/10.1007/s00220-014-2272-x}{{\em Commun. Math. Phys.}
  {\bfseries 336} no.~3, (2015) 1359--1433},
\href{http://arxiv.org/abs/1312.5344}{{\ttfamily arXiv:1312.5344 [hep-th]}}.

\bibitem{Xie:2016evu}
D.~Xie, W.~Yan, and S.-T. Yau, ``{Chiral algebra of Argyres-Douglas theory from
  M5 brane},''
\href{http://arxiv.org/abs/1604.02155}{{\ttfamily arXiv:1604.02155 [hep-th]}}.

\bibitem{Song:2017oew}
J.~Song, D.~Xie, and W.~Yan, ``{Vertex operator algebras of Argyres-Douglas
  theories from M5-branes},''
  \href{http://dx.doi.org/10.1007/JHEP12(2017)123}{{\em JHEP} {\bfseries 12}
  (2017) 123},
\href{http://arxiv.org/abs/1706.01607}{{\ttfamily arXiv:1706.01607 [hep-th]}}.

\bibitem{Xie:2019yds}
D.~Xie and W.~Yan, ``{$W$ Algebra, Cosets and Voas for 4D ${\mathcal{N}}\!=2$
  SCFT from M5 Branes},''
\href{http://arxiv.org/abs/1902.02838}{{\ttfamily arXiv:1902.02838 [hep-th]}}.

\bibitem{Xie:2019zlb}
D.~Xie and W.~Yan, ``{Schur Sector of Argyres-Douglas Theory and
  $W$-algebra},''
\href{http://arxiv.org/abs/1904.09094}{{\ttfamily arXiv:1904.09094 [hep-th]}}.

\bibitem{moore1989classical}
G.~Moore and N.~Seiberg, ``Classical and quantum conformal field theory,'' {\em
  Communications in Mathematical Physics} {\bfseries 123} no.~2, (1989)
  177--254.

\bibitem{kac2008rationality}
V.~G. Kac and M.~Wakimoto, ``On rationality of w-algebras,'' {\em
  Transformation Groups} {\bfseries 13} no.~3-4, (2008) 671--713.

\bibitem{witten1989quantum}
E.~Witten, ``Quantum field theory and the jones polynomial,'' {\em
  Communications in Mathematical Physics} {\bfseries 121} no.~3, (1989)
  351--399.

\bibitem{zhu1996modular}
Y.~Zhu, ``Modular invariance of characters of vertex operator algebras,'' {\em
  Journal of the American Mathematical Society} {\bfseries 9} no.~1, (1996)
  237--302.

\bibitem{arakawa2012remark}
T.~Arakawa, ``A remark on the c 2-cofiniteness condition on vertex algebras,''
  {\em Mathematische Zeitschrift} {\bfseries 270} no.~1-2, (2012) 559--575.

\bibitem{Beem:2017ooy}
C.~Beem and L.~Rastelli, ``{Vertex operator algebras, Higgs branches, and
  modular differential equations},''
  \href{http://dx.doi.org/10.1007/JHEP08(2018)114}{{\em JHEP} {\bfseries 08}
  (2018) 114},
\href{http://arxiv.org/abs/1707.07679}{{\ttfamily arXiv:1707.07679 [hep-th]}}.

\bibitem{Arakawa:2017fdq}
T.~Arakawa, ``{Representation theory of W-algebras and Higgs branch
  conjecture},'' in {\em {International Congress of Mathematicians (ICM 2018)
  Rio de Janeiro, Brazil, August 1-9, 2018}}.
\newblock 2017.
\newblock
\href{http://arxiv.org/abs/1712.07331}{{\ttfamily arXiv:1712.07331 [math.RT]}}.
\newblock

\bibitem{Gaiotto:2009we}
D.~Gaiotto, ``{N=2 dualities},''
  \href{http://dx.doi.org/10.1007/JHEP08(2012)034}{{\em JHEP} {\bfseries 1208}
  (2012) 034},
\href{http://arxiv.org/abs/0904.2715}{{\ttfamily arXiv:0904.2715 [hep-th]}}.

\bibitem{Xie:2012hs}
D.~Xie, ``{General Argyres-Douglas Theory},''
  \href{http://dx.doi.org/10.1007/JHEP01(2013)100}{{\em JHEP} {\bfseries 1301}
  (2013) 100},
\href{http://arxiv.org/abs/1204.2270}{{\ttfamily arXiv:1204.2270 [hep-th]}}.

\bibitem{Xie:2015rpa}
D.~Xie and S.-T. Yau, ``{4d N=2 SCFT and singularity theory Part I:
  Classification},''
\href{http://arxiv.org/abs/1510.01324}{{\ttfamily arXiv:1510.01324 [hep-th]}}.

\bibitem{Wang:2015mra}
Y.~Wang and D.~Xie, ``{Classification of Argyres-Douglas theories from M5
  branes},''
\href{http://arxiv.org/abs/1509.00847}{{\ttfamily arXiv:1509.00847 [hep-th]}}.

\bibitem{Wang:2018gvb}
Y.~Wang and D.~Xie, ``{Codimension-two defects and Argyres-Douglas theories
  from outer-automorphism twist in 6d $(2,0)$ theories},''
\href{http://arxiv.org/abs/1805.08839}{{\ttfamily arXiv:1805.08839 [hep-th]}}.

\bibitem{Chen:2017wkw}
B.~Chen, D.~Xie, S.~S.~T. Yau, S.-T. Yau, and H.~Zuo, ``{4d $\mathcal{N}=2$
  SCFT and singularity theory Part III: Rigid singularity},''
  \href{http://dx.doi.org/10.4310/ATMP.2018.v22.n8.a2}{{\em Adv. Theor. Math.
  Phys.} {\bfseries 22} (2018) 1885--1905},
\href{http://arxiv.org/abs/1712.00464}{{\ttfamily arXiv:1712.00464 [hep-th]}}.

\bibitem{MR3456698}
T.~Arakawa, ``Associated varieties of modules over {K}ac-{M}oody algebras and
  {$C_2$}-cofiniteness of {$W$}-algebras,'' {\em Int. Math. Res. Not. IMRN}
  no.~22, (2015) 11605--11666.

\bibitem{kac1998vertex}
V.~G. Kac, {\em Vertex algebras for beginners}.
\newblock No.~10. American Mathematical Soc., 1998.

\bibitem{Li2005}
H.~Li, ``Abelianizing vertex algebras,''
  \href{http://dx.doi.org/10.1007/s00220-005-1348-z}{{\em Communications in
  Mathematical Physics} {\bfseries 259} no.~2, (2005) 391--411}.
  \url{http://dx.doi.org/10.1007/s00220-005-1348-z}.

\bibitem{Arakawa:2016hkg}
T.~Arakawa and K.~Kawasetsu, ``{Quasi-lisse vertex algebras and modular linear
  differential equations},''
\href{http://arxiv.org/abs/1610.05865}{{\ttfamily arXiv:1610.05865 [math.QA]}}.

\bibitem{kac2017remark}
V.~G. Kac and M.~Wakimoto, ``A remark on boundary level admissible
  representations,'' {\em Comptes Rendus Mathematique} {\bfseries 355} no.~2,
  (2017) 128--132.

\bibitem{2019arXiv190511473A}
T.~{Arakawa} and J.~{van Ekeren}, ``{Rationality and Fusion Rules of
  Exceptional W-Algebras},'' {\em arXiv e-prints} (May, 2019) arXiv:1905.11473,
  \href{http://arxiv.org/abs/1905.11473}{{\ttfamily arXiv:1905.11473
  [math.RT]}}.

\bibitem{Beem:2014rza}
C.~Beem, W.~Peelaers, L.~Rastelli, and B.~C. van Rees, ``{Chiral algebras of
  class S},'' \href{http://dx.doi.org/10.1007/JHEP05(2015)020}{{\em JHEP}
  {\bfseries 1505} (2015) 020},
\href{http://arxiv.org/abs/1408.6522}{{\ttfamily arXiv:1408.6522 [hep-th]}}.

\bibitem{Lemos:2014lua}
M.~Lemos and W.~Peelaers, ``{Chiral Algebras for Trinion Theories},''
  \href{http://dx.doi.org/10.1007/JHEP02(2015)113}{{\em JHEP} {\bfseries 02}
  (2015) 113},
\href{http://arxiv.org/abs/1411.3252}{{\ttfamily arXiv:1411.3252 [hep-th]}}.

\bibitem{Buican:2015hsa}
M.~Buican and T.~Nishinaka, ``{Argyres--Douglas theories, S$^1$ reductions, and
  topological symmetries},''
  \href{http://dx.doi.org/10.1088/1751-8113/49/4/045401}{{\em J. Phys.}
  {\bfseries A49} no.~4, (2016) 045401},
\href{http://arxiv.org/abs/1505.06205}{{\ttfamily arXiv:1505.06205 [hep-th]}}.

\bibitem{Cordova:2015nma}
C.~Cordova and S.-H. Shao, ``{Schur Indices, BPS Particles, and Argyres-Douglas
  Theories},'' \href{http://dx.doi.org/10.1007/JHEP01(2016)040}{{\em JHEP}
  {\bfseries 01} (2016) 040},
\href{http://arxiv.org/abs/1506.00265}{{\ttfamily arXiv:1506.00265 [hep-th]}}.

\bibitem{Buican:2015tda}
M.~Buican and T.~Nishinaka, ``{Argyres-Douglas Theories, the Macdonald Index,
  and an RG Inequality},''
  \href{http://dx.doi.org/10.1007/JHEP02(2016)159}{{\em JHEP} {\bfseries 02}
  (2016) 159},
\href{http://arxiv.org/abs/1509.05402}{{\ttfamily arXiv:1509.05402 [hep-th]}}.

\bibitem{Song:2015wta}
J.~Song, ``{Superconformal indices of generalized Argyres-Douglas theories from
  2d TQFT},'' \href{http://dx.doi.org/10.1007/JHEP02(2016)045}{{\em JHEP}
  {\bfseries 02} (2016) 045},
\href{http://arxiv.org/abs/1509.06730}{{\ttfamily arXiv:1509.06730 [hep-th]}}.

\bibitem{Lemos:2015awa}
M.~Lemos and P.~Liendo, ``{Bootstrapping ${\mathcal N}=2$ chiral
  correlators},''
\href{http://arxiv.org/abs/1510.03866}{{\ttfamily arXiv:1510.03866 [hep-th]}}.

\bibitem{Cecotti:2015lab}
S.~Cecotti, J.~Song, C.~Vafa, and W.~Yan, ``{Superconformal Index, BPS
  Monodromy and Chiral Algebras},''
  \href{http://dx.doi.org/10.1007/JHEP11(2017)013}{{\em JHEP} {\bfseries 11}
  (2017) 013},
\href{http://arxiv.org/abs/1511.01516}{{\ttfamily arXiv:1511.01516 [hep-th]}}.

\bibitem{Lemos:2015orc}
M.~Lemos and P.~Liendo, ``{$\mathcal{N}=2$ central charge bounds from $2d$
  chiral algebras},'' \href{http://dx.doi.org/10.1007/JHEP04(2016)004}{{\em
  JHEP} {\bfseries 04} (2016) 004},
\href{http://arxiv.org/abs/1511.07449}{{\ttfamily arXiv:1511.07449 [hep-th]}}.

\bibitem{Nishinaka:2016hbw}
T.~Nishinaka and Y.~Tachikawa, ``{On 4d rank-one $ \mathcal{N}=3 $
  superconformal field theories},''
  \href{http://dx.doi.org/10.1007/JHEP09(2016)116}{{\em JHEP} {\bfseries 09}
  (2016) 116},
\href{http://arxiv.org/abs/1602.01503}{{\ttfamily arXiv:1602.01503 [hep-th]}}.

\bibitem{Buican:2016arp}
M.~Buican and T.~Nishinaka, ``{Conformal Manifolds in Four Dimensions and
  Chiral Algebras},''
  \href{http://dx.doi.org/10.1088/1751-8113/49/46/465401}{{\em J. Phys.}
  {\bfseries A49} no.~46, (2016) 465401},
\href{http://arxiv.org/abs/1603.00887}{{\ttfamily arXiv:1603.00887 [hep-th]}}.

\bibitem{Cordova:2016uwk}
C.~Cordova, D.~Gaiotto, and S.-H. Shao, ``{Infrared Computations of Defect
  Schur Indices},'' \href{http://dx.doi.org/10.1007/JHEP11(2016)106}{{\em JHEP}
  {\bfseries 11} (2016) 106},
\href{http://arxiv.org/abs/1606.08429}{{\ttfamily arXiv:1606.08429 [hep-th]}}.

\bibitem{Lemos:2016xke}
M.~Lemos, P.~Liendo, C.~Meneghelli, and V.~Mitev, ``{Bootstrapping
  $\mathcal{N}=3$ superconformal theories},''
  \href{http://dx.doi.org/10.1007/JHEP04(2017)032}{{\em JHEP} {\bfseries 04}
  (2017) 032},
\href{http://arxiv.org/abs/1612.01536}{{\ttfamily arXiv:1612.01536 [hep-th]}}.

\bibitem{Bonetti:2016nma}
F.~Bonetti and L.~Rastelli, ``{Supersymmetric localization in AdS$_{5}$ and the
  protected chiral algebra},''
  \href{http://dx.doi.org/10.1007/JHEP08(2018)098}{{\em JHEP} {\bfseries 08}
  (2018) 098},
\href{http://arxiv.org/abs/1612.06514}{{\ttfamily arXiv:1612.06514 [hep-th]}}.

\bibitem{Song:2016yfd}
J.~Song, ``{Macdonald Index and Chiral Algebra},''
  \href{http://dx.doi.org/10.1007/JHEP08(2017)044}{{\em JHEP} {\bfseries 08}
  (2017) 044},
\href{http://arxiv.org/abs/1612.08956}{{\ttfamily arXiv:1612.08956 [hep-th]}}.

\bibitem{Creutzig:2017qyf}
T.~Creutzig, ``{W-algebras for Argyres-Douglas theories},''
\href{http://arxiv.org/abs/1701.05926}{{\ttfamily arXiv:1701.05926 [hep-th]}}.

\bibitem{Fredrickson:2017yka}
L.~Fredrickson, D.~Pei, W.~Yan, and K.~Ye, ``{Argyres-Douglas Theories, Chiral
  Algebras and Wild Hitchin Characters},''
\href{http://arxiv.org/abs/1701.08782}{{\ttfamily arXiv:1701.08782 [hep-th]}}.

\bibitem{Cordova:2017ohl}
C.~Cordova, D.~Gaiotto, and S.-H. Shao, ``{Surface Defect Indices and 2d-4d BPS
  States},''
\href{http://arxiv.org/abs/1703.02525}{{\ttfamily arXiv:1703.02525 [hep-th]}}.

\bibitem{Cordova:2017mhb}
C.~Cordova, D.~Gaiotto, and S.-H. Shao, ``{Surface Defects and Chiral
  Algebras},'' \href{http://dx.doi.org/10.1007/JHEP05(2017)140}{{\em JHEP}
  {\bfseries 05} (2017) 140},
\href{http://arxiv.org/abs/1704.01955}{{\ttfamily arXiv:1704.01955 [hep-th]}}.

\bibitem{Dedushenko:2017tdw}
M.~Dedushenko, S.~Gukov, and P.~Putrov, ``{Vertex algebras and 4-manifold
  invariants},''
\href{http://arxiv.org/abs/1705.01645}{{\ttfamily arXiv:1705.01645 [hep-th]}}.

\bibitem{Buican:2017uka}
M.~Buican and T.~Nishinaka, ``{On Irregular Singularity Wave Functions and
  Superconformal Indices},''
  \href{http://dx.doi.org/10.1007/JHEP09(2017)066}{{\em JHEP} {\bfseries 09}
  (2017) 066},
\href{http://arxiv.org/abs/1705.07173}{{\ttfamily arXiv:1705.07173 [hep-th]}}.

\bibitem{Buican:2017fiq}
M.~Buican, Z.~Laczko, and T.~Nishinaka, ``{$ \mathcal{N} $ = 2 S-duality
  revisited},'' \href{http://dx.doi.org/10.1007/JHEP09(2017)087}{{\em JHEP}
  {\bfseries 09} (2017) 087},
\href{http://arxiv.org/abs/1706.03797}{{\ttfamily arXiv:1706.03797 [hep-th]}}.

\bibitem{Pan:2017zie}
Y.~Pan and W.~Peelaers, ``{Chiral Algebras, Localization and Surface
  Defects},'' \href{http://dx.doi.org/10.1007/JHEP02(2018)138}{{\em JHEP}
  {\bfseries 02} (2018) 138},
\href{http://arxiv.org/abs/1710.04306}{{\ttfamily arXiv:1710.04306 [hep-th]}}.

\bibitem{Fluder:2017oxm}
M.~Fluder and J.~Song, ``{Four-dimensional Lens Space Index from
  Two-dimensional Chiral Algebra},''
  \href{http://dx.doi.org/10.1007/JHEP07(2018)073}{{\em JHEP} {\bfseries 07}
  (2018) 073},
\href{http://arxiv.org/abs/1710.06029}{{\ttfamily arXiv:1710.06029 [hep-th]}}.

\bibitem{Buican:2017rya}
M.~Buican and Z.~Laczko, ``{Nonunitary Lagrangians and unitary non-Lagrangian
  conformal field theories},''
  \href{http://dx.doi.org/10.1103/PhysRevLett.120.081601}{{\em Phys. Rev.
  Lett.} {\bfseries 120} no.~8, (2018) 081601},
\href{http://arxiv.org/abs/1711.09949}{{\ttfamily arXiv:1711.09949 [hep-th]}}.

\bibitem{Choi:2017nur}
J.~Choi and T.~Nishinaka, ``{On the chiral algebra of Argyres-Douglas theories
  and S-duality},'' \href{http://dx.doi.org/10.1007/JHEP04(2018)004}{{\em JHEP}
  {\bfseries 04} (2018) 004},
\href{http://arxiv.org/abs/1711.07941}{{\ttfamily arXiv:1711.07941 [hep-th]}}.

\bibitem{Kozcaz:2018usv}
C.~Koz{\c{c}}az, S.~Shakirov, and W.~Yan, ``{Argyres-Douglas Theories,
  Modularity of Minimal Models and Refined Chern-Simons},''
\href{http://arxiv.org/abs/1801.08316}{{\ttfamily arXiv:1801.08316 [hep-th]}}.

\bibitem{Costello:2018fnz}
K.~Costello and D.~Gaiotto, ``{Vertex Operator Algebras and 3d $\mathcal N=4$
  gauge theories},''
\href{http://arxiv.org/abs/1804.06460}{{\ttfamily arXiv:1804.06460 [hep-th]}}.

\bibitem{Feigin:2018bkf}
B.~Feigin and S.~Gukov, ``{VOA[$M_4$]},''
\href{http://arxiv.org/abs/1806.02470}{{\ttfamily arXiv:1806.02470 [hep-th]}}.

\bibitem{Niarchos:2018mvl}
V.~Niarchos, ``{Geometry of Higgs-branch superconformal primary bundles},''
  \href{http://dx.doi.org/10.1103/PhysRevD.98.065012}{{\em Phys. Rev.}
  {\bfseries D98} no.~6, (2018) 065012},
\href{http://arxiv.org/abs/1807.04296}{{\ttfamily arXiv:1807.04296 [hep-th]}}.

\bibitem{Creutzig:2018lbc}
T.~Creutzig, ``{Logarithmic W-algebras and Argyres-Douglas theories at higher
  rank},'' \href{http://dx.doi.org/10.1007/JHEP11(2018)188}{{\em JHEP}
  {\bfseries 11} (2018) 188},
\href{http://arxiv.org/abs/1809.01725}{{\ttfamily arXiv:1809.01725 [hep-th]}}.

\bibitem{Dedushenko:2018bpp}
M.~Dedushenko, S.~Gukov, H.~Nakajima, D.~Pei, and K.~Ye, ``{3D Tqfts from
  Argyres-Douglas Theories},''
\href{http://arxiv.org/abs/1809.04638}{{\ttfamily arXiv:1809.04638 [hep-th]}}.

\bibitem{Bonetti:2018fqz}
F.~Bonetti, C.~Meneghelli, and L.~Rastelli, ``{VOAs labelled by complex
  reflection groups and $4d$ SCFTs},''
\href{http://arxiv.org/abs/1810.03612}{{\ttfamily arXiv:1810.03612 [hep-th]}}.

\bibitem{Arakawa:2018egx}
T.~Arakawa, ``{Chiral Algebras of Class $\mathcal{S}$ and Moore-Tachikawa
  Symplectic Varieties},''
\href{http://arxiv.org/abs/1811.01577}{{\ttfamily arXiv:1811.01577 [math.RT]}}.

\bibitem{Costello:2018swh}
K.~Costello, T.~Creutzig, and D.~Gaiotto, ``{Higgs and Coulomb branches from
  vertex operator algebras},''
\href{http://arxiv.org/abs/1811.03958}{{\ttfamily arXiv:1811.03958 [hep-th]}}.

\bibitem{Nishinaka:2018zwq}
T.~Nishinaka, S.~Sasa, and R.-D. Zhu, ``{On the Correspondence between Surface
  Operators in Argyres-Douglas Theories and Modules of Chiral Algebra},''
\href{http://arxiv.org/abs/1811.11772}{{\ttfamily arXiv:1811.11772 [hep-th]}}.

\bibitem{Agarwal:2018zqi}
P.~Agarwal, S.~Lee, and J.~Song, ``{Vanishing OPE Coefficients in 4d $N=2$
  SCFTs},''
\href{http://arxiv.org/abs/1812.04743}{{\ttfamily arXiv:1812.04743 [hep-th]}}.

\bibitem{Beem:2018duj}
C.~Beem, ``{Flavor symmetries and unitarity bounds in ${\mathcal N}=2$
  SCFTs},''
\href{http://arxiv.org/abs/1812.06099}{{\ttfamily arXiv:1812.06099 [hep-th]}}.

\bibitem{Kiyoshige:2018wol}
K.~Kiyoshige and T.~Nishinaka, ``{OPE Selection Rules for Schur Multiplets in
  4D $N=2$ Superconformal Field Theories},''
\href{http://arxiv.org/abs/1812.06394}{{\ttfamily arXiv:1812.06394 [hep-th]}}.

\bibitem{Buican:2019huq}
M.~Buican and Z.~Laczko, ``{Rationalizing CFTs and Anyonic Imprints on Higgs
  Branches},''
\href{http://arxiv.org/abs/1901.07591}{{\ttfamily arXiv:1901.07591 [hep-th]}}.

\bibitem{Pan:2019bor}
Y.~Pan and W.~Peelaers, ``{Schur Correlation Functions on $S^3\times S^1$},''
\href{http://arxiv.org/abs/1903.03623}{{\ttfamily arXiv:1903.03623 [hep-th]}}.

\bibitem{Beem:2019tfp}
C.~Beem, C.~Meneghelli, and L.~Rastelli, ``{Free Field Realizations from the
  Higgs Branch},''
\href{http://arxiv.org/abs/1903.07624}{{\ttfamily arXiv:1903.07624 [hep-th]}}.

\bibitem{Oh:2019bgz}
J.~Oh and J.~Yagi, ``{Chiral Algebras from $\Omega$-deformation},''
\href{http://arxiv.org/abs/1903.11123}{{\ttfamily arXiv:1903.11123 [hep-th]}}.

\bibitem{Dedushenko:2019yiw}
M.~Dedushenko and M.~Fluder, ``{Chiral Algebra, Localization, Modularity,
  Surface Defects, and All That},''
\href{http://arxiv.org/abs/1904.02704}{{\ttfamily arXiv:1904.02704 [hep-th]}}.

\bibitem{arakawa2018chiral}
T.~Arakawa, ``Chiral algebras of class s and moore-tachikawa symplectic
  varieties,'' {\em arXiv preprint arXiv:1811.01577} (2018) .

\bibitem{Dolan:2002zh}
F.~Dolan and H.~Osborn, ``{On short and semi-short representations for
  four-dimensional superconformal symmetry},''
  \href{http://dx.doi.org/10.1016/S0003-4916(03)00074-5}{{\em Annals Phys.}
  {\bfseries 307} (2003) 41--89},
\href{http://arxiv.org/abs/hep-th/0209056}{{\ttfamily arXiv:hep-th/0209056
  [hep-th]}}.

\bibitem{Seiberg:1994rs}
N.~Seiberg and E.~Witten, ``{Electric - magnetic duality, monopole
  condensation, and confinement in N=2 supersymmetric Yang-Mills theory},''
  \href{http://dx.doi.org/10.1016/0550-3213(94)90124-4,
  10.1016/0550-3213(94)00449-8}{{\em Nucl. Phys.} {\bfseries B426} (1994)
  19--52}, \href{http://arxiv.org/abs/hep-th/9407087}{{\ttfamily
  arXiv:hep-th/9407087 [hep-th]}}.
[Erratum: Nucl. Phys.B430,485(1994)].

\bibitem{Seiberg:1994aj}
N.~Seiberg and E.~Witten, ``{Monopoles, duality and chiral symmetry breaking in
  N=2 supersymmetric QCD},''
  \href{http://dx.doi.org/10.1016/0550-3213(94)90214-3}{{\em Nucl. Phys.}
  {\bfseries B431} (1994) 484--550},
\href{http://arxiv.org/abs/hep-th/9408099}{{\ttfamily arXiv:hep-th/9408099
  [hep-th]}}.

\bibitem{Argyres:2015ffa}
P.~Argyres, M.~Lotito, Y.~Lü, and M.~Martone, ``{Geometric constraints on the
  space of $ \mathcal{N} $ = 2 SCFTs. Part I: physical constraints on relevant
  deformations},'' \href{http://dx.doi.org/10.1007/JHEP02(2018)001}{{\em JHEP}
  {\bfseries 02} (2018) 001},
\href{http://arxiv.org/abs/1505.04814}{{\ttfamily arXiv:1505.04814 [hep-th]}}.

\bibitem{Gaiotto:2009hg}
D.~Gaiotto, G.~W. Moore, and A.~Neitzke, ``{Wall-crossing, Hitchin Systems, and
  the WKB Approximation},''
\href{http://arxiv.org/abs/0907.3987}{{\ttfamily arXiv:0907.3987 [hep-th]}}.

\bibitem{reeder2012gradings}
M.~Reeder, P.~Levy, J.-K. Yu, and B.~H. Gross, ``Gradings of positive rank on
  simple lie algebras,'' {\em Transformation Groups} {\bfseries 17} no.~4,
  (2012) 1123--1190.

\bibitem{Chacaltana:2012zy}
O.~Chacaltana, J.~Distler, and Y.~Tachikawa, ``{Nilpotent orbits and
  codimension-two defects of 6d N=(2,0) theories},''
  \href{http://dx.doi.org/10.1142/S0217751X1340006X}{{\em Int. J. Mod. Phys.}
  {\bfseries A28} (2013) 1340006},
\href{http://arxiv.org/abs/1203.2930}{{\ttfamily arXiv:1203.2930 [hep-th]}}.

\bibitem{hitchin1987stable}
N.~Hitchin {\em et~al.}, ``Stable bundles and integrable systems,'' {\em Duke
  mathematical journal} {\bfseries 54} no.~1, (1987) 91--114.

\bibitem{Chen:2016bzh}
B.~Chen, D.~Xie, S.-T. Yau, S.~S.~T. Yau, and H.~Zuo, ``{4D $\mathcal{N} = 2$
  SCFT and singularity theory. Part II: complete intersection},''
  \href{http://dx.doi.org/10.4310/ATMP.2017.v21.n1.a2}{{\em Adv. Theor. Math.
  Phys.} {\bfseries 21} (2017) 121--145},
\href{http://arxiv.org/abs/1604.07843}{{\ttfamily arXiv:1604.07843 [hep-th]}}.

\bibitem{Wang:2016yha}
Y.~Wang, D.~Xie, S.~S.~T. Yau, and S.-T. Yau, ``{$4d$ $\mathcal{N} = 2$ SCFT
  from complete intersection singularity},''
  \href{http://dx.doi.org/10.4310/ATMP.2017.v21.n3.a6}{{\em Adv. Theor. Math.
  Phys.} {\bfseries 21} (2017) 801--855},
\href{http://arxiv.org/abs/1606.06306}{{\ttfamily arXiv:1606.06306 [hep-th]}}.

\bibitem{Xie:2015xva}
D.~Xie and S.-T. Yau, ``{Semicontinuity of 4d N=2 spectrum under
  renormalization group flow},''
  \href{http://dx.doi.org/10.1007/JHEP03(2016)094}{{\em JHEP} {\bfseries 03}
  (2016) 094},
\href{http://arxiv.org/abs/1510.06036}{{\ttfamily arXiv:1510.06036 [hep-th]}}.

\bibitem{Li:2018rdd}
S.~Li, D.~Xie, and S.-T. Yau, ``{Seiberg–Witten Differential via Primitive
  Forms},'' \href{http://dx.doi.org/10.1007/s00220-019-03401-y}{{\em Commun.
  Math. Phys.} {\bfseries 367} no.~1, (2019) 193--214},
\href{http://arxiv.org/abs/1802.06751}{{\ttfamily arXiv:1802.06751 [hep-th]}}.

\bibitem{Xie:2014pua}
D.~Xie and K.~Yonekura, ``{The moduli space of vacua of $ \mathcal{N}=2 $ class
  $ \mathcal{S} $ theories},''
  \href{http://dx.doi.org/10.1007/JHEP10(2014)134}{{\em JHEP} {\bfseries 10}
  (2014) 134},
\href{http://arxiv.org/abs/1404.7521}{{\ttfamily arXiv:1404.7521 [hep-th]}}.

\bibitem{Xie:2013rsa}
D.~Xie and K.~Yonekura, ``{Generalized Hitchin system, Spectral curve and
  $\mathcal{N} = 1$ dynamics},''
  \href{http://dx.doi.org/10.1007/JHEP01(2014)001}{{\em JHEP} {\bfseries 01}
  (2014) 001},
\href{http://arxiv.org/abs/1310.0467}{{\ttfamily arXiv:1310.0467 [hep-th]}}.

\bibitem{kawamata1988crepant}
Y.~Kawamata, ``Crepant blowing-up of 3-dimensional canonical singularities and
  its application to degenerations of surfaces,'' {\em Annals of Mathematics}
  {\bfseries 127} no.~1, (1988) 93--163.

\bibitem{caibuar2003divisor}
M.~Caibar, ``On the divisor class group of 3-fold singularities,'' {\em
  International Journal of Mathematics} {\bfseries 14} no.~01, (2003) 105--117.

\bibitem{DelZotto:2015rca}
M.~Del~Zotto, C.~Vafa, and D.~Xie, ``{Geometric engineering, mirror symmetry
  and $ 6{\mathrm{d}}_{\left(1,0\right)}\to
  4{\mathrm{d}}_{\left(\mathcal{N}=2\right)} $},''
  \href{http://dx.doi.org/10.1007/JHEP11(2015)123}{{\em JHEP} {\bfseries 11}
  (2015) 123},
\href{http://arxiv.org/abs/1504.08348}{{\ttfamily arXiv:1504.08348 [hep-th]}}.

\bibitem{Chacaltana:2010ks}
O.~Chacaltana and J.~Distler, ``{Tinkertoys for Gaiotto Duality},''
  \href{http://dx.doi.org/10.1007/JHEP11(2010)099}{{\em JHEP} {\bfseries 11}
  (2010) 099},
\href{http://arxiv.org/abs/1008.5203}{{\ttfamily arXiv:1008.5203 [hep-th]}}.

\bibitem{Chacaltana:2011ze}
O.~Chacaltana and J.~Distler, ``{Tinkertoys for the $D_{N}$ Series},''
  \href{http://dx.doi.org/10.1007/JHEP02(2013)110}{{\em JHEP} {\bfseries 02}
  (2013) 110},
\href{http://arxiv.org/abs/1106.5410}{{\ttfamily arXiv:1106.5410 [hep-th]}}.

\bibitem{Chacaltana:2013oka}
O.~Chacaltana, J.~Distler, and A.~Trimm, ``{Tinkertoys for the Twisted
  D-Series},'' \href{http://dx.doi.org/10.1007/JHEP04(2015)173}{{\em JHEP}
  {\bfseries 04} (2015) 173},
\href{http://arxiv.org/abs/1309.2299}{{\ttfamily arXiv:1309.2299 [hep-th]}}.

\bibitem{Chacaltana:2014jba}
O.~Chacaltana, J.~Distler, and A.~Trimm, ``{Tinkertoys for the E$_{6}$
  Theory},'' \href{http://dx.doi.org/10.1007/JHEP09(2015)007}{{\em JHEP}
  {\bfseries 09} (2015) 007},
\href{http://arxiv.org/abs/1403.4604}{{\ttfamily arXiv:1403.4604 [hep-th]}}.

\bibitem{Chacaltana:2015bna}
O.~Chacaltana, J.~Distler, and A.~Trimm, ``{Tinkertoys for the Twisted $E_{6}$
  Theory},''
\href{http://arxiv.org/abs/1501.00357}{{\ttfamily arXiv:1501.00357 [hep-th]}}.

\bibitem{Chacaltana:2016shw}
O.~Chacaltana, J.~Distler, and A.~Trimm, ``{Tinkertoys for the Z3-Twisted D4
  Theory},''
\href{http://arxiv.org/abs/1601.02077}{{\ttfamily arXiv:1601.02077 [hep-th]}}.

\bibitem{Chacaltana:2017boe}
O.~Chacaltana, J.~Distler, A.~Trimm, and Y.~Zhu, ``{Tinkertoys for the E$_{7}$
  Theory},'' \href{http://dx.doi.org/10.1007/JHEP05(2018)031}{{\em JHEP}
  {\bfseries 05} (2018) 031},
\href{http://arxiv.org/abs/1704.07890}{{\ttfamily arXiv:1704.07890 [hep-th]}}.

\bibitem{Chcaltana:2018zag}
O.~Chacaltana, J.~Distler, A.~Trimm, and Y.~Zhu, ``{Tinkertoys for the $E_8$
  Theory},''
\href{http://arxiv.org/abs/1802.09626}{{\ttfamily arXiv:1802.09626 [hep-th]}}.

\bibitem{Collingwood:1993rr}
D.~Collingwood and W.~McGovern, ``{Nilpotent orbits in semisimple lie
  algebra},''. VanNostrand Reinhold Math.Series, New York, 1993.

\bibitem{Xie:2017aqx}
D.~Xie and K.~Ye, ``{Argyres-Douglas matter and S-duality: Part II},''
  \href{http://dx.doi.org/10.1007/JHEP03(2018)186}{{\em JHEP} {\bfseries 03}
  (2018) 186},
\href{http://arxiv.org/abs/1711.06684}{{\ttfamily arXiv:1711.06684 [hep-th]}}.

\bibitem{Benini:2010uu}
F.~Benini, Y.~Tachikawa, and D.~Xie, ``{Mirrors of 3d Sicilian theories},''
  \href{http://dx.doi.org/10.1007/JHEP09(2010)063}{{\em JHEP} {\bfseries 09}
  (2010) 063},
\href{http://arxiv.org/abs/1007.0992}{{\ttfamily arXiv:1007.0992 [hep-th]}}.

\bibitem{Arakawa:2016aa}
T.~Arakawa and A.~Moreau, ``On the irreducibility of associated varieties of
  w-algebras,'' \href{http://arxiv.org/abs/1608.03142}{{\ttfamily 1608.03142}}.
  \url{http://arxiv.org/abs/1608.03142}.

\bibitem{Shapere:2008zf}
A.~D. Shapere and Y.~Tachikawa, ``{Central charges of N=2 superconformal field
  theories in four dimensions},''
  \href{http://dx.doi.org/10.1088/1126-6708/2008/09/109}{{\em JHEP} {\bfseries
  09} (2008) 109},
\href{http://arxiv.org/abs/0804.1957}{{\ttfamily arXiv:0804.1957 [hep-th]}}.

\bibitem{Xie:2013jc}
D.~Xie and P.~Zhao, ``{Central charges and RG flow of strongly-coupled N=2
  theory},'' \href{http://dx.doi.org/10.1007/JHEP03(2013)006}{{\em JHEP}
  {\bfseries 03} (2013) 006},
\href{http://arxiv.org/abs/1301.0210}{{\ttfamily arXiv:1301.0210}}.

\bibitem{Shimizu:2017kzs}
H.~Shimizu, Y.~Tachikawa, and G.~Zafrir, ``{Anomaly matching on the Higgs
  branch},'' \href{http://dx.doi.org/10.1007/JHEP12(2017)127}{{\em JHEP}
  {\bfseries 12} (2017) 127},
\href{http://arxiv.org/abs/1703.01013}{{\ttfamily arXiv:1703.01013 [hep-th]}}.

\bibitem{Arakawa:2016ad}
T.~Arakawa and A.~Moreau, ``Sheets and associated varieties of affine vertex
  algebras,'' \href{http://arxiv.org/abs/1601.05906}{{\ttfamily 1601.05906}}.
  \url{http://arxiv.org/abs/1601.05906}.

\bibitem{Semikhatov:1993pr}
A.~M. Semikhatov, ``{The MFF singular vectors in topological conformal
  theories},'' \href{http://dx.doi.org/10.1142/S0217732394001738}{{\em Mod.
  Phys. Lett.} {\bfseries A9} (1994) 1867--1896},
  \href{http://arxiv.org/abs/hep-th/9311180}{{\ttfamily arXiv:hep-th/9311180
  [hep-th]}}.
[JETP Lett.58,860(1993)].

\bibitem{2004math1023A}
D.~{Adamovic}, ``{A construction of admissible $A_1^{(1)}$-modules of level
  $-{4/3}$},'' {\em arXiv Mathematics e-prints} (Jan, 2004) math/0401023,
  \href{http://arxiv.org/abs/math/0401023}{{\ttfamily arXiv:math/0401023
  [math.QA]}}.

\bibitem{Adamovic:2014lra}
D.~Adamovic, ``{A realization of certain modules for the $N=4$ superconformal
  algebra and the affine Lie algebra $A_2 ^{(1)}$},''
\href{http://arxiv.org/abs/1407.1527}{{\ttfamily arXiv:1407.1527 [math.QA]}}.

\bibitem{2017arXiv171111342A}
D.~{Adamovic}, ``{Realizations of simple affine vertex algebras and their
  modules: the cases $\widehat{sl(2)}$ and $\widehat{osp(1,2)}$},'' {\em arXiv
  e-prints} (Nov, 2017) arXiv:1711.11342,
  \href{http://arxiv.org/abs/1711.11342}{{\ttfamily arXiv:1711.11342
  [math.QA]}}.

\bibitem{Xie:2019}
P.~Shan, D.~Xie, and W.~Yan, ``{To appear},''.

\bibitem{Gaiotto:2008ak}
D.~Gaiotto and E.~Witten, ``{S-Duality of Boundary Conditions In N=4 Super
  Yang-Mills Theory},''
  \href{http://dx.doi.org/10.4310/ATMP.2009.v13.n3.a5}{{\em Adv. Theor. Math.
  Phys.} {\bfseries 13} no.~3, (2009) 721--896},
\href{http://arxiv.org/abs/0807.3720}{{\ttfamily arXiv:0807.3720 [hep-th]}}.

\end{thebibliography}\endgroup

\end{document}